\documentclass[fleqn,usenatbib]{mnras}

\usepackage[T1]{fontenc}

\DeclareRobustCommand{\VAN}[3]{#2}
\let\VANthebibliography\thebibliography
\def\thebibliography{\DeclareRobustCommand{\VAN}[3]{##3}\VANthebibliography}

\usepackage{graphicx}	
\usepackage{amsmath}	
\usepackage{amssymb}	
\usepackage{adjustbox}
\usepackage{verbatim}
\usepackage{multirow}
\usepackage{multicol}
\usepackage{pdflscape}
\usepackage[table,xcdraw]{xcolor}
\usepackage{placeins}
\usepackage{orcidlink}
\usepackage{makecell}

\usepackage{newtxtext,newtxmath}

\DeclareMathAlphabet{\mathsc}{OT1}{cmr}{m}{sc}
\def\testbx{bx}%
\DeclareRobustCommand{\ion}[2]{%
\relax\ifmmode
\ifx\testbx\f@series
{\mathbf{#1\,\mathsc{#2}}}\else
{\mathrm{#1\,\mathsc{#2}}}\fi
\else\textup{#1\,{\mdseries\textsc{#2}}}%
\fi}

\newcommand{\Heii}{\ion{He}{ii}}
\newcommand{\Oi}{[\ion{O}{i}]}
\newcommand{\Oiii}{[\ion{O}{iii}]}

\newcommand{\Nii}{[\ion{N}{ii}]}

\newcommand{\Fevii}{[\ion{Fe}{vii}]}

\newcommand{\Fex}{[\ion{Fe}{x}]}
\newcommand{\Fexi}{[\ion{Fe}{xi}]}

\newcommand{\Fexiv}{[\ion{Fe}{xiv}]}

\newcommand{\Neiii}{[\ion{Ne}{iii}]}

\newcommand{\msol}{\mbox{M$_{\odot}$}}

\newcommand{\kms}{\mbox{$\rm{km}\,s^{-1}$}}

\title[Long-term observations of ECLEs]{Long-term follow-up observations of extreme coronal line emitting galaxies}

\author[P.~Clark~et~al.]{Peter~Clark$^{1}$\thanks{E-mail:~icg.pclark@gmail.com}\orcidlink{0000-0002-6576-7400},
Or~Graur$^{1,2}$\orcidlink{0000-0002-4391-6137},
Joseph~Callow$^{1}$\orcidlink{0000-0002-0804-9533},
Jessica~Aguilar$^{3}$\orcidlink{0000-0001-6286-1744},
Steven~Ahlen$^{4}$\orcidlink{0000-0001-6098-7247},
Joseph~P.~Anderson$^{5,6}$\orcidlink{0000-0003-0227-3451},\newauthor
Edo~Berger$^{7}$\orcidlink{0000-0002-9392-9681},
Tom\'as~E.~M\"uller-Bravo$^{8,9}$\orcidlink{0000-0003-3939-7167},
Thomas~G.~Brink$^{10}$\orcidlink{0000-0001-5955-2502},
David~Brooks$^{11}$\orcidlink{0000-0002-8458-5047},
Ting-Wan~Chen$^{12,13}$\orcidlink{0000-0002-1066-6098},\newauthor
Todd~Claybaugh$^{3}$\orcidlink{0000-0002-5024-6987},
Axel~de la Macorra$^{14}$\orcidlink{0000-0002-1769-1640},
Peter~Doel$^{11}$\orcidlink{0000-0002-6397-4457},
Alexei~V.~Filippenko$^{10}$\orcidlink{0000-0003-3460-0103},\newauthor
Jamie~E.~Forero-Romero$^{15,16}$\orcidlink{0000-0002-2890-3725},
Sebastian~Gomez$^{17}$\orcidlink{0000-0001-6395-6702},
Mariusz~Gromadzki$^{18}$\orcidlink{0000-0002-1650-1518},
Klaus~Honscheid$^{19,20}$\orcidlink{0000-0002-6550-2023},\newauthor
Cosimo~Inserra$^{21}$\orcidlink{0000-0002-3968-4409},
Theodore~Kisner$^{3}$\orcidlink{0000-0003-3510-7134},
Martin~Landriau$^{3}$\orcidlink{0000-0003-1838-8528},
Lydia~Makrygianni$^{22}$\orcidlink{0000-0002-7466-4868},
Marc~Manera$^{23,24}$\orcidlink{0000-0003-4962-8934},\newauthor
Aaron~Meisner$^{25}$\orcidlink{0000-0002-1125-7384},
Ramon~Miquel$^{24,26}$\orcidlink{0000-0002-6610-4836},
John~Moustakas$^{27}$\orcidlink{0000-0002-2733-4559},
Matt~Nicholl$^{28}$\orcidlink{0000-0002-2555-3192},
Jundan~Nie$^{29}$\orcidlink{0000-0001-6590-8122},\newauthor
Francesca~Onori$^{30}$\orcidlink{0000-0001-6286-1744},
Antonella~Palmese$^{31}$\orcidlink{0000-0002-6011-0530},
Claire~Poppett$^{3,32}$\orcidlink{0000-0003-0512-5489},
Thomas~Reynolds$^{33,34}$\orcidlink{0000-0002-1022-6463},
Mehdi~Rezaie$^{35}$\orcidlink{0000-0001-5589-7116},\newauthor
Graziano~Rossi$^{36}$\orcidlink{0000-0002-5102-5019},
Eusebio~Sanchez$^{37}$\orcidlink{0000-0002-9646-8198},
Michael~Schubnell$^{38}$\orcidlink{0000-0001-9504-2059},
Gregory~Tarl\'{e}$^{38}$\orcidlink{0000-0003-1704-0781},
Benjamin~A.~Weaver$^{25}$,
\newauthor
Thomas~Wevers$^{5}$\orcidlink{0000-0002-4043-9400},
David~R.~Young$^{28}$\orcidlink{0000-0002-1229-2499},
WeiKang~Zheng$^{10}$\orcidlink{0000-0002-2636-6508}
and Zhimin~Zhou$^{29}$\orcidlink{0000-0002-4135-0977}
\\ \\
\parbox{\textwidth}{
$^{1}$Institute of Cosmology and Gravitation, University of Portsmouth, Portsmouth, PO1 3FX, UK \\
$^{2}$Department of Astrophysics, American Museum of Natural History, New York, NY 10024, USA \\
$^{3}$ Lawrence Berkeley National Laboratory, 1 Cyclotron Road, Berkeley, CA 94720, USA\\
$^{4}$ Physics Dept., Boston University, 590 Commonwealth Avenue, Boston, MA 02215, USA\\
$^{5}$ European Southern Observatory, Alonso de C\'ordova 3107, Casilla 19, Santiago, Chile\\
$^{6}$ Millennium Institute of Astrophysics MAS, Nuncio Monsenor Sotero Sanz 100, Off. 104, Providencia, Santiago, Chile\\
$^{7}$ Center for Astrophysics | Harvard \& Smithsonian, 60 Garden Street, Cambridge, MA 02138, USA\\
$^{8}$ Institute of Space Sciences (ICE, CSIC), Campus UAB, Carrer de Can Magrans, s/n, E-08193 Barcelona, Spain\\
$^{9}$ Institut d'Estudis Espacials de Catalunya (IEEC), E-08034 Barcelona, Spain\\
$^{10}$ Department of Astronomy, University of California, Berkeley, CA 94720-3411, USA\\
$^{11}$ Department of Physics \& Astronomy, University College London, Gower Street, London, WC1E 6BT, UK\\
$^{12}$ Technische Universit{\"a}t M{\"u}nchen, TUM School of Natural Sciences, Physik-Department, James-Franck-Stra{\ss}e 1, 85748 Garching, Germany\\
$^{13}$ Max-Planck-Institut f{\"u}r Astrophysik, Karl-Schwarzschild Stra{\ss}e 1, 85748 Garching, Germany\\
$^{14}$ Instituto de F\'{\i}sica, Universidad Nacional Aut\'{o}noma de M\'{e}xico,  Cd. de M\'{e}xico  C.P. 04510,  M\'{e}xico\\
$^{15}$ Departamento de F\'isica, Universidad de los Andes, Cra. 1 No. 18A-10, Edificio Ip, CP 111711, Bogot\'a, Colombia\\
$^{16}$ Observatorio Astron\'omico, Universidad de los Andes, Cra. 1 No. 18A-10, Edificio H, CP 111711 Bogot\'a, Colombia\\
$^{17}$ Space Telescope Science Institute, 3700 San Martin Dr, Baltimore, MD 21218, USA\\
$^{18}$ Astronomical Observatory, University of Warsaw, Al. Ujazdowskie 4, 00-478 Warszawa, Poland\\
$^{19}$ Center for Cosmology and AstroParticle Physics, The Ohio State University, 191 West Woodruff Avenue, Columbus, OH 43210, USA\\
$^{20}$ Department of Physics, The Ohio State University, 191 West Woodruff Avenue, Columbus, OH 43210, USA\\
$^{21}$ Cardiff Hub for Astrophysics Research and Technology, School of Physics \& Astronomy, Cardiff University, Queens Buildings, The Parade, Cardiff, CF24 3AA, UK\\
$^{22}$ The School of Physics and Astronomy, Tel Aviv University, Tel Aviv 69978, Israel\\
$^{23}$ Departament de F\'{i}sica, Universitat Aut\`{o}noma de Barcelona, 08193 Bellaterra (Barcelona), Spain\\
$^{24}$ Institut de F\'{i}sica d’Altes Energies (IFAE), The Barcelona Institute of Science and Technology, Campus UAB, 08193 Bellaterra Barcelona, Spain\\
$^{25}$ NSF's NOIRLab, 950 N. Cherry Ave., Tucson, AZ 85719, USA\\
$^{26}$ Instituci\'{o} Catalana de Recerca i Estudis Avan\c{c}ats, Passeig de Llu\'{\i}s Companys, 23, 08010 Barcelona, Spain\\
$^{27}$ Department of Physics and Astronomy, Siena College, 515 Loudon Road, Loudonville, NY 12211, USA\\
$^{28}$ Astrophysics Research Centre, School of Mathematics and Physics, Queens University Belfast, Belfast BT7 1NN, UK\\
$^{29}$ National Astronomical Observatories, Chinese Academy of Sciences, A20 Datun Rd., Chaoyang District, Beijing, 100012, P.R. China\\
$^{30}$ INAF-Osservatorio Astronomico d'Abruzzo, via M. Maggini snc, I-64100 Teramo, Italy\\
$^{31}$ Department of Physics, Carnegie Mellon University, 5000 Forbes Avenue, Pittsburgh, PA 15213, USA\\
$^{32}$ Space Sciences Laboratory, University of California, Berkeley, 7 Gauss Way, Berkeley, CA  94720, USA\\
$^{33}$Tuorla Observatory, Department of Physics and Astronomy, University of Turku, FI-20014 Turku, Finland\\
$^{34}$ Cosmic Dawn Center (DAWN), Niels Bohr Institute, University of Copenhagen, Jagtvej 128, DK-2200 København N, Denmark\\
$^{35}$ Department of Physics, Kansas State University, 116 Cardwell Hall, Manhattan, KS 66506, USA\\
$^{36}$ Department of Physics and Astronomy, Sejong University, Seoul, 143-747, Korea\\
$^{37}$ CIEMAT, Avenida Complutense 40, E-28040 Madrid, Spain\\
$^{38}$ Department of Physics, University of Michigan, Ann Arbor, MI 48109, USA\\
}
}

\date{Accepted XXX. Received YYY; in original form ZZZ}

\pubyear{2023}

\begin{document}
\label{firstpage}
\pagerange{\pageref{firstpage}--\pageref{lastpage}}
\maketitle

\clearpage

\begin{abstract}
We present new spectroscopic and photometric follow-up observations of the known sample of extreme coronal line emitting galaxies (ECLEs) identified in the Sloan Digital Sky Survey (SDSS). With these new data, observations of the ECLE sample now span a period of two decades following their initial SDSS detections. We confirm the non-recurrence of the iron coronal line signatures in five of the seven objects, further supporting their identification as the transient light echoes of tidal disruption events (TDEs). Photometric observations of these objects in optical bands show little overall evolution. In contrast, mid-infrared (MIR) observations show ongoing long-term declines consistent with power law decay. The remaining two objects had been classified as active galactic nuclei (AGN) with unusually strong coronal lines rather than being TDE related, given the persistence of the coronal lines in earlier follow-up spectra. We confirm this classification, with our spectra continuing to show the presence of strong, unchanged coronal-line features and AGN-like MIR colours and behaviour. We have constructed spectral templates of both subtypes of ECLE to aid in distinguishing the likely origin of newly discovered ECLEs. We highlight the need for higher cadence, and more rapid, follow-up observations of such objects to better constrain their properties and evolution. We also discuss the relationships between ECLEs, TDEs, and other identified transients having significant MIR variability.

\end{abstract}

\begin{keywords}
transients: tidal disruption events, galaxies: active 
\end{keywords}



\section{Introduction}
\label{Sec:Introduction}

Tidal disruption events (TDEs) are luminous flaring transients produced by the gravitational shredding of a star that passes too close to its galaxy's central supermassive black hole (SMBH). This leads to a portion of the star's mass being accreted onto the disrupting SMBH via an accretion disk, with the remaining material becoming unbound and ejected from the system. Whilst around half of the disrupted star's mass is initially gravitationally bound to the black hole following the disruption \citep{ulmer_1999_FlaresTidalDisruption}, the actual amount of material accreted is significantly less as more material becomes unbound as the event evolves \citep{ayal_2000_TidalDisruptionSolarType}. It is thought that either the circularisation of the accretion disk or collisions within the infalling material streams (or a combination of both) releases the energy observed as the flaring TDE (e.g., \citealt{lacy_1982_NatureCentralParsec, rees_1988_TidalDisruptionStars, phinney_1989_CosmicMergerMania, evans_1989_TidalDisruptionStar}), though the specifics of the processes are still under debate. In the case of non-rotating SMBHs, only those $<10^8$~\msol\ are expected to be responsible for producing TDEs, as at larger SMBH masses the Roche limit (the radius within which a star will be tidally disrupted) is within the event horizon and so the star is absorbed whole prior to disruption, thereby not producing a visible transient. For the more physically realistic situation of a rotating Kerr SMBH, this mass limit for visible disruptions is increased as black hole spin increases \citep{hills_1975_PossiblePowerSource, kesden_2012_TidaldisruptionRateStars}.

TDEs have been observed with a wide range of properties and have been detected through numerous methods across the electromagnetic spectrum. The first events were identified in the 1990s by X-ray surveys, at energies where overall TDE emission is predicted to peak \citep{bade_1996_DetectionExtremelySoft}. TDEs are now routinely detected by wide-field optical surveys. Examples of such surveys, from which we utilise data in this work, are the Asteroid Terrestrial-impact Last Alert System \citep[ATLAS;][]{tonry_2018_ATLASHighcadenceAllsky, smith_2020_DesignOperationATLAS}, Pan-STARRS1 \citep[PS1;][]{chambers_2016_PanSTARRS1Surveys}, and the Zwicky Transient Facility \citep[ZTF;][]{bellm_2019_ZwickyTransientFacility}. Subsequent follow-up observations have also detected TDEs at radio and infrared wavelengths --- for example, \cite{alexander_2017_RadioObservationsTidal} and \cite{dou_2017_DiscoveryMidinfraredEcho}, respectively.

A literature search reveals that upward of 100 TDE candidates have been identified (e.g., \citealt{hinkle_2021_SwiftFixNuclear, vanvelzen_2021_SeventeenTidalDisruption, charalampopoulos_2022_DetailedSpectroscopicStudy, hammerstein_2023_FinalSeasonReimagined}). However, given the wide range of properties observed, and the varied methods used in their detection, it is still debated whether all candidates identified so far are genuine TDEs or are the result of more than one kind of accretion activity onto an SMBH e.g., flares from temporary increases in the accretion rate of active galactic nuclei (AGN).

A small subset of TDE candidates have been identified from residual signatures in the spectra of their host galaxies. Nuclear spectra of these galaxies exhibit strong, narrow emission lines of ionic species more commonly associated with the high-temperature environment of the Solar corona, most notably emission lines produced by high-ionisation states of iron (\Fevii--\Fexiv). As a result, these objects have been termed `extreme coronal line emitters' or `extreme coronal line emitting galaxies' (ECLEs) \citep{wang_2012_EXTREMECORONALLINE}.

The first ECLE (SDSS~J095209.56+214313.3, which we refer to as SDSS~J0952) was identified by \cite{komossa_2008_DiscoverySuperstrongFading, komossa_2009_NTTSpitzerChandra}, who noted that the object changed in brightness and overall spectral energy distribution (SED) between photometric observations by the Sloan Digital Sky Survey \citep[SDSS;][]{york_2000_SloanDigitalSky} in 2004 and subsequent spectroscopic observations the following year. During this time, the object dimmed to be more consistent with that of near-infrared (NIR) photometry obtained in 1998 by the Two Micron All-Sky Survey \citep[2MASS;][]{skrutskie_2006_TwoMicronAll}, whilst displaying a continuum best described by a combination of underlying starlight and an additional power-law component. This spectrum also presented the strong emission lines of highly ionised Fe that subsequently became the hallmark spectral features for the identification of ECLEs. These Fe emission lines had both broad and narrow components and were accompanied by multipeaked Balmer emission lines.

Additionally, ultraviolet (UV) observations obtained two months after the SDSS spectrum by the \textit{Galaxy Evolution Explorer} \citep[\textit{GALEX};][]{martin_2005_GalaxyEvolutionExplorer} were found to be significantly brighter than would be expected from host-galaxy starlight alone yet consistent with an extrapolation of the power-law component identified in the continuum of the SDSS spectrum. Follow-up photometry and spectroscopy tracked a decline in luminosity across the electromagnetic spectrum and fading of the observed Fe coronal lines, with the higher ionisation state lines fading more quickly.

\cite{wang_2011_TransientSuperstrongCoronala} later identified a second similar object (SDSS~J074820.67+471214.3~:~SDSS~J0748). A systematic survey of the seventh data release of the SDSS \citep{abazajian_2009_SeventhDataRelease} conducted by \cite{wang_2012_EXTREMECORONALLINE} recovered five additional objects showing similar, though not identical, properties, bringing the total number of known ECLEs to seven. 

The connection between the appearance of the Fe coronal lines and TDE light echoes was first made by \cite{komossa_2008_DiscoverySuperstrongFading} through their observations of SDSS~J0952. The high ionisation potentials of the highly ionised states of Fe (358~eV for \Fexiv) require the presence of a soft X-ray continuum. Whilst the process that generates this X-ray continuum in a TDE remains somewhat unclear, modelling indicates the likely source is the resulting accretion disk after the material removed from the disrupted star has circularised around the SMBH \citep[e.g.,][]{hayasaki_2021_OriginLatetimeXRay}. This continuum may be obscured by the presence of dense circumnuclear material, which once ionised generates the observed coronal lines.

Several other possible explanations for ECLEs have been suggested, including a new form of AGN variability or an exotic form of supernova (SN). The TDE light-echo explanation for ECLEs has been supported by the long duration of the events. ECLEs have been seen to leave detectable emission-line signatures in their host spectra for at least several years post discovery and continue to display mid-infrared (MIR) evolution over the course of more than a decade, longer than would be expected of other forms of astrophysical transients, such as supernovae \citep[SNe;][]{palaversa_2016_REVEALINGNATUREEXTREME}. The spectroscopic and MIR photometric evolution of ECLEs are both less erratic and larger in amplitude than what is observed in most AGN variability, which is normally seen to be $\sim 0.1$~mag in amplitude on timescales of weeks to months \citep{hawkins_2002_VariabilityActiveGalactic}. 

Previously, the most clear connection between ECLEs and TDEs was the discovery spectrum of SDSS~J0748. This object was first observed with a broad, strong \Heii\ feature along with broad H$\alpha$ emission commonly associated with conventional, optically selected TDEs \citep{gromadzki_2017_EPESSTOTransientClassification}. Additionally, a further two objects (SDSS~J0952 and J1350) were also initially observed with clearly broad and complex H$\alpha$ emission features comprised of multiple components, with the broad components fading over time \citep{yang_2013_LONGTERMSPECTRALEVOLUTION}. Recently, the connection between ECLEs and the wider group of optically selected TDEs has become much more evident through observations of a small number of optically selected and spectroscopically confirmed TDEs that have developed coronal line emission features following their classification. We discuss these in more depth in Section~\ref{subsec:Coronal_Line_TDEs}.

The long duration of the ECLE spectroscopic signatures, as well as their occurrence not being limited to a specific type of galaxy \citep{graur_2018_DependenceTidalDisruptiona}, allows them to serve as a window into the long-term behaviour of the environments surrounding both active and quiescent black holes. This includes cases where the initial TDE was not directly observed --- the coronal line signatures of ECLEs can persist long after the TDE is no longer photometrically detectable and after any broad H or He features have faded \citep{onori_2022_NuclearTransient2017ggea}.

Despite the limited sample size, two spectroscopic subclasses of ECLE were suggested by \cite{wang_2012_EXTREMECORONALLINE}: those objects showing \Fevii\ emission features (4/7) and those without (3/7). Two scenarios for this were proposed, with either \Fevii\ being collisionally de-excited in some objects having a higher density of circumnuclear material (with the higher ionisation states of Fe not being affected owing to their significantly higher critical densities) or that the soft X-ray radiation field was of sufficient intensity to overionise the line-emitting material, preventing the formation of the \Fevii\ lines.

Follow-up spectra obtained by \cite{yang_2013_LONGTERMSPECTRALEVOLUTION} up to 9~yr after the initial SDSS observations found that four objects displayed significant evolution over this period, with the remaining three being spectroscopically non-variable. We note here that this classification does not divide the sample into the same two groups as the initial detection/non detection of \Fevii\ put forward by \cite{wang_2012_EXTREMECORONALLINE}. \cite{yang_2013_LONGTERMSPECTRALEVOLUTION} suggested that the three non-variable objects were the result of AGN activity or the tidal disruption of giant stars rather than the disruption of a main-sequence star as is usually suggested for observed TDEs. A subset of AGN have been observed with Fe coronal lines, though these lines in AGN are observed at lower intensities than in ECLEs \citep{nagao_2000_HighIonizationNuclearEmissionLine, komossa_2008_DiscoverySuperstrongFading}. ECLEs have also been observed with line ratios expected of star-forming galaxies rather than those of typical AGN. However, these ratios have been seen to shift to more AGN-like values as the ECLEs evolve.

X-ray observations of SDSS~J134244.41+053056.1 (SDSS~J1342), one of the original ECLEs, obtained with the \textit{Neil Gehrels Swift Observatory} \citep[\textit{Swift};][]{gehrels_2004_SwiftGammaRayBursta} and \textit{XMM-Newton} \citep{jansen_2001_XMMNewtonObservatorySpacecraft}, revealed a long-term decline consistent with the $t^{-5/3}$ power law expected from accretion events. The authors concluded that this object was consistent with a long-duration TDE by a $10^5$~\msol SMBH \citep{he_2021_LongtermXrayEvolution}.

Here we present new spectroscopic follow-up observations of all seven of the ECLEs in the \cite{wang_2012_EXTREMECORONALLINE} sample with the time between the initial observations and these new spectra now approaching two decades. Summary information for all seven objects in this work (including both the full and abbreviated names used along with their coordinates) is provided in Table~\ref{Tab:Object_Summary_Info}.

This paper is organised in the following manner. Section~\ref{Sec:ObservationsAndData} outlines the observations and reduction techniques. In Section~\ref{Sec:Analysis_and_Results} we detail the analysis of the new set of follow-up spectra, including updated Baldwin-Phillips-Terlevich \citep[BPT;][]{baldwin_1981_ClassificationParametersEmissionline} diagnostics for each object. We use the SDSS discovery spectra of the sample to construct template spectra of both the variable and non-variable ECLEs, and we use these to compare against other SDSS galaxy templates. Additionally, we present updated optical and MIR photometric analyses of the evolution of all ECLEs. Whilst there has been little overall evolution across the sample at optical wavelengths, the majority of ECLEs with variable coronal lines show ongoing MIR declines. In Section~\ref{Sec:Discussion}, we discuss the links between ECLEs and other types of transient identified with coronal lines. Finally, in Section~\ref{Sec:Conclusions}, we present a summary of our main findings. 

Throughout, we assume a Hubble-Lema\^ itre constant H$_0 = 73$ \kms\ Mpc$^{-1}$ and adopt a standard cosmological model with $\Omega_M=0.27$ and $\Omega_{\Lambda}=0.73$.

\section{Observations and Data Reduction}
\label{Sec:ObservationsAndData}

\subsection{Optical Spectroscopy}
\label{subsec:Optical_Spectroscopy}

We obtained optical spectra with a combination of the 6.5~m MMT \citep{blanco_2004_NewMMT} using the Binospec spectrograph \citep{fabricant_2019_BinospecWidefieldImaging}; the European Southern Observatory (ESO) 4~m New Technology Telescope (NTT) using the ESO Faint Object Spectrograph and Camera (EFOSC2) \citep{buzzoni_1984_ESOFaintObject} as part of the advanced Public ESO Spectroscopic Survey of Transient Objects \citep[ePESSTO+;][]{smartt_2015_PESSTOSurveyDescription}; the Shane 3~m telescope at Lick Observatory making use of the Kast Double Spectrograph \citep[Kast;][]{miller_1994_KASTDoubleSpectrograph}; and the Dark Energy Spectroscopic Instrument (DESI) mounted on the Mayall 4~m telescope \citep{desicollaboration_2016_DESIExperimentPart, desicollaboration_2016_DESIExperimentParta}.

The MMT spectrum was reduced using a He-Ne-Ar comparison lamp and flat-field taken immediately after the spectrum, and flux-calibrated using a standard star observed during the night. 

NTT + EFOSC2 spectra were obtained through the ePESSTO+ collaboration and reduced using a custom pipeline, applying bias-subtraction, flat-fielding, wavelength and flux calibration, and telluric correction, as described by \cite{smartt_2015_PESSTOSurveyDescription}.

The Kast observations utilised a $2''$-wide slit, 600/4310 grism, and 300/7500 grating. This configuration has a combined wavelength range of $\sim 3500$--10,500~\AA, and a spectral resolving power of $R \approx 800$. To minimise slit losses caused by atmospheric dispersion \citep{filippenko_1982_ImportanceAtmosphericDifferential}, the spectra were acquired with the slit oriented at or near the parallactic angle and were reduced following standard techniques for CCD processing and spectrum extraction \citep{silverman_2012_BerkeleySupernovaIaa} utilising IRAF \citep{tody_1986_IRAFDataReduction} routines and custom Python and IDL codes.\footnote{\url{https://github.com/ishivvers/TheKastShiv}} Low-order polynomial fits to comparison-lamp spectra were wavelength calibration, and small adjustments derived from night-sky lines in the target frames were applied. The spectra were flux-calibrated using observations of appropriate spectrophotometric standard stars observed on the same night, at similar airmasses, and with an identical instrument configuration.

The DESI spectrum of SDSS~J1342 was obtained as part of survey validation \citep{desicollaboration_2023_ValidationScientificProgram, desicollaboration_2023_EarlyDataRelease}, whilst those of SDSS~J0938 and J0952 were taken as part of the bright galaxy survey (BGS) during main survey operations \citep{hahn_2023_DESIBrightGalaxy}. All DESI spectra were processed by the custom DESI spectroscopic pipeline, which includes a full suite of processing and correction steps to provide fully flux- and wavelength-calibrated spectra \citep{guy_2023_SpectroscopicDataProcessing}.

These new observations were combined with spectra previously analysed by \cite{wang_2012_EXTREMECORONALLINE} and \cite{yang_2013_LONGTERMSPECTRALEVOLUTION} obtained as part of the SDSS and with the MMT respectively.

The details of the full set of optical spectra obtained for all ECLE's explored in this work are given in Appendix Table~\ref{Tab:Spectra_Info}.

As these spectra were obtained over long durations from multiple instruments, there are a few caveats to be aware of. The SDSS and DESI spectra were obtained via fibres placed on the nuclei of the galaxies, whereas the MMT, NTT, and Shane 3~m telescopes obtained long-slit spectra. Furthermore, the SDSS fibres had diameters of $3''$, while DESI fibres are smaller at $1.5''$ in diameter \citep{kent_2016_ImpactOpticalDistortions}. Consequently, DESI spectra will contain less light from the outer regions of the host galaxies despite being centred on the same location. This may act to introduce artificial changes in line fluxes and ratios depending on the line-emitting regions included or excluded by the fibres. The same is true for the long-slit spectra that have been extracted using apertures smaller in area than the SDSS fiber spectra. As described by \cite{yang_2013_LONGTERMSPECTRALEVOLUTION}, this will primarily affect starlight contributions and low-ionisation (narrow) lines from any extended star-forming regions rather than the centrally located coronal lines.

The varying resolutions of instruments (in particular the lower resolution of the NTT and Kast spectra) also leads to the artificial broadening of narrow lines which must be considered when making comparisons between the spectra.

\subsection{Optical Photometry}
\label{subsec:Optical_Photometry}

Whilst there has been no dedicated long-term photometric follow-up program of the ECLEs, all-sky surveys provide an opportunity to obtain repeated coincidental observations across multiple filters and over an extended period of time. We have explored observations of our sample obtained by the ATLAS, Catalina Real-Time Transient Survey \citep[CRTS;][]{drake_2009_FIRSTRESULTSCATALINA}, PS1, and ZTF sky surveys. 

Throughout this work, unless stated otherwise, apparent magnitudes are listed as observed, with no additional corrections. Wherever we note absolute magnitudes, a correction for Milky Way extinction has been applied using the appropriate photometric extinction coefficient. Unless specified otherwise, coefficients have been retrieved from \cite{schlafly_2011_MeasuringReddeningSloan}. To match the preferred extinction parameters of \cite{schlafly_2011_MeasuringReddeningSloan}, we apply the extinction law of \cite{fitzpatrick_1999_CorrectingEffectsInterstellar} throughout and assume $R_{V} = 3.1$. A summary of all photometric datasets used here is provided in Table~\ref{Tab:Photom_Basic}. 

The photometry for all optical sky-surveys has been post-processed in a similar way prior to inclusion in this work, with any exceptions detailed in the specific text for that survey. Following retrieval and conversion to absolute magnitudes, the data were processed on a per filter basis by fist applying a rolling three sigma clipping cut to remove any likely spurious data-points suffering from observational artefacts. Additionally, given the large multi-year timescales over which each ECLE has been observed, we bin the photometry on a 14 day cadence to focus on any long-term trends rather than stochastic variability between closely spaced observations.

The ATLAS data were retrieved using the ATLAS forced-photometry server \citep{shingles_2021_ReleaseATLASForced}.\footnote{\url{https://fallingstar-data.com/forcedphot/}} Observations were made using the ATLAS broad-band filters `cyan' (\textit{c}; approximately equivalent to \textit{g} + \textit{r}) and `orange' (\textit{o}; approximately equivalent to \textit{r} + \textit{i}). ATLAS observations are available in the range MJD 57230--60277. Whilst the forced-photometry server can provide template-subtracted difference photometry, we do not make use of this option, instead using the direct source photometry to allow for like-to-like comparisons between photometry from other sources for which difference photometry is not available. As ATLAS-specific photometric extinction coefficients are unavailable, the photometry has been corrected using a mean of the corresponding SDSS filter pairs that cover the same approximate filter range of the ATLAS broad-band filters. Additionally, as the forced-photometry server provides the raw flux information for each of the observations, the post-processing sigma clipping and stacking was conducted in flux rather than magnitude space.

The CRTS dataset was compiled from the second public data release,\footnote{\url{http://nesssi.cacr.caltech.edu/DataRelease/}} and consists of CRTS \textit{V}-band observations, covering an MJD range of 53464--56656 \citep{drake_2009_FIRSTRESULTSCATALINA}. 

PS1 observations were retrieved from the second public data release \citep{flewelling_2018_PanSTARRSDataRelease} available through the Mikulski Archive for Space Telescopes (MAST)\footnote{\url{https://archive.stsci.edu/}} across all available filters (\textit{grizy}) and cover an MJD range of 54996--57009. Due to the small number of PS1 observations, no sigma clipping or stacking has been applied.

ZTF observations were made using the \textit{gri} filters, and retrieved from the nineteenth public data release accessed through the NASA/IPAC infrared science archive (IRSA).\footnote{\url{https://irsa.ipac.caltech.edu/}} These observations cover an MJD range of 58198--60132 \citep{masci_2018_ZwickyTransientFacility}. 

In addition to the datasets available for the full ECLE sample, we make use of Lincoln Near-Earth Asteroid Program \citep[LINEAR;][]{stokes_2000_LincolnNearEarthAsteroida} observations of SDSS~J0952 first published by \cite{palaversa_2016_REVEALINGNATUREEXTREME}. These data were obtained without a specific photometric filter, with the instrument's response function covering the approximate range of the SDSS \textit{griz} filters. 

\subsection{Infrared Photometry}
\label{subsec:IR_Photometry}

To explore the behaviour of each ECLE well before its initial outburst, we retrieve near-infrared (NIR) \textit{JHK} photometry obtained by the Two Micron All-Sky Survey \citep[2MASS;][]{skrutskie_2006_TwoMicronAll} from IRSA. This analysis is described in Section~\ref{subsec:2MASS_Analysis}.

In a similar manner to \cite{dou_2016_LONGFADINGMIDINFRARED}, we also retrieve the available MIR photometry obtained by the \textit{Wide-field Infrared Survey Explorer} (\textit{WISE}), from both the ALLWISE \citep{wright_2010_WIDEFIELDINFRAREDSURVEY} and NEOWISE Reactivation Releases \citep[NEOWISE-R;][]{mainzer_2011_NEOWISEOBSERVATIONSNEAREARTH, mainzer_2014_INITIALPERFORMANCENEOWISE} from IRSA. Specifically, ALLWISE data was retrieved from the `AllWISE Multiepoch Photometry Table' and NEOWISE-R data from the corresponding `NEOWISE-R Single Exposure (L1b) Source Table'.
Given the time between this work and that of \cite{dou_2016_LONGFADINGMIDINFRARED}, an additional $\sim 6$~yr of NEOWISE-R data are available, providing a means to further explore the long-term evolution in the \textit{W1} and \textit{W2} bands. The start of \textit{WISE} observations is several years following the initial spectral observations of the ECLEs and so cannot be used to infer their early-time behaviour.

As \textit{WISE} obtains several images of each object during each observing phase (once every six months), we process the observations using a custom Python script. This script filters out any observation marked as an upper limit, was observed when the spacecraft was close to the South Atlantic Anomaly (saa\textunderscore sep $< 5.0$) or the sky position of the Moon. Additionally, any observation with a low frame quality or that suffered from potential `contamination or confusion' as flagged by the \textit{WISE} pipeline was also removed, with the exception of \textit{W1} observations flagged as potentially contaminated, but not dominated by, a nearby bright star halo. This choice was made to prevent the removal of the vast majority of \textit{W1} observations of SDSS~J1350, which visual inspection showed to be unlikely to be significantly affected by the presence of a nearby star. A weighted average is then used to provide a single magnitude value per filter for each observation block. \cite{dou_2016_LONGFADINGMIDINFRARED} previously explored whether the variable ECLEs displayed variability during each observation block, with no such variability being detected. As such, combining the individual observations allows for any long-term trends to be seen more easily. \textit{WISE} magnitudes are presented in the Vega magnitude system as reported in the stated databases.

\begin{table}
\centering
\caption{Details of the photometric datasets used in this work.}
\begin{adjustbox}{width=1\columnwidth}
\begin{tabular}{llll}
\hline
Survey & Filters & MJD Range & Reference\\ \hline
\textbf{Optical} & & &\\
ATLAS & \textit{c \& o}$^1$ & 57230--60277 &\cite{tonry_2018_ATLASHighcadenceAllsky}\\
CRTS & \textit{V}$^2$ & 53464--56656 &\cite{drake_2009_FIRSTRESULTSCATALINA} \\
LINEAR & Clear$^3$ & 52614--54613 &\cite{palaversa_2016_REVEALINGNATUREEXTREME} \\
PS1 & \textit{g, r, i, z, y} & 54996--57009 &\cite{chambers_2016_PanSTARRS1Surveys}\\
ZTF & \textit{g, r, i} & 58198--59889 &\cite{bellm_2019_ZwickyTransientFacility}\\
\hline
\textbf{Infrared} & & &\\
2MASS & \textit{J,H,K} & 50836--51928 &\cite{skrutskie_2006_TwoMicronAll}\\
WISE & \textit{W1, W2, W3} & 55204--55573 &\cite{wright_2010_WIDEFIELDINFRAREDSURVEY}\\
NEOWISE & \textit{W1, W2} & 56663--59926 &\cite{mainzer_2014_INITIALPERFORMANCENEOWISE}\\
\hline
\end{tabular}
\end{adjustbox}
\begin{flushleft}
$^1$ATLAS observations were made using two broad-band filters; \textit{c} (cyan) is approximately equivalent to \textit{g} + \textit{r} and \textit{o} (orange) is roughly \textit{r} + \textit{i}. \\
$^2$CRTS observations were made with an unfiltered optical CCD and calibrated to an approximation of the \textit{V} band. \\
$^3$LINEAR observations are available only for SDSS~J0952 and were made with an unfiltered optical CCD with a response covering the approximate range of the SDSS \textit{griz} filters. \\
\textbf{Note}: Where possible, data for each object have been retrieved across all filters and surveys.
\end{flushleft}
\label{Tab:Photom_Basic}
\end{table}

\raggedbottom

\subsection{Search for Additional Transient Activity}
\label{subsec:TransientSearch}

As well as retrieving archival photometric data, we performed a search of the Transient Name Server (TNS)\footnote{\url{https://www.wis-tns.org/}} at the coordinates of each ECLE to confirm that no other survey (i.e., those whose data are not explored in detail here) had reported new transient activity of any of the ECLEs over the last several years. No such reports were found for the five TDE-related ECLEs. The lack of such reports supports the assumption that members of the variable subclass of ECLE are produced by a single-epoch event, rather than a recurring process.

One report was located at the position of SDSS~J1055 - AT~2023jke by ZTF \citep{fremling_2023_ZTFTransientDiscovery}. Whilst this newly reported transient lacks any spectroscopic follow-up, given the nature of the host galaxy we attribute it to AGN variability.

\section{Analysis and Results}
\label{Sec:Analysis_and_Results}

\subsection{Overall Optical Spectral Evolution}
\label{subsec:Overall_Spectral_Evolution}

We now explore the observed spectroscopic evolution of each ECLE in turn. In all of the following figures, the spectra are shown with the earliest at the top of the plot with progressively more recent spectra displayed below. All spectra are colour coded based on the telescope and instrument with which they were obtained. The spectroscopic sequences of the ECLE sample are shown in Figure~\ref{fig:ECLE_Evolution_General_0}.

\subsubsection{SDSS~J0748}
The initial 2004 SDSS spectrum of SDSS~J0748 displayed \Fex\--\Fexiv\ emission lines along with broad \Heii\ and Balmer lines that typify the H+He subclass of optically selected, active TDEs \citep{arcavi_2014_CONTINUUMHeRICHTIDAL}. All of these features had faded prior to the 2011 \cite{yang_2013_LONGTERMSPECTRALEVOLUTION} MMT spectrum and are also absent in our 2019 MMT spectrum. The spectral shapes of both MMT spectra are consistent. This indicates that the initial flaring event was a single epoch rather than a recurring transient, with the optical spectrum having now most likely returned to a quiescent state.

\subsubsection{SDSS~J0938}
SDSS~J0938 was reclassified by \cite{yang_2013_LONGTERMSPECTRALEVOLUTION} as a Seyfert 2 AGN with star-forming regions rather than being related to a transient TDE. This reclassification was based on the lack of variability in the coronal emission lines between the original 2006 SDSS spectrum and their 2011 MMT follow-up spectrum. Our 2021 NTT and 2022 DESI spectra show no detectable changes in any of the coronal lines (beyond the expected width changes as a result of instrumental resolution) or in overall spectral shape. Based on these findings, we concur with this AGN classification. However, the processes involved in generating such strong coronal lines relative to the rest of the AGN population over timescales of at least two decades are still unclear. 

\subsubsection{SDSS~J0952}
Between the 2005 SDSS spectrum and the 2011 \cite{yang_2013_LONGTERMSPECTRALEVOLUTION} MMT spectrum, the Fe coronal lines displayed by SDSS~J0952 faded significantly though remained detectable. These features have continued to fade and are no longer present in either our 2021 NTT or 2022 DESI spectra. A broad H$\alpha$ component was also seen in the initial SDSS spectrum which, like the Fe lines, had faded between the SDSS and MMT follow-up spectra. Whilst challenging observing conditions mean the NTT spectrum has a low signal-to-noise ratio (S/N), the most recent DESI spectrum confirms the presence of only narrow H$\alpha$. Likewise, a narrow \Heii\ feature was visible in the initial SDSS spectrum but is absent from the follow-up spectra.

\subsubsection{SDSS~J1055}
In a similar manner to SDSS~J0938, SDSS~J1055 was reclassified by \cite{yang_2013_LONGTERMSPECTRALEVOLUTION} as a Seyfert 1 AGN based on its spectral invariance between 2002 and 2011. Our 2021 Kast spectrum confirms this lack of evolution, supporting the AGN reclassification.

\subsubsection{SDSS~J1241}
This object was originally identified by \cite{yang_2013_LONGTERMSPECTRALEVOLUTION} as non-variable. In their analysis they lacked the red component of the spectrum owing to observational issues, with the blue component showing that the \Fevii~3759~\AA\ and \Neiii~3896~\AA\ emission lines remained prominent and that there were no significant changes to the continuum or overall spectral shape within the blue region of the spectrum. Our follow-up spectrum of SDSS~J1241 covers the full range of the original SDSS observation and reveals that the object has in fact displayed spectral variability consistent with the other variable ECLEs. Specifically, the coronal lines can now be seen to have faded, with none detected in our 2021 Kast spectrum. \Fevii\ lines have been seen to persist or develop with time in other ECLEs relative to the other Fe coronal lines, so it is possible that the higher ionisation lines had faded at the time of the 2011 MMT spectrum, though this is not possible to confirm. Whilst the lower resolution of the Kast spectrum makes it difficult to confirm, the \Neiii~3896~\AA\ emission line also appears to have reduced in strength significantly compared to both the 2004 SDSS and 2011 MMT spectra. 

\subsubsection{SDSS~J1342}
The initial 2002 SDSS spectrum of SDSS~J1342 displayed \Fex, \Fexi, and \Fexiv\ but lacked any \Fevii\ lines. By the time of the MMT spectrum in 2011, the higher ionisation lines were no longer detectable, but \Fevii\ lines were now clearly observable. The higher-resolution 2022 DESI spectrum reveals a persistence of the \Fevii\ emission features first seen in the 2011 MMT follow-up spectrum, with no indication of recurrence of the higher ionisation state lines. The NTT spectrum of SDSS~J1342 obtained around one month after the DESI spectrum does display some apparent \Fevii\ coronal emission features, though this spectrum is of too low resolution and S/N for any additional confirmation. This highlights the necessity of high-S/N and high-resolution follow-up spectra to fully capture the evolution of ECLEs.  
This object is most interesting for the very large increase in the line flux of \Oiii~$\lambda\lambda$4959, 5007 observed in both the DESI and NTT spectra. Whilst \cite{yang_2013_LONGTERMSPECTRALEVOLUTION} note the increase in \Oiii\ emission strength in all four of the ECLEs they identify as variable between the intial SDSS spectra and their 2011 MMT spectra, the increase displayed by SDSS~J1342 after 2011 is much more extreme, and unique among the current ECLE sample. We discuss this further in Section~\ref{subsec:BPT_Diagnostics}.

\subsubsection{SDSS~J1350}
SDSS~J1350 initially exhibited \Fex--\Fexiv\ emission lines which faded between the 2006 SDSS spectrum and the follow-up 2011 MMT spectrum obtained by \cite{yang_2013_LONGTERMSPECTRALEVOLUTION}, with \Fevii\ emission lines developing over the same period. Like the higher ionisation state lines before them, these \Fevii\ lines have now faded; with no remaining coronal emission present in our 2021 NTT follow-up spectrum, with the possible exception of a low-S/N \Fexi\ feature. Given its low S/N and the lack of lower ionisation state lines, we do not claim its detection.

\begin{figure*}
    \centering
    \includegraphics[width=0.94\textwidth]{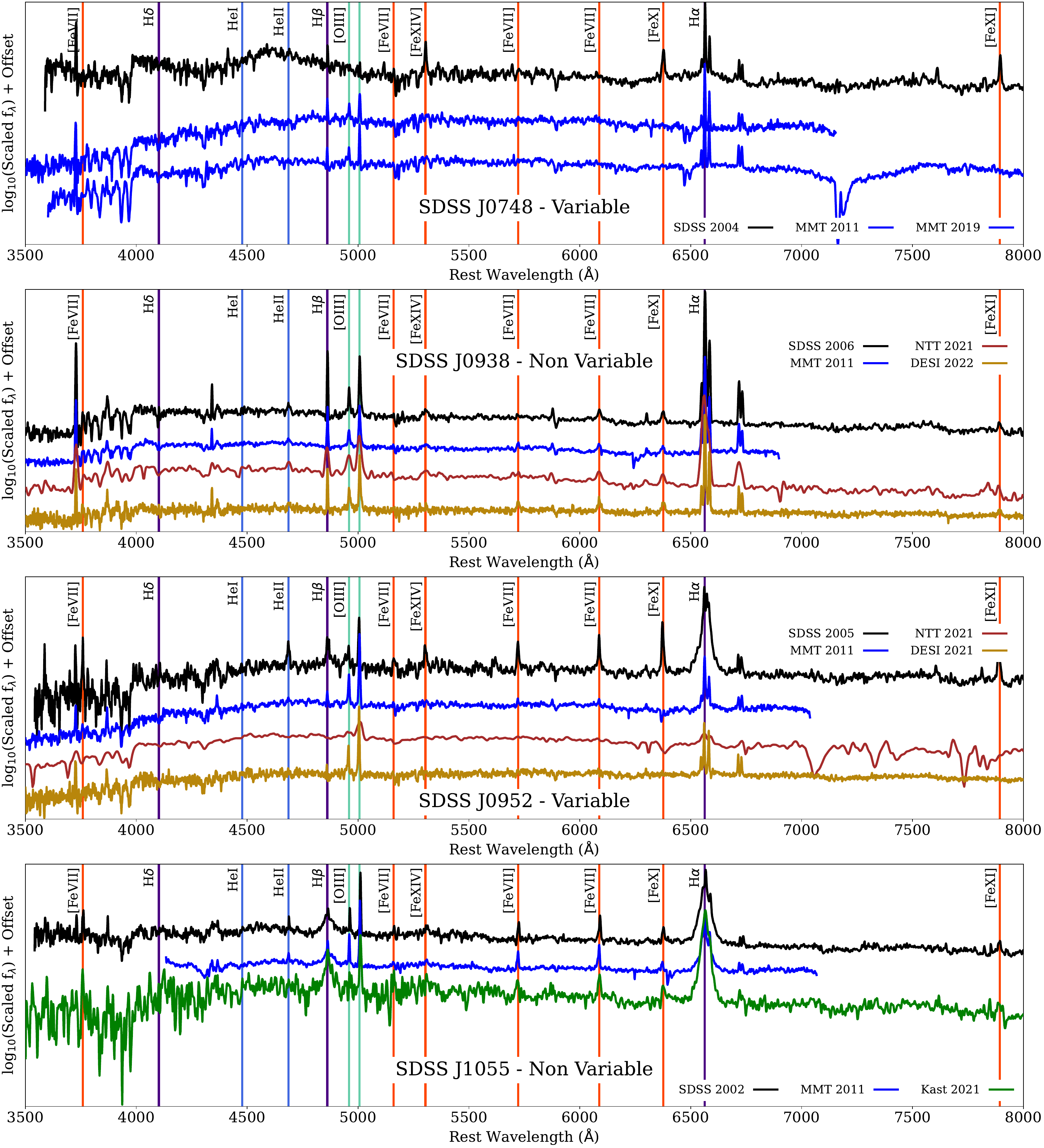}
    \caption{Spectral sequences showing the original SDSS spectra (black) for each ECLE along with the corresponding MMT follow-up spectrum obtained by (\citealt{yang_2013_LONGTERMSPECTRALEVOLUTION}; blue) and the new follow-up spectra obtained through this work (other colours depending on source). Objects are ordered based on their SDSS identification. Labels indicate whether each object has shown spectral variability - in particular of the coronal Fe lines - over the observation period. All spectra are presented following mild Gaussian smoothing ($\sigma = 1$) to improve visual clarity.
    }
    \label{fig:ECLE_Evolution_General_0}
\end{figure*}

\begin{figure*}
    \centering
    \includegraphics[width=0.94\textwidth]{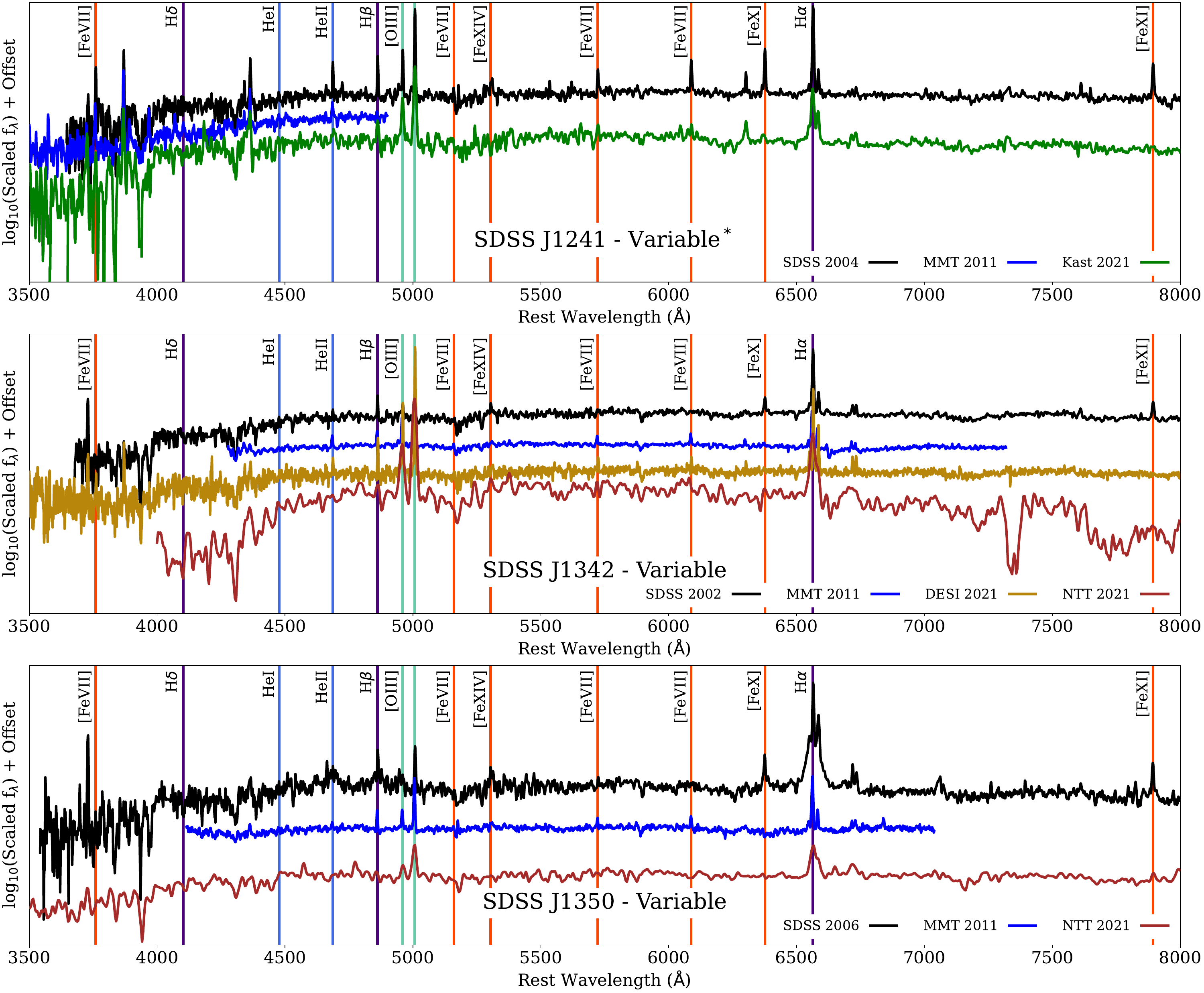}
    \contcaption{Spectral sequences showing the original SDSS spectra (black) for each ECLE along with the corresponding MMT follow-up spectrum obtained by (\citealt{yang_2013_LONGTERMSPECTRALEVOLUTION}; blue) and the new follow-up spectra obtained through this work (other colours depending on source). Objects are ordered based on their SDSS identification. Labels indicate whether each object has shown spectral variability - in particular of the coronal Fe lines - over the observation period. Note that SDSS~J1241 was originally classified as non-variable by \cite{yang_2013_LONGTERMSPECTRALEVOLUTION} but shown to be variable by our observations. All spectra are presented following mild Gaussian smoothing ($\sigma = 1$) to improve visual clarity.
    }
    \label{fig:ECLE_Evolution_General_1}
\end{figure*}

\subsection{BPT Diagnostics and Narrow Line Analysis}
\label{subsec:BPT_Diagnostics}

In the original SDSS observations analysed by \cite{wang_2012_EXTREMECORONALLINE}, the ECLEs were seen to display emission-line intensity ratios consistent with star-forming galaxies and did not meet the diagnostic thresholds of AGN activity when plotted on the usual set of BPT line-diagnostic diagrams \citep{baldwin_1981_ClassificationParametersEmissionline, veilleux_1987_SpectralClassificationEmissionLine, kewley_2001_TheoreticalModelingStarburst, kauffmann_2003_HostGalaxiesActive}. As the objects evolved, their emission-line ratios were seen to change over time. Follow-up observations by \cite{yang_2013_LONGTERMSPECTRALEVOLUTION} revealed a tendency for these ratios to drift to values more indicative of AGN. They note that this evolution is largely due to a combination of the increasing \Oiii\ line strength and an observational effect resulting from the smaller aperture size of the MMT spectra obtained by \cite{yang_2013_LONGTERMSPECTRALEVOLUTION} compared to the original SDSS observations --- that is, the spectra were more restricted to the nuclear region with a reduction in the light obtained from more distant star-forming regions.

\begin{figure*}
    \centering
    \includegraphics[width=\textwidth]{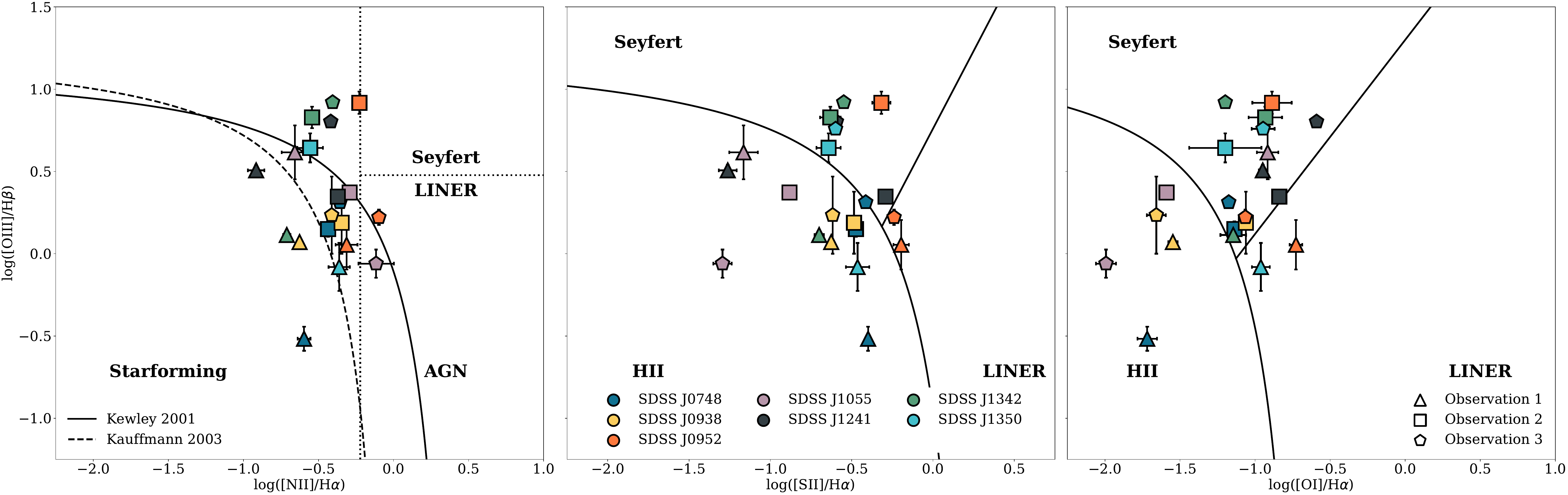}
    \caption{Comparison of the AGN line diagnostic results for each object between the original SDSS spectra (shown by triangles), the 2011 MMT spectra (shown by squares), and our new set of follow-up spectra (shown by pentagons). Those ECLEs with non-variable coronal lines can be seen to display only small changes in measured line ratios, whilst those with variable coronal lines are observed to have much larger changes in their measured ratios tending to evolve into the AGN regions.}
    \label{fig:BPT_ECLE_Evolution}
\end{figure*}

We retrieve the data used to construct these BPT diagrams from Table~3 of \cite{wang_2012_EXTREMECORONALLINE} and Table~2 of \cite{yang_2013_LONGTERMSPECTRALEVOLUTION} to produce a full comparison of the behaviour of the ECLE sample given in Figure~\ref{fig:BPT_ECLE_Evolution}. Table~3 of \cite{wang_2012_EXTREMECORONALLINE} does not include flux measurements for the diagnostic \Oi\ line which we measure here. Likewise, \cite{yang_2013_LONGTERMSPECTRALEVOLUTION} do not include line fluxes for the two objects with non-variable coronal lines, which we include using our own measurements of the 2013 MMT spectral dataset. Furthermore, as the lower resolution of the NTT spectra obtained using the EFOSC2 instrument makes accurate emission-line flux measurements very difficult, we opt to use only the higher resolution spectra from Kast and DESI, with the exception of SDSS~J1350, for which only an NTT spectrum is available.

While the non-variable objects show some changes in the measured line ratios between observations --- likely the result of differences in the exact regions of the host galaxy explored in each observation and measurement differences introduced by the varying resolutions of the instruments --- they remain within the star-forming or composite region in each set of spectra. We further explore the spectral evolution of the objects by showing a comparison of each of these line regions in Figure~\ref{fig:AGN_Line_Diagnostic_Zoom_General_0}. As with the full spectral comparisons previously presented, each spectrum has been scaled to have the same mean flux density in the range 5925--6000~\AA\ as the original SDSS spectrum; however, in this case the spectra are directly over-plotted to highlight relative changes rather than offset to display an evolutionary sequence.

\begin{figure*}
    \centering
    \includegraphics[width=0.93\textwidth]{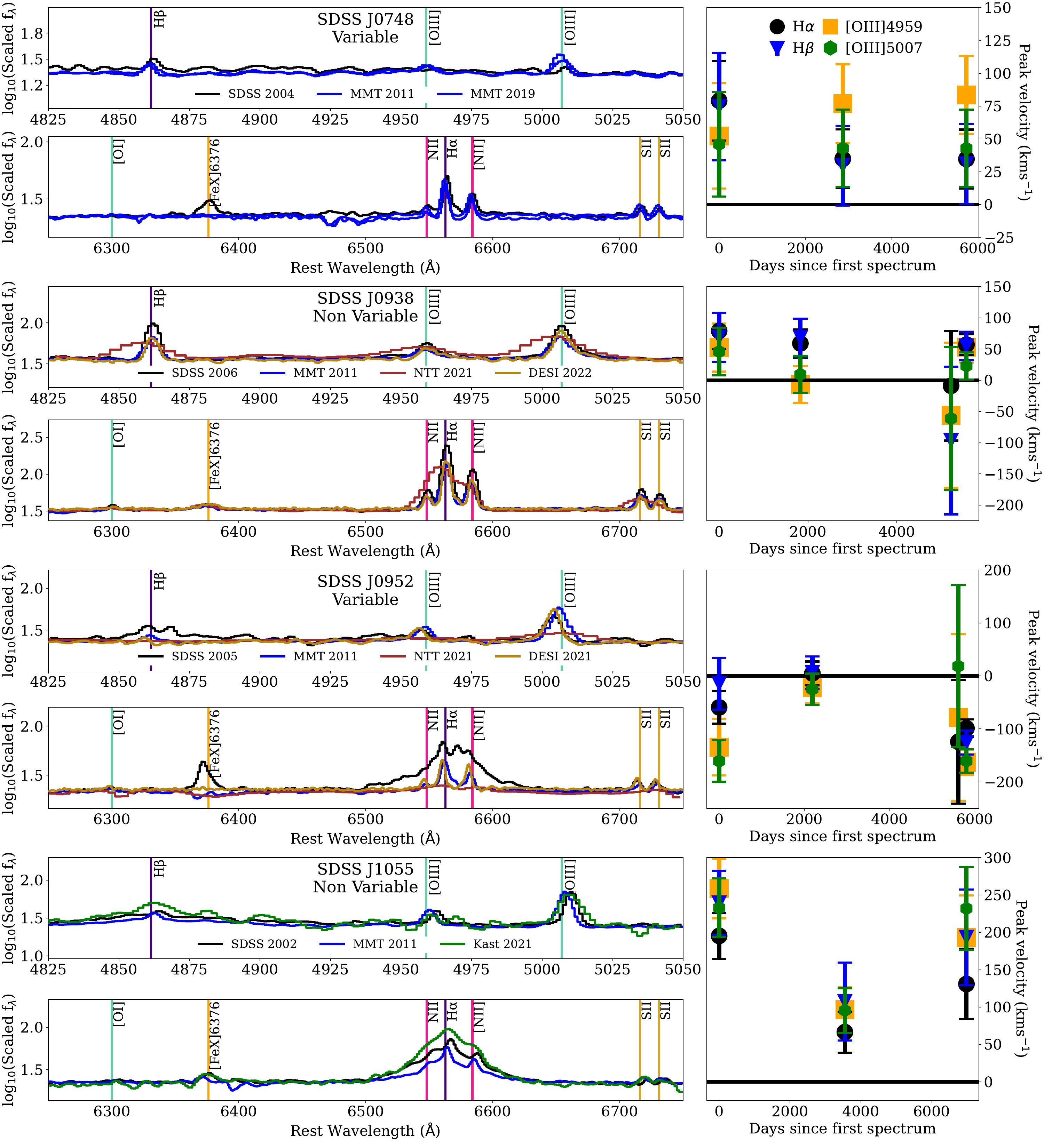}
    \caption{\textit{Left:} Focused plots of the AGN diagnostic lines of the ECLEs. Objects are presented in the same order as Figure~\ref{fig:ECLE_Evolution_General_0} with the colours used for each spectral source also matching this previous figure. Note: The Kast and NTT spectra are of significantly lower resolution than the SDSS, MMT, and DESI spectra. The flux density of each spectrum has been scaled to match in the featureless region 5925--6000~\AA, and has been presented on a log scale to better show the range of line strengths. All spectra are presented following mild Gaussian smoothing ($\sigma = 1$) to improve visual clarity. \textit{Right:} Evolution of the peak velocity for H$\alpha$, H$\beta$, \Oiii~$\lambda\lambda$4959, 5007 lines. NB: Velocity measurements made prior to smoothing.}
    \label{fig:AGN_Line_Diagnostic_Zoom_General_0}
\end{figure*}

\begin{figure*}
    \centering
    \includegraphics[width=0.93\textwidth]{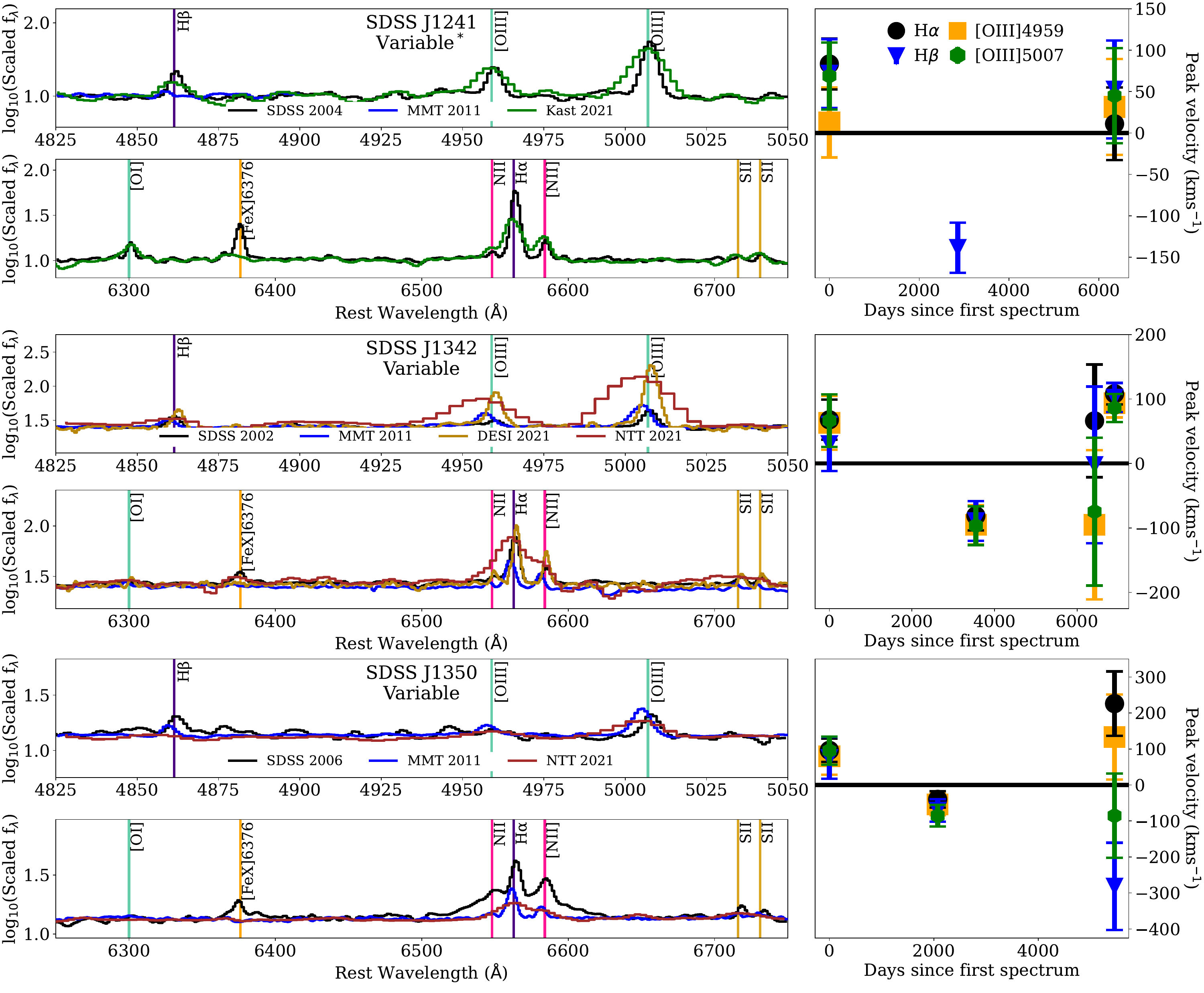}
    \contcaption{Focused plot of the AGN diagnostic lines of the ECLEs. Objects are presented in the same order as Figure~\ref{fig:ECLE_Evolution_General_0} with the colours used for each spectral source also matching this previous figure. Note: The Kast and NTT spectra are of lower resolution than the SDSS, MMT, and DESI spectra. The flux density of each spectrum has been scaled to match in the featureless region 5925--6000~\AA, and has been presented on a log scale to better show the range of line strengths. All spectra are presented following mild Gaussian smoothing ($\sigma = 1$) to improve visual clarity. \textit{Right:} Evolution of the peak velocity for H$\alpha$, H$\beta$, \Oiii~$\lambda\lambda$4959, 5007. NB: Velocity measurements made prior to smoothing.}
    \label{fig:AGN_Line_Diagnostic_Zoom_General_1}
\end{figure*}

\cite{yang_2013_LONGTERMSPECTRALEVOLUTION} observed strengthening of the \Oiii\ lines in all of their variable objects, with a proportion of this strengthening attributed to the more nuclear-focused nature of their spectra compared to the original SDSS observations. These smaller spectral footprints reduced the contribution of starlight from the outskirts of each galaxy, increasing the proportion of the spectrum contributed by the narrow-line region. Between the spectra obtained by \cite{yang_2013_LONGTERMSPECTRALEVOLUTION} in 2011 and our recent spectra, continued strengthening of \Oiii\ line emission (relative to the continuum flux) is observed in two objects: SDSS~J0748, and most significantly SDSS~J1342. The sharp increase in \Oiii~$\lambda$5007 emission strength in SDSS~J1342 (it is now the spectrum's dominant feature) may be the result of either the TDE triggering AGN activity by increasing the accretion rate onto the SMBH in the form of a `turn-on' event (e.g., \citealt{gezari_2017_IPTFDiscoveryRapid}), or due to the delayed response of more distant low-density gas to the TDE flare as proposed by \cite{yang_2013_LONGTERMSPECTRALEVOLUTION}. Balmer emission has also increased in strength, though with no associated higher Doppler broadened line velocities which has been seen in other such events. Further observations will be required to determine if this behaviour is a temporary change, or indicative of a more permanent alteration in the behaviour of the galaxy's SMBH. The low S/N ratio of the NTT spectrum of SDSS~J0952 makes line measurements impossible, but the DESI spectrum obtained at approximately the same time shows the relative strength of the \Oiii, Balmer, and \Nii\ lines to be largely unchanged when compared to the 2011 MMT spectrum.

Several of the objects appear to display changes in line velocities over time (see the right panels of Fig~\ref{fig:AGN_Line_Diagnostic_Zoom_General_0}). Whilst changes in the full width at half-maximum line velocities are dominated by the range of resolutions (as determined using the quoted $R$ values -- means where appropriate -- in Table~\ref{Tab:Spectra_Info}), changes to the peak velocity of some lines is seen in SDSS~J0952, SDSS~J1055, SDSS~J1350 and SDSS~J1342. We discount the observed change in the H$\beta$ line SDSS~J1241 as an observational artefact due to the described observational issues with the 2011 MMT spectrum of this object. None of the objects show consistent changes, with the most common evolution being an apparent blueshift in the 2011 observations before a return to SDSS-consistent values or further redshifting in later observations, with the exception of the H$\beta$ line of SDSS~J1350 which does show consistently increasing blueshift but is only detected in the most recent NTT spectrum at low S/N. Alterations in line peak velocities could indicate the formation of dust as a result of a TDE or changes in AGN obscuration. A higher cadence of observation -- particularly at early phases -- would allow this to be explored in more depth.

\subsection{Spectral Templates}
In order to explore if there are differences in the spectral properties between two subgroups of ECLE, that would allow for better classification between the two first observation in the future, we construct template spectra using the original SDSS spectra and perform comparisons both between the ECLE templates and to other SDSS galaxy templates. Whilst limited by the small sample size and the variable and unknown phase of the existing sample upon observation, the analysis does reveal tentative hints that those ECLEs with variable coronal lines are less blue overall and have stronger \Fex and \Fexi emission on first observation. Additional refinement of these templates will be undertaken as more ECLEs are identified. The construction of the templates and the comparisons themselves are outlined in greater detail within Appendix~\ref{Appendix:Spectral_Templates}.

\subsection{Optical Photometric Evolution}
\label{subsec:Optical_Photometric_Evolution}

As described in Section~\ref{subsec:Optical_Photometry}, we have used data from a number of all-sky surveys to explore the optical photometric evolution of each ECLE. The long period over which the ECLE sample has been observed, whilst invaluable for monitoring their long-term behaviour, presents a number of issues. Individual photometric surveys have not operated consistently over this extended multi-decade duration, which changes the sources of photometry (and the filters available) over time. Even where notionally the same filters are available in differing surveys, differences between the filter responses, calibration and processing introduces additional systematics. A specific example of this is the CRTS-\textit{V} observations. These were obtained using an unfiltered CCD and converted to effective \textit{V} observations using comparisons with reference stars. The conversion process introduces a source of systematic offset from other observations, highlighting that care must be taken when examining the behaviour of objects across differing surveys. 

As such, providing a fully consistent picture across the full range of observations is impossible. Instead, we instead focus on long-term trends. As a result of the large period over which the original SDSS spectra were collected, the time between the start of spectral coverage and when regularly spaced photometric observations began varies significantly. For three ECLEs (SDSS~J0938, J0952, and J1350), the SDSS spectrum was obtained during CRTS \textit{V}-band observations, whilst for the other four ECLEs, a period of at least a year separates the initial spectrum from such photometric observations. The start time of each ECLE flare is also poorly constrained (with the exception of SDSS~J0952 which is constrained by the LINEAR observations), making their evolutionary phases uncertain. The combined optical light curves for each are shown in Figure~\ref{fig:Optical_Photometric_Evolution_0}. In these plots the ECLE sample has been divided into the same three groupings used to present their optical spectra and are shown in the same order. Where eight or more epochs of data are available in a given filter from the same source, a cubic polynomial fit is also shown as a visual aid. 

\subsubsection{SDSS~J0748}
\cite{wang_2012_EXTREMECORONALLINE} found that SDSS~J0748 had brightened between the SDSS photometric and spectroscopic observations, indicating that the TDE likely occurred in the gap between the two sets of observations. This is supported by the SDSS spectrum being the only ECLE spectrum observed thus far with a distinct broad \Heii\ feature typical of conventional optically selected TDEs. The presence of such a feature indicates that SDSS~J0748 was likely spectroscopically observed during the active TDE phase of its evolution. Unfortunately, there is no photometry of SDSS~J0748 during this period to provide additional context for its early evolution.

Over the full period of observation, it has displayed a largely stable brightness. However, in more recent ATLAS observations it has displayed a long-term decline of $\sim 0.2$~mag in \textit{c} observations and an undulation in its \textit{o}-band light curve, first fading by 0.2~mag between the start of observations until $\sim$~MJD~59000, before rebrightening by $\sim 0.15$~mag over the following period to its current value. We note here that these changes in brightness are primarily seen between visibility periods rather than between individual observations and in particular, there are two discontinuities in \textit{o-band} observations between MJD $\sim$~58750 and 59250 for all objects except SDSS~J1350. Given the similarity and shared timing of this discontinuity across the sample along with the lack of observed variability in the difference imaging light curves, or in contemporaneous ZTF photometry, leads us to the the conclusion that the bulk of this observed variability is likely a calibration effect rather than intrinsic to the object.

The only PS1 data available for SDSS~J0748 are two epochs of \textit{g}-band observations, which do appear to show a significant decline between the two observations. This is not observed in the CRTS data during the same time period, which (whilst a much broader filter) do cover the full PS1 \textit{g}-band wavelength range.

\subsubsection{SDSS~J0938}
SDSS~J0938 has likewise had a largely stable brightness, baring slight changes of amplitudes consistent with expected low-level long-term AGN variability.

\subsubsection{SDSS~J0952}
SDSS~J0952 was noted by \cite{wang_2012_EXTREMECORONALLINE} to have faded between SDSS observations. This fading is supported by the contemporaneous CRTS observations which also show a slow decline for the first few years of observations (approximate MJD range: 53460--56660). The LINEAR observations over this period also capture the TDE flaring behaviour, though as described by \cite{palaversa_2016_REVEALINGNATUREEXTREME} the exact time of peak flare brightness was not observed. Following this initial decline, SDSS~J0952 remained at a stable brightness in the remaining CRTS observations. Whilst the PS1 observations of SDSS~J0952 display some level of scatter, it is important to note that this level (up to 0.2~mag) is similar to the range of scatter observed in the unbinned CRTS observations. Given the lack of clear trends in the PS1 observations, we attribute this scatter to stochastic variability between observations.

More recent ATLAS observations of SDSS~J0952 have shown a 0.15~mag decline in the \textit{c} band between the start of observations and MJD~58750 before stabilising at the current value of $\sim -19.9$~mag. \textit{o}-band observations have been largely stable with the exception of the aforementioned systematic offset. ZTF observations of SDSS~J0952 have shown no evolution, with the only observed variation being stochastic in nature. 

\subsubsection{SDSS~J1055}
In the study conducted by \cite{wang_2012_EXTREMECORONALLINE}, SDSS~J1055 was observed to have faded between the SDSS photometric and spectroscopic observations. This behaviour is consistent with both its AGN classification and its more recent photometric evolution, which, like SDSS~J0938, has consisted of long-term undulations.

\subsubsection{SDSS~J1241}
SDSS~J1241 was not found to have varied in brightness between the two epochs of SDSS observations. Likewise, no photometric evolution was observed during CRTS observations, with a stable brightness measured over the full duration of the survey. The two epochs of PS1 photometry from this period are from two different filters so reveal nothing further about its evolution. 

Interestingly, during the first three years of ATLAS observations, SDSS~J1241 faded from $\sim -19.25$~mag to $\sim -18.5$~mag in the \textit{c} (with significant scatter among individual observations) and \textit{o} bands before largely stabilising after accounting for the systematic \textit{o}-band systematic, though with significant scatter. In contrast, ZTF \textit{gri} observations of this object, whilst not covering the time period of the decline observed by ATLAS, are stable across all three bands for the full duration of available observations, with the possible exception of the most recent set of ZTF \textit{r}-band observations which could indicate a brightening, though the small number of recent observations makes this difficult to conclude reliably.

We note again here that the ATLAS photometry used in this work is based on observed images rather than difference imaging for direct comparison with other surveys. When difference imaging is used, this decline is not observed in the ATLAS data. As such, we do not view this decline as physical and treat the late-time optical behaviour of SDSS~J1241 as largely constant. 

\subsubsection{SDSS~J1342}
\cite{wang_2012_EXTREMECORONALLINE} found the brightness of SDSS~J1342 to be unchanged between its photometric and spectroscopic observations. Likewise, CRTS \textit{V}-band observations of SDSS~J1342 exhibit a stable brightness across the survey, with PS1 observations during the same period also showing no overall evolution (beyond stochastic variability, as seen in SDSS~J0952). The ATLAS observations of SDSS~J1342 reveal a decline of 0.15~mag between the start of ATLAS observations and MJD~58250 in the \textit{c} band, after which the object stabilised in brightness, as well as a similar decline of $\sim 0.2$~mag in ATLAS \textit{o}-band observations. Later ATLAS and ZTF \textit{gi} data display consistent brightness. 

\subsubsection{SDSS~J1350}
In contrast to the previously described objects, SDSS~J1350 has shown more significant optical variability. An increase in brightness was observed by \cite{wang_2012_EXTREMECORONALLINE} to have occurred between the SDSS spectroscopic and photometric observations. A sharp increase in brightness in the CRTS \textit{V} band by $\sim 1.5$~mag was also observed around MJD~54600 before plateauing for several years. We note that there are significantly fewer CRTS observations of this object, 33 (pre-binning) compared to a mean of 389 observations for the other objects in the ECLE sample. As such, the CRTS light curve of SDSS~J1350 is much less well sampled than the other objects. Two epochs of CRTS data have been removed from this light curve, with both being $\sim 3$~mag brighter than the previous and subsequent observations. We attribute these anomalies to a nearby bright (\textit{r} = 8.75~mag) star rather than astrophysical behaviour.

SDSS~J1350's later behaviour in PS1, ATLAS, and ZTF observations is more stable, with no additional such changes observed, though early ATLAS observations in both bands have larger uncertainties.

\begin{figure*}
    \centering
    \includegraphics[width=0.94\textwidth]{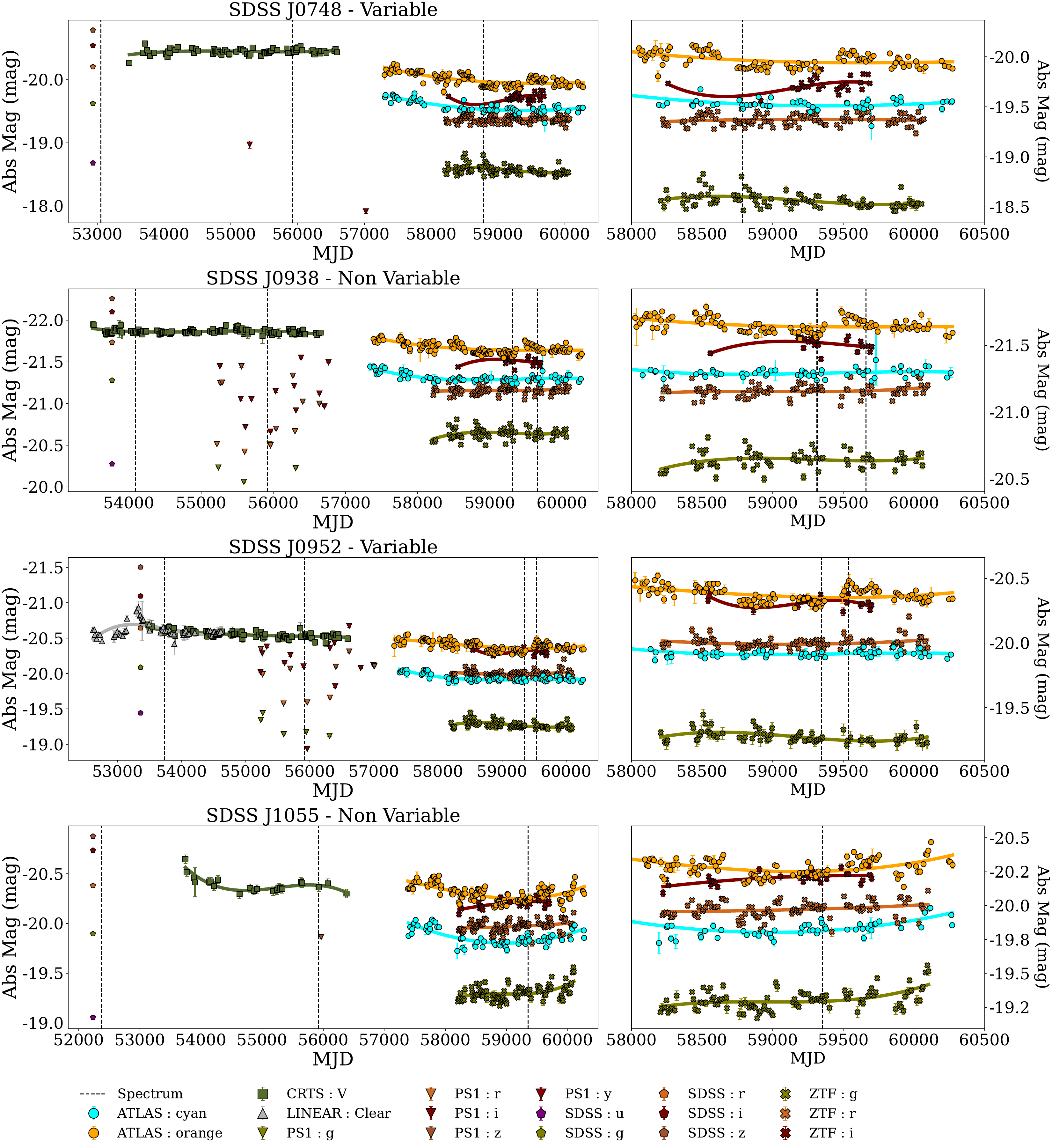}
    \caption{Optical photometric evolution of the ECLE sample. The colours of each point indicate the filter used, with symbol shape representing the source. Where sufficient data are available, a cubic polynomial fit (per filter) is included to guide the eye to any long-term trends. The full range of photometric observations is included in the left panels, with the right panels showing only data obtained more recently than MJD = 58000. As described in Section~\ref{subsec:Optical_Photometry}, the photometry from all sources has undergone a 5-$\sigma$ clipping procedure and has been binned to a 14 day cadence.}
    \label{fig:Optical_Photometric_Evolution_0}
\end{figure*}

\begin{figure*}
    \centering
    \includegraphics[width=0.94\textwidth]{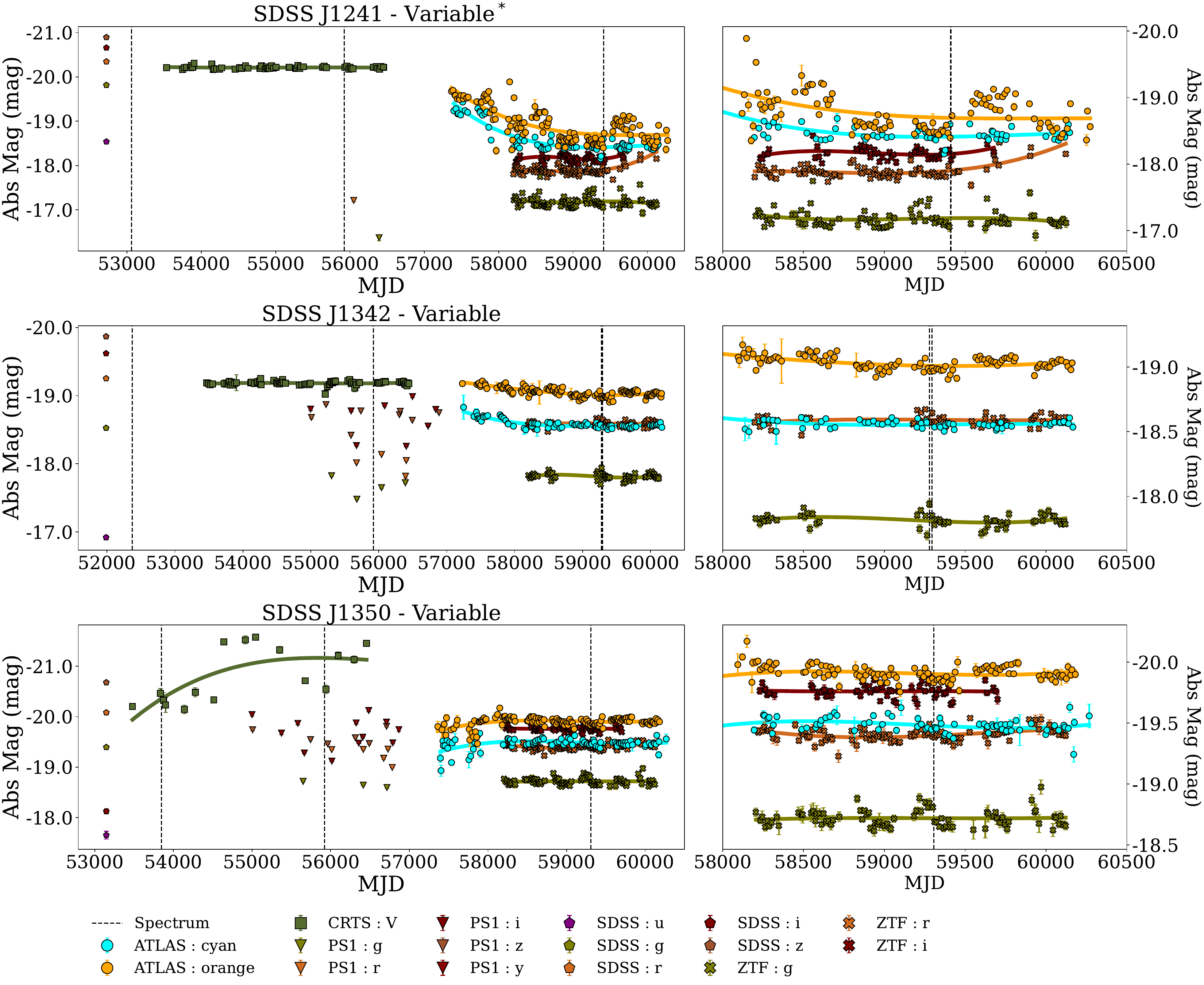}
    \contcaption{Optical photometric evolution of the ECLE sample. The colours of each point indicate the filter used, with symbol shape representing the source. Where sufficient data are available, a cubic polynomial fit (per filter) is included to guide the eye to any long-term trends. The full range of photometric observations is included in the left panels, with the right panels showing only data obtained more recently than MJD = 58000. As described in Section~\ref{subsec:Optical_Photometry}, the photometry from all sources has undergone a 5-$\sigma$ clipping procedure and has been binned to a 14 day cadence.}
    \label{fig:Optical_Photometric_Evolution_1}
\end{figure*}

\subsection{MIR Photometric Evolution}
\label{subsec:MIR_Evolution}

\begin{figure*}
    \includegraphics[width=0.94\textwidth]{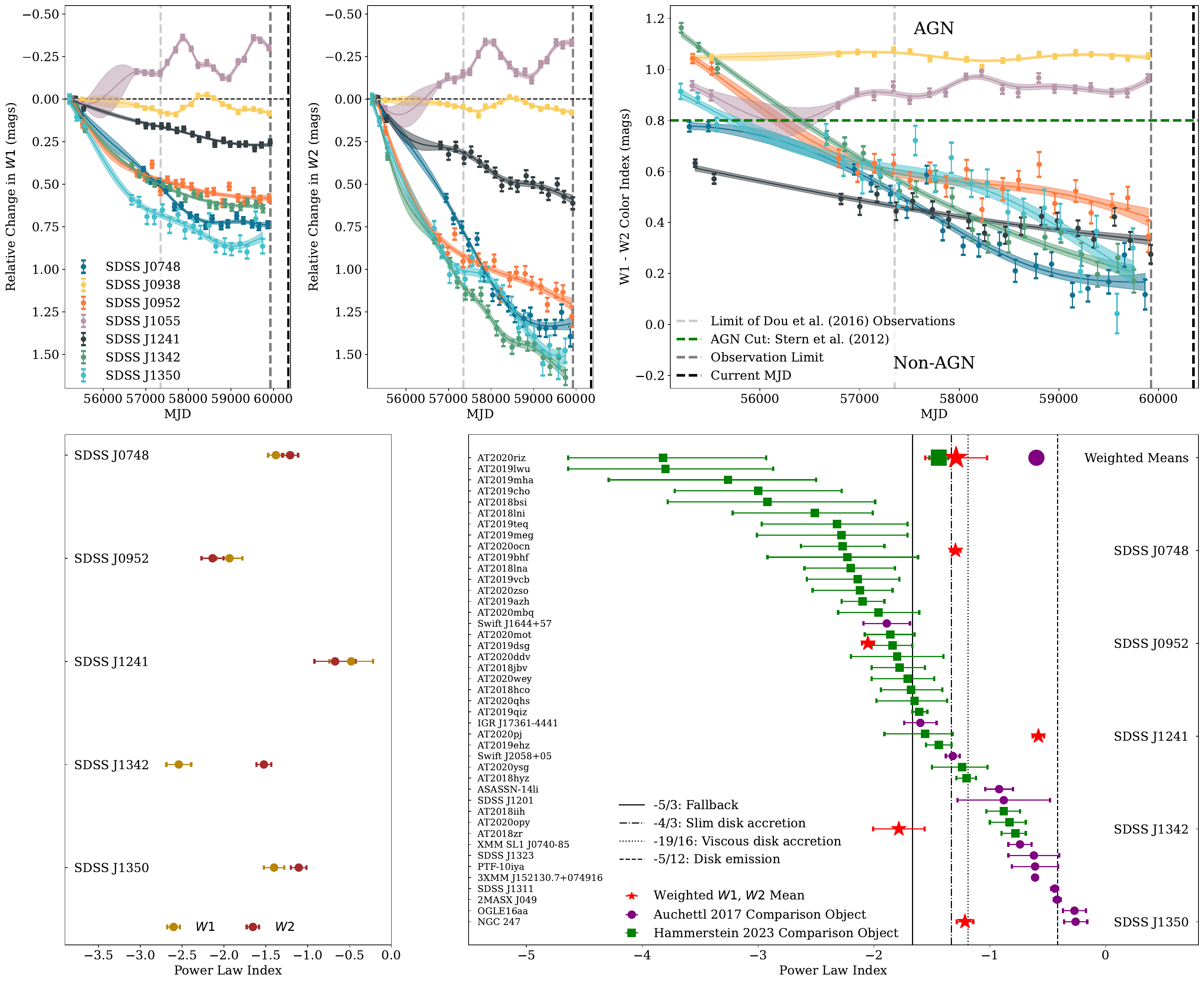}
    \caption{{\it Top left:} Relative change in \textit{W1} and \textit{W2} brightness compared to first \textit{WISE} observation. {\it Top right:} \textit{W1}--\textit{W2} colour evolution. The dashed horizontal line shown is the AGN/non-AGN dividing line from \protect\cite{stern_2012_MIDINFRAREDSELECTIONACTIVE}. Objects with a \textit{W1}--\textit{W2} colour greater than this value display AGN-like behaviour. The difference between the two non-variable ECLEs and the five variable ECLEs is clear. In all panels, fits displayed are obtained through Gaussian process regression \protect\citep{ambikasaran_2015_FastDirectMethods}, with the shaded regions indicating the 1$\sigma$ fitting uncertainties.
    {\it Lower Left:} Comparison between the measured values of the power-law indices ($B$) for both the \textit{W1} and \textit{W2} bands. {\it Lower Right:} Comparison between the MIR power-law decline indices for this sample (red stars), X-ray power-law decline indices of the objects of \protect\cite{auchettl_2017_NewPhysicalInsights} (purple circles) and the optical decline indices of the \protect\cite{hammerstein_2023_FinalSeasonReimagined} sample (green squares). Weighted means for each of the groups is shown, with the ECLE and optical value being consistent, and larger in magnitude than the value measured for X-ray events. Vertical lines indicate the expected values for a range of accretion models, standard fallback \protect\citep[e.g.,][]{evans_1989_TidalDisruptionStar,phinney_1989_CosmicMergerMania}, viscous disk accretion \protect\citep{cannizzo_1990_DiskAccretionTidally}, disk emission \protect\citep{lodato_2011_MultibandLightCurves}, and advective super-Eddington thin-disk accretion \protect\citep{cannizzo_2009_NewParadigmGammaray, cannizzo_2011_GRB110328ASwift}.}
    \label{fig:MIR_Comparison}
\end{figure*}

The MIR evolution of four of the seven ECLEs first identified by \cite{wang_2012_EXTREMECORONALLINE} was previously described by \cite{dou_2016_LONGFADINGMIDINFRARED} using data obtained with \textit{WISE} \citep{wright_2010_WIDEFIELDINFRAREDSURVEY}. The objects included in this analysis had been identified by \cite{yang_2013_LONGTERMSPECTRALEVOLUTION} as showing long-term variability (i.e., SDSS~J0748, J0952, J1342, and J1350), with the remaining three objects (SDSS~J0938, J1055, and J1241) excluded. The \textit{WISE} observations were obtained between early 2010 and late 2015 ($\sim$~MJD~55200--57350).  Here we present a continuation of this analysis using the more recent data obtained through the NEOWISE Reactivation survey \citep[NEOWISE-R;][]{mainzer_2011_NEOWISEOBSERVATIONSNEAREARTH, mainzer_2014_INITIALPERFORMANCENEOWISE}\footnote{\url{https://wise2.ipac.caltech.edu/docs/release/neowise/neowise_2022_release_intro.html}}, and extend this to include the remaining three objects in the original \cite{wang_2012_EXTREMECORONALLINE} sample. A summary of the MIR evolution of the full sample of ECLEs is given in the top row of Figure~\ref{fig:MIR_Comparison}.

The MIR evolution of the four objects previously identified as displaying long-term variation all showed declines in their \textit{W1} and \textit{W2} luminosities over timescales of years, with all also trending toward bluer \textit{W1}--\textit{W2} colours. All the ECLEs studied by \cite{dou_2016_LONGFADINGMIDINFRARED} were seen to share a similar overall behaviour, despite the original flaring events occurring at different times, with declines of $\sim 0.5$--1.1~mag over the course of $\sim 5.5$~yr of \textit{WISE} observations. This MIR evolution is also shared by SDSS~J1241, which our follow-up optical spectroscopy has revealed to also display fading coronal-line emission. All of the ECLEs with variable coronal lines, with the exception of SDSS~J0748, continue to show declines in one or both \textit{W1} and \textit{W2} bands, along with colour evolution. This ongoing evolution indicates that these objects have not yet completely faded back to their quiescent pre-event states. 

This behaviour differs from that used by \cite{dou_2016_LONGFADINGMIDINFRARED} to determine the host-galaxy contributions to the MIR transient light curves. Their models (with caveats) were constructed assuming the objects had reached a plateau consistent with the flaring transient event having faded and the light of the host galaxy now dominating in the final epoch of NEOWISE data available to them. With the additional $\sim 7$~yr of data, we can see that this was not in fact the case, with the objects all showing continuing MIR declines in the intervening years. As such, the galactic contribution differs from their values which were, by and large, overestimated. 

Whilst the trend toward bluer \textit{W1}--\textit{W2} colours with time is seen across the full sample of variable objects, this trend has not been as smooth in recent years, with the variable objects displaying some scatter around the overall evolutionary trend in both bands. SDSS~J0952 was noted by \cite{dou_2016_LONGFADINGMIDINFRARED} as potentially displaying `non-monotonous [sic] variability' which they attributed to complexities in the object's dust formation. Whilst this variability is not clear in its \textit{W1}--\textit{W2} colour evolution, it is apparent in the individual filter light curves, with several instances of observations being several standard deviations above or below the smoothed overall trend.

SDSS~J0748 and J0952 have displayed smooth evolution over the course of observations in both bands, with the exception of some stochastic variability in the case of SDSS~J0952, and a single epoch of significant colour deviation for SDSS~J0748 at MJD~57674. The \textit{W1}-band evolution of J1241 has also been remarkably consistent, though its \textit{W2}-band curve displays a two-phase decline, with a reduction in the rate of decline after MJD~58000.

The most significant deviations from smooth overall evolution amongst the variable coronal line objects are seen in \textit{W2}-band observations of SDSS~J1342 and SDSS~J1350, both of which display shoulders. During these shoulders their \textit{W2} luminosity remains constant or even rises slightly before the long-term overall decline resumes. SDSS~J1350 has displayed one such shoulder beginning around MJD~57000 and lasting until $\sim$~MJD~5800, where as SDSS~J1342 has shown two shoulders, at MJD~57200--57770 and 58500--59020.

A \textit{W1}--\textit{W2} colour cut can be used to effectively differentiate between AGN hosting and nonhosting galaxies. This cut was developed specifically for \textit{WISE} observations by \cite{stern_2012_MIDINFRAREDSELECTIONACTIVE}, with AGN activity indicated by \textit{W1}--\textit{W2} $\ge 0.8$~mag. When applied to the ECLE sample, all variable ECLEs with the exception of SDSS~J1241 are initially observed to be at, or above, this colour cut indicating AGN-like activity. SDSS~J1241, which was not included in the \cite{dou_2016_LONGFADINGMIDINFRARED} analysis, had an initial \textit{W1}--\textit{W2} colour index of $0.63 \pm 0.01$~mag in the first epoch of ALLWISE observations, suggesting no dominant AGN activity. This AGN activity colour cut was initially selected as it was shown to have both good completeness (78\%) and reliability (95\%) in dividing AGN from non-AGN in \textit{WISE} data. Evolution is observed in the \textit{W1}--\textit{W2} colours of all five variable objects, with all trending toward bluer colour indices over time and now falling well below the \textit{W1}--\textit{W2} AGN colour cut with values in the range 0.10--0.52~mag. 

If the continued MIR flux evolution is taken as the result of the accretion of residual material from the TDE, the observed shoulders in brightness could arise from periods where the accretion rates have stabilised. This is perhaps in turn related to the density of the material being accreted, with the overall reduction seen as the overall mass of material available to the SMBH reduces as an individual TDE can only provide a fixed mass of material to the system. Alternatively, if there is an underlying, weak, or obscured AGN within the galaxy, these light-curve features could be the result of increases in the accretion rate from material not necessarily produced in the initial transient flare, as AGN themselves are known to display MIR variability \citep{hawkins_2002_VariabilityActiveGalactic}. These trends in MIR brightness and \textit{W1}--\textit{W2} colour evolution in the variable objects are in stark contrast to the two objects with non-variable coronal lines (i.e., the AGN-related SDSS~J0938 and J1055), which do not display such long-duration reddening.

SDSS~J0938 has been observed with multiyear undulations in both bands and an overall range in brightness of $\sim 0.1$~mag, the object being slightly dimmer on average compared to its first observation during these undulations. SDSS~J0938 brightened slightly more in the \textit{W1} band during this undulation, with its colour is seen to evolve blueward during the brightening period, though this change is small overall and remains well within the expected AGN colour region.

SDSS~J1055 is observed to have brightened in both bands over the course of \textit{WISE} observations by 0.3--0.4~mag, though this evolution has included several epochs of brightening and fading, with the object observed to be fading in the most recent observations, though remaining significantly brighter than when first observed. The \textit{W1}--\textit{W2} colour of SDSS~J1055 has also shown variability, though at a lower level than in the individual bands, with a value close to 0.9~mag seen across the observation period.

It is clear that the ECLE sample with variable coronal lines has continued to decline in the MIR during the time period covered by the NEOWISE observations. \cite{dou_2016_LONGFADINGMIDINFRARED} fitted the available data in flux space using both a power-law and exponential model, finding power-law decay to be preferable. We extend this modelling to include SDSS~J1241 and utilise the new NEOWISE photometry employing the same power-law model, given by

\begin{equation}
    f(t)=A t^{B}+C\, .
\label{Equation1}
\end{equation}
\noindent

Fitting was initially conducted independently for each band, with the times of outburst (which are poorly constrained for the sample) being the same as those used by \cite{dou_2016_LONGFADINGMIDINFRARED}. The exception is SDSS~J1241, included here for the first time; for which we adopt a value of 1~yr prior to its SDSS photometric observation as an approximate outburst time, based on it already declining by the time of its SDSS spectrum. The most important values in this fitting are the power-law index, given by $B$, and the quiescent flux of the galaxy, given by $C$. The remaining term, $A$, is a constant scaling factor.

In all cases, the power-law index $B$ is well constrained, though the quiescent galaxy flux $C$ is poorly constrained in the \textit{W2} band for objects other than SDSS~J0952 and SDSS~J1350, likely owing to the presence of more deviation from a smooth decline in the \textit{W2} band compared to \textit{W1}. The quiescent flux of SDSS~J1241 is also poorly constrained in the \textit{W1} band, with the decline of this object in both bands significantly shallower than the rest of the variable coronal-line ECLEs. When compared, the measured power-law indices for each object are found to be consistent between the \textit{W1} and \textit{W2} bands with the exception of SDSS~J1342, where the decline is significantly steeper in the \textit{W1} band ($C = -2.54 \pm 0.15$) than in the \textit{W2} band ($C = -0.74 \pm 0.23$).

Following the initial fitting, the power-law index of the \textit{W2} data was set to match the best-fitting \textit{W1} index to explore if the host quiescent flux in \textit{W2} could be better constrained. For all but SDSS~J1241 (which has a poorly constrained host component in \textit{W1}), a \textit{W2} host contribution can now be obtained, with the overall fitted light-curve shapes remaining largely unchanged. A comparison of the power-law indices obtained in this fitting is shown in the lower left panel of Figure~\ref{fig:MIR_Comparison}. Additionally, The results of this fitting are given in full within Appendix~\ref{Appndix:MIR_Power_Law}, presented graphically in Figure~\ref{fig:ECLE_MIR_Power_Law_Fit} with the parameters of the fits given in Table~\ref{tab:MIR_Power_Law_Fits}.

Using the weighted average of the \textit{W1} and freely fitted \textit{W2} results, we compare to the power-law decline indices measured by \cite{auchettl_2017_NewPhysicalInsights} for a sample of TDEs in X-rays and a similar sample measured in the optical from \cite{hammerstein_2023_FinalSeasonReimagined}. We note the caveats that both of these comparison samples explore TDEs much closer to maximum light - i.e., at a different phase of evolution - than the ECLEs and in different wavelength bands. This comparison is presented in the lower right panel of Figure~\ref{fig:MIR_Comparison}. For our sample, SDSS~J0952 and J1342 fall within the region expected of the standard fallback accretion, which as modelled by \cite{guillochon_2013_HydrodynamicalSimulationsDetermine}) extends to values steeper than $-5/3$. SDSS~J0748 and J1350 have shallower declines consistent with disk accretion. The shallowest declining object in our sample is SDSS~J1241, whose individual-band values are consistent with disk emission. We also compare the weighted averages of the three samples. The weighted ECLE power-law decay index was measured to be -1.29 $\pm$ 0.27 consistent with the optical weighted mean of -1.44 $\pm$ 0.08 measured by \cite{hammerstein_2023_FinalSeasonReimagined}, both of which are steeper declines than mean X-Ray value of the \cite{auchettl_2017_NewPhysicalInsights} sample which was found to be -0.60 $\pm$ 0.05. We note here that uncertainties given on the power law indices in \cite{hammerstein_2023_FinalSeasonReimagined} were asymmetric in most cases though for this comparison we have treated them all as symmetric in nature for consistency. We explored if this was a suitable simplification using both forms of Equation~17 from \cite{barlow_2003_AsymmetricErrors} and found no significant difference in the determined values of the weighted mean for those events with asymmetric uncertainties. Given that this formalism is not suitable in cases where the uncertainties are symmetric, to include all the objects of the \cite{hammerstein_2023_FinalSeasonReimagined} sample we adopted the stated simplification.

\begin{figure}
    \includegraphics[width=0.9\columnwidth]{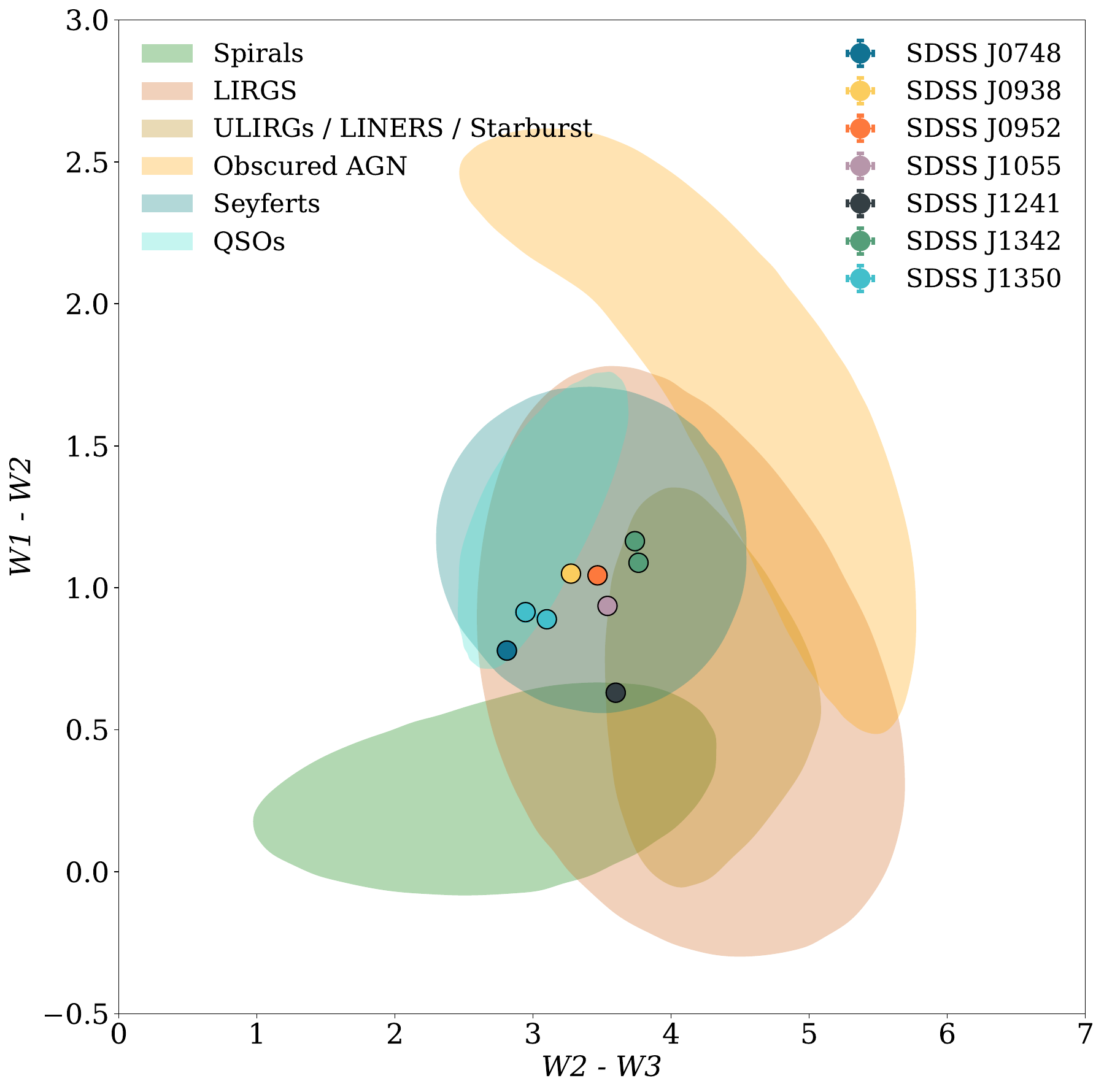}
    \caption{ALLWISE colour-colour plot showing the location of all ECLEs within the regions covered by various forms of AGN. Whilst no additional \textit{W3} data are available, the trend in decreasing \textit{W1}--\textit{W2} colours would move all variable ECLEs toward the star-forming/spiral region over time. Regions have been sourced from \protect\cite{wright_2010_WIDEFIELDINFRAREDSURVEY}. Uncertainties are included but are generally smaller than the points}
    \label{fig:ALLWISE_Colour_Index_Plot}
\end{figure}

We provide the observed \textit{W1}--\textit{W2} and \textit{W2}--\textit{W3} colours obtained during the initial ALLWISE sky survey in Table~\ref{tab:ALLWISE_Colour_Indices}. The use of a second colour index expands the parameter space and allows for a better identification of different classes of object. Owing to the limited duration of the ALLWISE mission, most objects have only one observation epoch where all data from all three filters are available. SDSS~J1342 and J1350 do have two epochs of data available with both included in Figure~\ref{fig:ALLWISE_Colour_Index_Plot}, and are shown connected.

\begin{table}
\caption{ALLWISE colour indices for the ECLE sample. All objects fall within the QSO/Seyfert region of the parameter space per \protect\cite{wright_2010_WIDEFIELDINFRAREDSURVEY}. Whilst no additional \textit{W3} data are available, the trend in \textit{W1}--\textit{W2} colours would move all variable ECLEs toward the star-forming/spiral region of the parameter space over time.}
\label{tab:ALLWISE_Colour_Indices}
\begin{tabular}{llll}
\hline
\multicolumn{1}{c}{Object} & \multicolumn{1}{c}{MJD} & \multicolumn{1}{c}{\textit{W1}--\textit{W2} (mag)} & \multicolumn{1}{c}{\textit{W2}--\textit{W3} (mag)} \\ \hline
\textbf{Variable ECLEs} &  &  &  \\
SDSS~J0748 &  &  &  \\
 & 55291 & 0.78~$\pm$~0.46 & 2.81~$\pm$~0.34 \\
SDSS~J0952 &  &  &  \\
 & 55324 & 1.04~$\pm$~0.38 & 3.47~$\pm$~0.27 \\
SDSS~J1241 &  &  &  \\
 & 55350 & 0.63~$\pm$~0.55 & 3.60~$\pm$~0.40 \\
SDSS~J1342 &  &  &  \\
 & 55211 & 1.16~$\pm$~0.37 & 3.74~$\pm$~0.37 \\
 & 55386 & 1.09~$\pm$~0.44 & 3.76~$\pm$~0.32 \\
SDSS~J1350 &  &  &  \\
 & 55204 & 0.92~$\pm$~0.33 & 2.94~$\pm$~0.33 \\
 & 55377 & 0.89~$\pm$~0.46 & 3.09~$\pm$~0.36 \\
 &  &  &  \\
\textbf{Non-Variable ECLEs} &  &  &  \\
SDSS~J0938 &  &  &  \\
 & 55324 & 1.05~$\pm$~0.41 & 3.28~$\pm$~0.29 \\
SDSS~J1055 &  &  &  \\
 & 55320 & 0.93~$\pm$~0.54 & 3.54~$\pm$~0.39\\ \hline
\end{tabular}
\end{table}

\cite{dou_2016_LONGFADINGMIDINFRARED} noted that the measured values of \textit{W2}--\textit{W3} vs. \textit{W1}--\textit{W2} for the four objects they examined fell within the region of the parameter space from \cite{wright_2010_WIDEFIELDINFRAREDSURVEY} expected of Seyfert galaxies, QSOs, and luminous infrared galaxies (LIRGs), and removed from the part of the parameter space occupied by elliptical and spiral galaxies not hosting AGN. We can extend this to the remaining three objects and confirm that these are also found within the same region of parameter space.

Whilst no later \textit{W2}--\textit{W3} observations are available, the observed evolution of decreasing \textit{W1}--\textit{W2} colour of all five variable objects would move them into the parameter space occupied by non-AGN hosting star-forming galaxies (assuming no change in \textit{W2}--\textit{W3} colour), with the two spectroscopically non-variable ECLEs remaining within the Seyfert/QSO/LIRG region. 

The MIR and optical spectroscopic evolution of the variable ECLEs thus appear to be in conflict. In the years after the initial flare, optical spectra reveal line ratios trending from non-AGN regions of BPT diagrams to those consistent with Seyfert-type AGN --- in particular, SDSS~J1342 with the drastic increase in \Oiii~$\lambda$5007 emission. Yet, at the same time, their MIR evolution is seen to trend away from AGN-like colours. A possible explanation for this conflict would be the delayed response of more distant low-density gas to the initial TDE flare being responsible for the generation of the increased \Oiii\ emission. This would then not require an ongoing elevation in the accretion rate onto the SMBH, which has in fact been returning to quiescent values shown through the long-term MIR decline. Differences in the line evolution across the sample would therefore indicate differences in the environments close to the SMBHs. 

\subsection{Pre-Outburst NIR Analysis}
\label{subsec:2MASS_Analysis}

\begin{figure}
    \includegraphics[width=0.92\columnwidth]{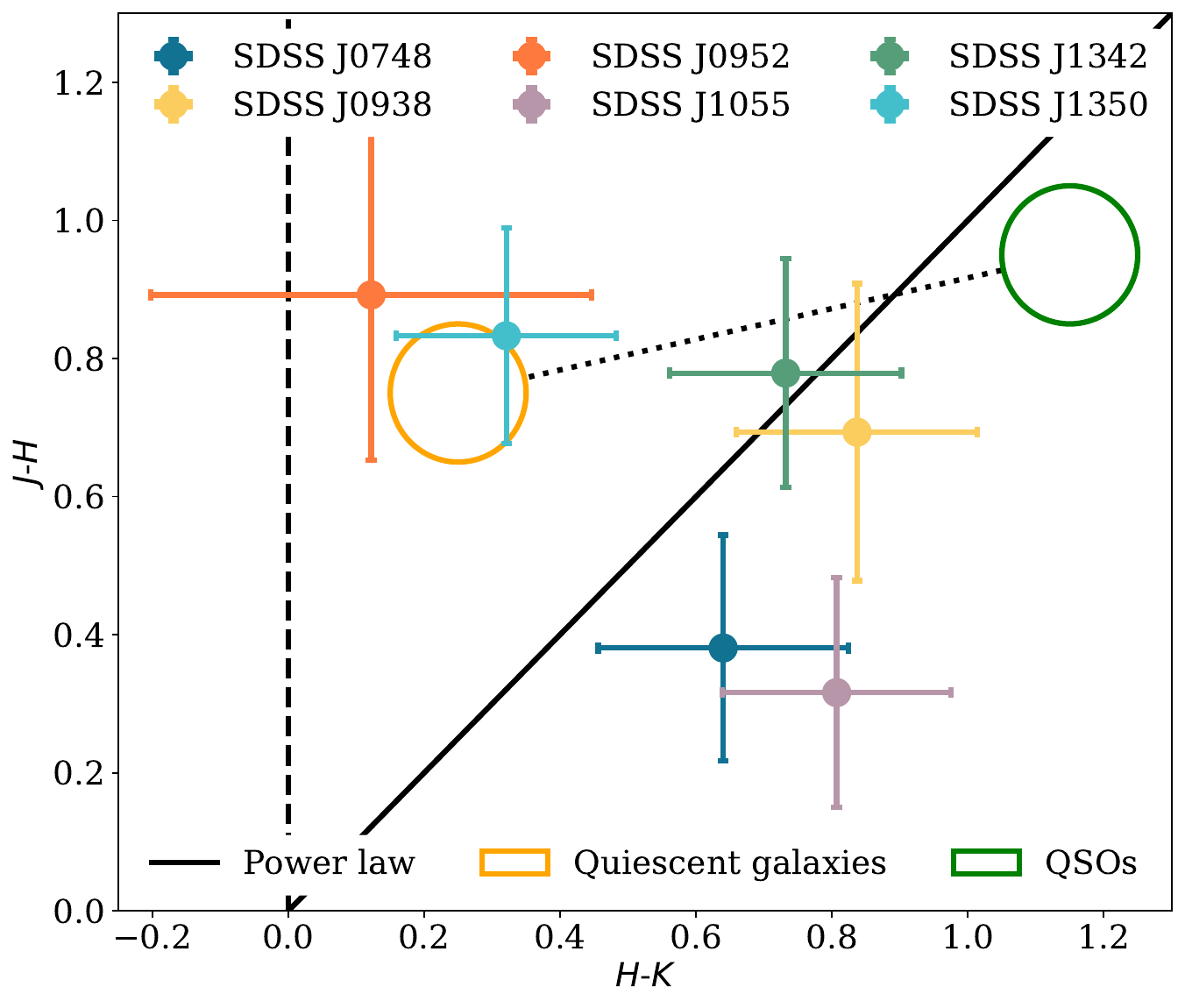}
    \caption{Available pre-outburst 2MASS IR photometry of the ECLE sample. SDSS~J0952 and 1350 fall within the region expected of quiescent, non-AGN-hosting galaxies, whilst the remaining objects are more consistent with a combination of AGN and starlight. However, SDSS~J0748 and J1055 have bluer \textit{J}--\textit{H} colours than the main locus of either quiescent or AGN-hosting galaxies.}
    \label{fig:2MASS_AGN_or_Starlight_Plot}
\end{figure}

Six of the ECLEs (the exception being SDSS~J1241) were observed as part of the 2MASS All-Sky Survey in the \textit{JHK} bands (we include this photometry here in Table~\ref{tab:ECLE_2MASS}). These observations were obtained between January 1998 and January 2001, well before the expected time of the initial ECLE flaring activity. This presents the opportunity to explore the quiescent behaviour of the ECLE galaxies. Figure~\ref{fig:2MASS_AGN_or_Starlight_Plot} shows the objects with available data in \textit{J}--\textit{H} vs. \textit{H}--\textit{K} parameter space. This allows for the separation of objects where IR luminosity is primarily the result of starlight from those where the IR flux is driven by AGN activity \citep{hyland_1982_InfraredStudyQuasars, komossa_2009_NTTSpitzerChandra}. We note that given the wide range of NIR colours displayed by galaxies of the same spectroscopic classification, they cannot be distinguished effectively using these NIR colours alone.

\begin{table}
\caption{Pre-outburst 2MASS IR photometry of the ECLE sample.}
\label{tab:ECLE_2MASS}
\begin{adjustbox}{width=1\columnwidth}
\begin{tabular}{llllll}
\hline
\multicolumn{1}{c}{Object} & \multicolumn{1}{c}{MJD} & \multicolumn{1}{c}{J (mag)} & \multicolumn{1}{c}{H (mag)} & \multicolumn{1}{c}{K (mag)} & \multicolumn{1}{c}{Type *} \\ \hline
\multicolumn{5}{l}{\textbf{Variable ECLEs}} &  \\
SDSS~J0748 & 51229 & -21.09~$\pm$~0.09 & -21.47~$\pm$~0.14 & -22.11~$\pm$~0.13 & Point \\
SDSS~J0952 & 50836 & -22.31~$\pm$~0.16 & -23.20~$\pm$~0.18 & -23.33~$\pm$~0.27 & Extended \\
SDSS~J1241 & --- & --- & --- & --- & --- \\
SDSS~J1342 & 51928 & -20.76~$\pm$~0.11 & -21.53~$\pm$~0.12 & -22.26~$\pm$~0.12 & Extended \\
SDSS~J1350 & 51645 & -21.15~$\pm$~0.11 & -21.99~$\pm$~0.11 & -22.31~$\pm$~0.12 & Point \\
\\
\multicolumn{5}{l}{\textbf{Non-Variable ECLEs}} &  \\
SDSS~J0938 & 51669 & -23.57~$\pm$~0.14 & -24.26~$\pm$~0.17 & -25.10~$\pm$~0.07 & Extended \\
SDSS~J1055 & 51507 & -21.61~$\pm$~0.09 & -21.93~$\pm$~0.14 & -22.74~$\pm$~0.10 & Point \\ \hline
\end{tabular}
\end{adjustbox}
\begin{flushleft}
*Type indicates from which 2MASS catalogue we retrieved the data shown here.\\ As these sources are extended galaxies, where both extended and point-source measurements were available we selected the extended-source measurements.
\end{flushleft}
\end{table}

\cite{komossa_2009_NTTSpitzerChandra} made use of these observations of SDSS~J0952 and found it to be located in the region expected of quiescent galaxies, as would be expected prior to a transient flaring event. We now extend this to the remaining five objects with available data. Similarly to the results of \cite{komossa_2009_NTTSpitzerChandra}, SDSS~J1350 is found in this region of nonactive galaxies. However, the picture for the remaining four is more complicated. 

The non-variable SDSS~J0938 and variable ECLE SDSS~J1342 are located in the region consistent with a combination of starlight and AGN activity. The final two objects, SDSS~J0748 and J1055, also have \textit{H}--\textit{K} colours indicative of a combination of starlight and AGN activity, but they are separated from the rest of the sample by their bluer than expected \textit{J}--\textit{H} colours. The weighted mean \textit{J}--\textit{H} colour of SDSS~J0748 and J1055 is 0.35~$\pm$~0.11~mag compared to 0.80~$\pm$~0.09~mag for the other four ECLEs. Given the large uncertainties in each object's photometry ($\sim 0.17$~mag) and in the weighted means, the statistical significance of this offset is small.

\raggedbottom

\section{Discussion}
\label{Sec:Discussion}

ECLE behaviour is complex, with even the limited sized sample showing two populations. Owing to the discovery of the known sample already in their declining phases, the early-time evolution of these objects has not yet been well observed. Here we explore the connections between ECLEs and other related classifications of objects: optically selected TDEs, which have been seen to develop coronal emission lines, and galaxies identified as showing flares/outbursts at MIR wavelengths.

\subsection{Coronal-Line TDEs}
\label{subsec:Coronal_Line_TDEs}

As mentioned, a small group of TDEs has recently been seen to develop coronal line signatures at varied times following their classification. This group includes AT~2017gge (17gge), AT~2019qiz (19qiz) and AT~2022upj (22upj) which we discuss in turn. 

17gge was classified as a centrally located optical transient with broad H and He spectral features consistent with a TDE \citep{fraser_2017_EPESSTOSpectroscopicClassification}. It was observed to display a delayed X-ray flare ($\sim 200$~d post optical discovery) coincident with the emergence of Fe coronal lines (\Fevii--\Fexiv) that have persisted (with altering line ratios) until at least 1700~d post-discovery \citep{onori_2022_NuclearTransient2017ggea}. It has also been seen with an outburst in the MIR followed by an ongoing, multiyear decline consistent with a high covering factor of reprocessing material. 17gge was included as a `Mid-infrared Outburst in a Nearby Galaxy' (MIRONG) in the sample of \cite{jiang_2021_MidinfraredOutburstsNearby}, with its MIR behaviour found to be consistent with a TDE by \cite{wang_2022_MidinfraredOutburstsNearby}. We further discuss MIRONG in Section~\ref{subsec:MIRONG}.

Similarly to 17gge, 19qiz initially lacked coronal emission lines and was observed with both broad H and He features with significant blueshifts (indicating an outflow), before developing Bowen fluorescence lines \citep{nicholl_2020_OutflowPowersOptical, hung_2021_DiscoveryFastIron}. Fe coronal emission was first observed at +428~d (in rest-frame) after the optical peak by \cite{short_2023_DelayedAppearanceEvolution}, with the lines persisting until at least +828~d. Interestingly, whist the other Fe lines increased in flux before beginning to decline, the flux of \Fexiv\ was still seen to be increasing by the time of the last spectrum. The authors attribute this (and other properties of 19qiz) to a complexly structured local SMBH environment, with material located at differing distances and a high overall covering factor to enable reprocessing (both in the form of coronal emission lines and a large MIR outburst) but where the direct line-of-sight to the SMBH from Earth is largely unobscured to allow for the rapid observed rise to peak.

Finally 22upj has also been observed with coronal Fe emission lines however, as opposed to 17gge and 19qiz, these lines are present in spectra obtained near maximum light rather than being significantly delayed in their development \citep{newsome_2022_2022upjZTF22abegjtxDiscovery}.

Until the identification of coronal lines in the $> 200$~d spectra of 17gge, SDSS~J0748 was the clearest example of an object with both conventional TDE features and the ECLE-defining Fe coronal lines. Now that a group of optically selected TDEs has been observed to develop Fe coronal lines, the link between the two is unambiguous. It remains to be determined what percentage of the overall TDE population displays coronal lines at some phase of their evolution, a determination which has not been helped by the lag between discovery and follow-up observations of the original ECLE sample. It is already clear from these examples, where the time between the triggering TDE event and the development of the coronal lines can be better constrained ($\sim 200$~d for 17gge, $\sim$ 200 -- 428~d for 17azh and $\sim 2$~months in the case of 22upj), that the timescales of such events and thus the nature of their environments are varied. This will provide the opportunity to utilise the techniques developed in AGN reverberation mapping \citep[e.g.,][]{cackett_2021_ReverberationMappingActive} to better model physical properties of the SMBH systems involved in these events. Further study into both newly identified and existing TDE host galaxies (the spectra of which may now be displaying residual or delayed coronal-line signatures) will be crucial in furthering our understanding of these events.

\subsection{Comparison with Mid-InfraRed Outbursts in Nearby Galaxies - MIRONG}
\label{subsec:MIRONG}

Further analysis of the MIR evolution of the objects identified as variable by \cite{yang_2013_LONGTERMSPECTRALEVOLUTION} was conducted by \cite{dou_2016_LONGFADINGMIDINFRARED} and revealed long-term declines in all four. The study Mid-infrared Outbursts in Nearby Galaxies (MIRONG) \citep{jiang_2021_MidinfraredOutburstsNearby, wang_2022_MidinfraredOutburstsNearby} --- galaxies displaying flaring behaviour of at least 0.5~mag in the MIR that is not necessarily associated with observed optical variability --- revealed that several objects displayed transient Fe coronal lines. However, the timescales of the coronal line and MIR evolution in MIRONG and ECLEs appear to differ. Both display increases in luminosity via outbursts, followed by long-term declines \citep{jiang_2021_MidinfraredOutburstsNearby}. 
MIRONG differ from ECLEs in that their observed MIR outburst from a quiescent state was the primary selection criterion for their initial identification, whereas the quiescent state of ECLEs was not observed in the MIR prior to their flaring events. Owing to the timing of the \textit{WISE} mission, the available MIR light curves for the ECLE sample begin 5--9~yr following the outburst event \citep{dou_2016_LONGFADINGMIDINFRARED}. 53 of the 137 galaxies in the MIRONG sample have had high-quality  follow-up spectroscopy described by \cite{wang_2022_MidinfraredOutburstsNearby}. Of this subset, 22 (42\%) displayed emission-line variability, and most interestingly nine have been detected with variable Fe coronal lines (17\% of the overall MIRONG sample and 42\% of those with emission-line variability).

All but one of these objects have also shown reductions in their H$\alpha$ line flux over the course of the follow-up spectroscopy, with two having H$\alpha$ fluxes now consistent with a quiescent state. The coronal lines in each of these objects were weak and short lived, fading after the first follow-up spectrum.
The exception to both of these behaviours is SDSS~J1442+5558. This object has maintained strong H$\alpha$ flux consistent with an AGN state-change (specifically a `turn-on' event) with consistently increased H$\alpha$ flux for at least 5~yr, with the Fe coronal lines developing in the most recent spectra available (years post state-change).

One spectroscopically variable ECLE (SDSS~J1342) still displays coronal emission lines (though only of \Fevii) over more than a decade following its discovery spectrum, in contrast to the short-duration coronal lines observed in some, but not all, TDE-associated MIRONG. The differences between ECLE and MIRONG MIR behaviour could be the result of the differing local environments (e.g., dust content and composition), with MIRONG observed to have much larger dust covering fractions than optically selected TDEs. Differences in the mass or structure of stars undergoing disruption between both groups could also play a role in their differing timescales.

The two groups of objects could be related --- a large subset of the MIRONG sample have been identified as TDE candidates --- with the differences in observed properties associated with environments in which they occur (e.g., local dust mass and composition). As described, ECLEs are a similar combination of multiple populations (TDE and AGN produced).

\section{Conclusions}
\label{Sec:Conclusions}

We have explored the long-term evolution, both spectroscopically and photometrically, of the ECLE sample of seven objects first identified in the SDSS by \cite{wang_2012_EXTREMECORONALLINE}. Through this analysis, we conclude the following.

\begin{itemize}
    \item The coronal-line persistence of two objects within the sample, first described by \cite{yang_2013_LONGTERMSPECTRALEVOLUTION}, is confirmed, showing that the coronal lines in these two objects (SDSS~J0938 and J1055) are persistent over a time-span of two decades.
    \item The third object classified by \cite{yang_2013_LONGTERMSPECTRALEVOLUTION} as invariable (SDSS~J1241) does in fact exhibit diminishing coronal-line emission. It also displays MIR evolution consistent with the other previously identified variable coronal-line ECLEs.
    \item Follow-up spectroscopy of objects where coronal lines have previously faded shows that these lines have not recurred (subject to caveats on the limited cadence), supporting their generation in single transient events rather than ongoing or recurring processes.
    \item We demonstrated a significant increase in the \Oiii\ flux of SDSS~J1342 since the previous follow-up spectrum in 2011, with the line having evolved to be the most dominant spectral feature.
    \item The long-duration MIR fading displayed by those with variable coronal-line emission as first identified by \cite{dou_2016_LONGFADINGMIDINFRARED} has continued for at least an additional 6~yr. The declines of all variable coronal lines remain consistent with ongoing power-law declines.
    \item The optical evolution of the variable ECLEs appears to indicate AGN-like activity. BPT line-ratio diagnostics of the most recent spectra continue to be more indicative of AGN values than was observed in the initial SDSS spectra. In contrast, the MIR colour evolution of these objects displays a continued trend away from the values expected of AGN. More modelling will be required to conclusively understand the behaviour of this class of object, though the delayed response of gas more distant from the SMBH could be used to explain the altered line ratios without the requirement of increased accretion activity not indicated by their MIR evolution. The range of behaviour displayed also highlights the importance of observing ECLEs over a wide wavelength range.
    \item High-resolution and high-S/N spectra are necessary to confirm the presence of weak and narrow coronal lines -- \Fevii -- which have persisted in SDSS~J1342 for two decades. These features would have been missed or gone unconfirmed if relying on lower resolution data.
    \item Spectral templates of variable and non-variable ECLEs were constructed using the original SDSS spectra. Whilst tentative, given the small sample size and variable phase of the objects at observation, the templates reveal potentially distinguishing properties between the two subgroups. Those objects with non-variable coronal-line signatures appear to be bluer overall than those with variable coronal lines (subject to poorly constrained phases). Whilst all Fe coronal lines can be observed in both the variable and non-variable objects, those with variable coronal lines display relatively stronger \Fex\ and \Fexi\ lines at early phases of their evolution. The same is likely also true of \Fexiv\ though harder to confirm owing to differences in the underlying continuum.
    \end{itemize}

The analysis undertaken here has strengthened the identification of five of the seven currently identified ECLEs as the light echoes of nonrecurring TDEs; no ECLE with variable coronal lines shows a resurgence in coronal-line emission. The work also highlights the importance of observing TDEs and their hosts across a large wavelength regime (i.e., optical observations alone are insufficient) so that a complete picture of their behaviour, which has been seen to be conflicting between optical and MIR evolution (where ongoing AGN-like behaviour is observed in the optical but a continued return to quiescence is observed in the MIR), can be further developed. It is likely that the local environments of the SMBHs involved in ECLEs is both gas-rich and complex with material at differing distances and structures that give rise to the varying signatures observed. Such a determination is supported by other work into these and similar objects \citep{short_2023_DelayedAppearanceEvolution, hinkle_2023_CoronalLineEmitters}.

Identification and monitoring of new ECLEs will be required to explore the full range of parameters displayed by the group and provide more rigid constraints for physical modelling. Given the recent discovery of coronal-line TDEs, and the varied, but long, duration of ECLE behaviour, additional late-time observations of known TDE host galaxies are clearly required. Such observations are needed to determine how common ECLE behaviour is following a TDE, along with placing better constraints on the timescales for both the onset and duration of such behaviour. In turn, this will improve our understanding of the local environments of SMBHs; the diversity of behaviour observed in ECLEs is likely to be strongly linked to the location and composition of material close to the SMBH responsible for the initial stellar disruption.

\section*{Acknowledgements}

This work was supported by the Science \& Technology Facilities Council [grants ST/S000550/1 and ST/W001225/1].
It was also funded by ANID, Millennium Science Initiative, ICN12\_009.
T.E.M.B. acknowledges financial support from the Spanish Ministerio de Ciencia e Innovaci\'on (MCIN), the Agencia Estatal de Investigaci\'on (AEI) 10.13039/501100011033, and the European Union Next Generation EU/PRTR funds under the 2021 Juan de la Cierva program FJC2021-04.
M.N. is supported by the European Research Council (ERC) under the European Union’s Horizon 2020 research and innovation programme (grant agreement No.~948381) and by funding from the UK Space Agency.
A.V.F.'s group at U.C. Berkeley has received financial assistance from the Christopher R. Redlich Fund, Alan Eustace (W.Z. is a Eustace Specialist in Astronomy), Briggs and Kathleen Wood (T.G.B. is a Wood Specialist in Astronomy), and many other donors.
This material is based upon work supported by the U.S. Department of Energy (DOE), Office of Science, Office of High-Energy Physics, under Contract No. DE-AC02-05CH11231, and by the National Energy Research Scientific Computing Center, a DOE Office of Science User Facility under the same contract. Additional support for DESI was provided by the U.S. National Science Foundation (NSF), Division of Astronomical Sciences under Contract No. AST-0950945 to the NSF's National Optical-Infrared Astronomy Research Laboratory; the Science and Technology Facilities Council of the United Kingdom; the Gordon and Betty Moore Foundation; the Heising-Simons Foundation; the French Alternative Energies and Atomic Energy Commission (CEA); the National Council of Science and Technology of Mexico (CONACYT); the Ministry of Science and Innovation of Spain (MICINN), and by the DESI Member Institutions: \url{https://www.desi.lbl.gov/collaborating-institutions}. Any opinions, findings, and conclusions or recommendations expressed in this material are those of the author(s) and do not necessarily reflect the views of the U. S. National Science Foundation, the U. S. Department of Energy, or any of the listed funding agencies.
The authors are honored to be permitted to conduct scientific research on Iolkam Du'ag (Kitt Peak), a mountain with particular significance to the Tohono O'odham Nation.
Based in part on observations collected at the European Organisation for Astronomical Research in the Southern Hemisphere, Chile, as part of ePESSTO+ (the advanced Public ESO Spectroscopic Survey for Transient Objects Survey). ePESSTO+ observations were obtained under ESO program IDs 1103.D-0328 and 106.216C (PI Inserra).
Some of the observations reported here were obtained at the MMT Observatory, a joint facility of the Smithsonian Institution and the University of Arizona.
The Kast red CCD detector upgrade on the Shane 3~m telescope at Lick Observatory, led by B. Holden, was made possible by the Heising–Simons Foundation, William and Marina Kast, and the University of California Observatories. Research at Lick Observatory is partially supported by a generous gift from Google.    

Funding for the Sloan Digital Sky Survey IV has been provided by the Alfred P. Sloan Foundation, the U.S. Department of Energy Office of Science, and the Participating Institutions. SDSS acknowledges support and resources from the Center for High-Performance Computing at the University of Utah. The SDSS website is www.sdss.org.

This research has made use of the NASA/IPAC Infrared Science Archive, which is funded by the National Aeronautics and Space Administration (NASA) and operated by the California Institute of Technology.
This publication also makes use of data products from NEOWISE, which is a project of the Jet Propulsion Laboratory/California Institute of Technology, funded by the Planetary Science Division of NASA.
The CRTS survey is supported by the U.S. National Science Foundation (NSF) under grants AST-0909182 and AST-1313422.

This work has made use of data from the Asteroid Terrestrial-impact Last Alert System (ATLAS) project. The ATLAS project is primarily funded to search for near-Earth objects through NASA grants NN12AR55G, 80NSSC18K0284, and 80NSSC18K1575; byproducts of the NEO search include images and catalogues from the survey area. This work was partially funded by Kepler/K2 grant J1944/80NSSC19K0112 and HST GO-15889, and STFC grants ST/T000198/1 and ST/S006109/1. The ATLAS science products have been made possible through the contributions of the University of Hawaii Institute for Astronomy, the Queen’s University Belfast, the Space Telescope Science Institute, the South African Astronomical Observatory, and The Millennium Institute of Astrophysics (MAS), Chile.

The Pan-STARRS1 Surveys (PS1) and the PS1 public science archive have been made possible through contributions by the Institute for Astronomy, the University of Hawaii, the Pan-STARRS Project Office, the Max-Planck Society and its participating institutes, the Max Planck Institute for Astronomy, Heidelberg and the Max Planck Institute for Extraterrestrial Physics, Garching, The Johns Hopkins University, Durham University, the University of Edinburgh, the Queen's University Belfast, the Harvard-Smithsonian Center for Astrophysics, the Las Cumbres Observatory Global Telescope Network Incorporated, the National Central University of Taiwan, the Space Telescope Science Institute, NASA under grant  NNX08AR22G issued through the Planetary Science Division of the NASA Science Mission Directorate, NSF grant AST-1238877, the University of Maryland, Eotvos Lorand University (ELTE), the Los Alamos National Laboratory, and the Gordon and Betty Moore Foundation.

Based in part on observations obtained with the Samuel Oschin Telescope 48-inch and the 60-inch Telescope at the Palomar Observatory as part of the Zwicky Transient Facility project. ZTF is supported by the NSF under grants AST-1440341 and AST-2034437, and a collaboration including current partners Caltech, IPAC, the Weizmann Institute for Science, the Oskar Klein Center at Stockholm University, the University of Maryland, Deutsches Elektronen-Synchrotron and Humboldt University, the TANGO Consortium of Taiwan, the University of Wisconsin at Milwaukee, Trinity College Dublin, Lawrence Livermore National Laboratories, IN2P3, University of Warwick, Ruhr University Bochum, Northwestern University and former partners the University of Washington, Los Alamos National Laboratories, and Lawrence Berkeley National Laboratories. Operations are conducted by COO, IPAC, and UW.

The authors thank Chenwei Yang for providing the spectra used in the 2013 analysis for comparative use in this work, and for clarification of the previous BPT analysis.
We also thank the anonymous reviewer for their helpful comments and suggestions.

\section*{Data Availability}
The data underlying this work are available in the article and in its online supplementary material available through Zenodo \citep{clark_2023_LongtermFollowupObservations}. Previously unpublished spectra will be made available through the Weizmann Interactive Supernova Data Repository (WISeREP) online archive.



\bibliographystyle{mnras}
\bibliography{Long-term_Follow-up_of_ECLEs.bib}

\begin{thebibliography}{}
\makeatletter
\relax
\def\mn@urlcharsother{\let\do\@makeother \do\$\do\&\do\#\do\^\do\_\do\%\do\~}
\def\mn@doi{\begingroup\mn@urlcharsother \@ifnextchar [ {\mn@doi@} {\mn@doi@[]}}
\def\mn@doi@[#1]#2{\def\@tempa{#1}\ifx\@tempa\@empty \href {http://dx.doi.org/#2} {doi:#2}\else \href {http://dx.doi.org/#2} {#1}\fi \endgroup}
\def\mn@eprint#1#2{\mn@eprint@#1:#2::\@nil}
\def\mn@eprint@arXiv#1{\href {http://arxiv.org/abs/#1} {{\tt arXiv:#1}}}
\def\mn@eprint@dblp#1{\href {http://dblp.uni-trier.de/rec/bibtex/#1.xml} {dblp:#1}}
\def\mn@eprint@#1:#2:#3:#4\@nil{\def\@tempa {#1}\def\@tempb {#2}\def\@tempc {#3}\ifx \@tempc \@empty \let \@tempc \@tempb \let \@tempb \@tempa \fi \ifx \@tempb \@empty \def\@tempb {arXiv}\fi \@ifundefined {mn@eprint@\@tempb}{\@tempb:\@tempc}{\expandafter \expandafter \csname mn@eprint@\@tempb\endcsname \expandafter{\@tempc}}}

\bibitem[\protect\citeauthoryear{Abazajian et~al.,}{Abazajian et~al.}{2009}]{abazajian_2009_SeventhDataRelease}
Abazajian K.~N.,  et~al., 2009, \mn@doi [The Astrophysical Journal Supplement Series] {10.1088/0067-0049/182/2/543}, 182, 543

\bibitem[\protect\citeauthoryear{Akaike}{Akaike}{1974}]{akaike_1974_NewLookStatistical}
Akaike H.,  1974, \mn@doi [IEEE Transactions on Automatic Control] {10.1109/TAC.1974.1100705}, 19, 716

\bibitem[\protect\citeauthoryear{Alexander, Wieringa, Berger, Saxton  \& Komossa}{Alexander et~al.}{2017}]{alexander_2017_RadioObservationsTidal}
Alexander K.~D.,  Wieringa M.~H.,  Berger E.,  Saxton R.~D.,   Komossa S.,  2017, \mn@doi [ApJ] {10.3847/1538-4357/aa6192}, 837, 153

\bibitem[\protect\citeauthoryear{Ambikasaran, {Foreman-Mackey}, Greengard, Hogg  \& O'Neil}{Ambikasaran et~al.}{2015}]{ambikasaran_2015_FastDirectMethods}
Ambikasaran S.,  {Foreman-Mackey} D.,  Greengard L.,  Hogg D.~W.,   O'Neil M.,  2015, \mn@doi [IEEE Transactions on Pattern Analysis and Machine Intelligence] {10.1109/TPAMI.2015.2448083}, 38, 252

\bibitem[\protect\citeauthoryear{Arcavi et~al.,}{Arcavi et~al.}{2014}]{arcavi_2014_CONTINUUMHeRICHTIDAL}
Arcavi I.,  et~al., 2014, \mn@doi [The Astrophysical Journal] {10.1088/0004-637X/793/1/38}, 793, 38

\bibitem[\protect\citeauthoryear{Auchettl, Guillochon  \& {Ramirez-Ruiz}}{Auchettl et~al.}{2017}]{auchettl_2017_NewPhysicalInsights}
Auchettl K.,  Guillochon J.,   {Ramirez-Ruiz} E.,  2017, \mn@doi [ApJ] {10.3847/1538-4357/aa633b}, 838, 149

\bibitem[\protect\citeauthoryear{Ayal, Livio  \& Piran}{Ayal et~al.}{2000}]{ayal_2000_TidalDisruptionSolarType}
Ayal S.,  Livio M.,   Piran T.,  2000, \mn@doi [The Astrophysical Journal] {10.1086/317835}, 545, 772

\bibitem[\protect\citeauthoryear{Bade, Komossa  \& Dahlem}{Bade et~al.}{1996}]{bade_1996_DetectionExtremelySoft}
Bade N.,  Komossa S.,   Dahlem M.,  1996, Astronomy and Astrophysics, 309, L35

\bibitem[\protect\citeauthoryear{Baldwin, Phillips  \& Terlevich}{Baldwin et~al.}{1981}]{baldwin_1981_ClassificationParametersEmissionline}
Baldwin J.~A.,  Phillips M.~M.,   Terlevich R.,  1981, \mn@doi [Publications of the Astronomical Society of the Pacific] {10.1086/130766}, 93, 5

\bibitem[\protect\citeauthoryear{Barlow}{Barlow}{2003}]{barlow_2003_AsymmetricErrors}
Barlow R.,  2003, in Statistical {{Problems}} in {{Particle Physics}}, {{Astrophysics}}, and {{Cosmology}}. {arXiv}, {Stanford Linear Accelerator Center}, p.~250, \mn@doi{10.48550/ARXIV.PHYSICS/0401042}

\bibitem[\protect\citeauthoryear{Bellm et~al.,}{Bellm et~al.}{2019}]{bellm_2019_ZwickyTransientFacility}
Bellm E.~C.,  et~al., 2019, \mn@doi [Publications of the Astronomical Society of the Pacific] {10.1088/1538-3873/aaecbe}, 131, 18002

\bibitem[\protect\citeauthoryear{Blanco et~al.,}{Blanco et~al.}{2004}]{blanco_2004_NewMMT}
Blanco D.,  et~al., 2004, in Ground-Based {{Telescopes}}. {SPIE}, pp 300--311, \mn@doi{10.1117/12.551963}

\bibitem[\protect\citeauthoryear{Buzzoni et~al.,}{Buzzoni et~al.}{1984}]{buzzoni_1984_ESOFaintObject}
Buzzoni B.,  et~al., 1984, ESO Messenger (ISSN 0722-6691), pp 9--13

\bibitem[\protect\citeauthoryear{Cackett, Bentz  \& Kara}{Cackett et~al.}{2021}]{cackett_2021_ReverberationMappingActive}
Cackett E.~M.,  Bentz M.~C.,   Kara E.,  2021, \mn@doi [iScience] {10.1016/j.isci.2021.102557}, 24, 102557

\bibitem[\protect\citeauthoryear{Cannizzo \& Gehrels}{Cannizzo \& Gehrels}{2009}]{cannizzo_2009_NewParadigmGammaray}
Cannizzo J.~K.,  Gehrels N.,  2009, \mn@doi [The Astrophysical Journal] {10.1088/0004-637X/700/2/1047}, 700, 1047

\bibitem[\protect\citeauthoryear{Cannizzo, Lee  \& Goodman}{Cannizzo et~al.}{1990}]{cannizzo_1990_DiskAccretionTidally}
Cannizzo J.~K.,  Lee H.~M.,   Goodman J.,  1990, \mn@doi [The Astrophysical Journal] {10.1086/168442}, 351, 38

\bibitem[\protect\citeauthoryear{Cannizzo, Troja  \& Lodato}{Cannizzo et~al.}{2011}]{cannizzo_2011_GRB110328ASwift}
Cannizzo J.~K.,  Troja E.,   Lodato G.,  2011, \mn@doi [The Astrophysical Journal] {10.1088/0004-637X/742/1/32}, 742, 32

\bibitem[\protect\citeauthoryear{Chambers et~al.,}{Chambers et~al.}{2016}]{chambers_2016_PanSTARRS1Surveys}
Chambers K.~C.,  et~al., 2016, In Prep

\bibitem[\protect\citeauthoryear{Charalampopoulos et~al.,}{Charalampopoulos et~al.}{2022}]{charalampopoulos_2022_DetailedSpectroscopicStudy}
Charalampopoulos P.,  et~al., 2022, \mn@doi [Astronomy and Astrophysics] {10.1051/0004-6361/202142122}, 659, A34

\bibitem[\protect\citeauthoryear{Clark, Graur, Callow  \& {al}}{Clark et~al.}{2023}]{clark_2023_LongtermFollowupObservations}
Clark P.,  Graur O.,  Callow J.,   {al} e.,  2023, \mn@doi [Zenodo] {10.5281/zenodo.8109559}

\bibitem[\protect\citeauthoryear{{DESI Collaboration} et~al.,}{{DESI Collaboration} et~al.}{2016a}]{desicollaboration_2016_DESIExperimentPart}
{DESI Collaboration} et~al., 2016a, The {{DESI Experiment Part I}}: {{Science}},{{Targeting}}, and {{Survey Design}}, \mn@doi{10.48550/arXiv.1611.00036}

\bibitem[\protect\citeauthoryear{{DESI Collaboration} et~al.,}{{DESI Collaboration} et~al.}{2016b}]{desicollaboration_2016_DESIExperimentParta}
{DESI Collaboration} et~al., 2016b, The {{DESI Experiment Part II}}: {{Instrument Design}}, \mn@doi{10.48550/arXiv.1611.00037}

\bibitem[\protect\citeauthoryear{{DESI Collaboration} et~al.,}{{DESI Collaboration} et~al.}{2023b}]{desicollaboration_2023_EarlyDataRelease}
{DESI Collaboration} et~al., 2023b, The {{Early Data Release}} of the {{Dark Energy Spectroscopic Instrument}}, \mn@doi{10.48550/arXiv.2306.06308}

\bibitem[\protect\citeauthoryear{{DESI Collaboration} et~al.,}{{DESI Collaboration} et~al.}{2023a}]{desicollaboration_2023_ValidationScientificProgram}
{DESI Collaboration} et~al., 2023a, Validation of the {{Scientific Program}} for the {{Dark Energy Spectroscopic Instrument}}, \mn@doi{10.48550/arXiv.2306.06307}

\bibitem[\protect\citeauthoryear{Dou, Wang, Jiang, Yang, Lyu  \& Zhou}{Dou et~al.}{2016}]{dou_2016_LONGFADINGMIDINFRARED}
Dou L.,  Wang T.-g.,  Jiang N.,  Yang C.,  Lyu J.,   Zhou H.,  2016, \mn@doi [ApJ] {10.3847/0004-637X/832/2/188}, 832, 188

\bibitem[\protect\citeauthoryear{Dou, Wang, Yan, Jiang, Yang, Cutri, Mainzer  \& Peng}{Dou et~al.}{2017}]{dou_2017_DiscoveryMidinfraredEcho}
Dou L.,  Wang T.,  Yan L.,  Jiang N.,  Yang C.,  Cutri R.~M.,  Mainzer A.,   Peng B.,  2017, \mn@doi [ApJ] {10.3847/2041-8213/aa7130}, 841, L8

\bibitem[\protect\citeauthoryear{Drake et~al.,}{Drake et~al.}{2009}]{drake_2009_FIRSTRESULTSCATALINA}
Drake A.~J.,  et~al., 2009, \mn@doi [ApJ] {10.1088/0004-637X/696/1/870}, 696, 870

\bibitem[\protect\citeauthoryear{Evans \& Kochanek}{Evans \& Kochanek}{1989}]{evans_1989_TidalDisruptionStar}
Evans C.~R.,  Kochanek C.~S.,  1989, \mn@doi [The Astrophysical Journal] {10.1086/185567}, 346, L13

\bibitem[\protect\citeauthoryear{Fabricant et~al.,}{Fabricant et~al.}{2019}]{fabricant_2019_BinospecWidefieldImaging}
Fabricant D.,  et~al., 2019, \mn@doi [PASP] {10.1088/1538-3873/ab1d78}, 131, 075004

\bibitem[\protect\citeauthoryear{Filippenko}{Filippenko}{1982}]{filippenko_1982_ImportanceAtmosphericDifferential}
Filippenko A.~V.,  1982, \mn@doi [Publications of the Astronomical Society of the Pacific] {10.1086/131052}, 94, 715

\bibitem[\protect\citeauthoryear{Fitzpatrick}{Fitzpatrick}{1999}]{fitzpatrick_1999_CorrectingEffectsInterstellar}
Fitzpatrick E.~L.,  1999, \mn@doi [Publications of the Astronomical Society of the Pacific] {10.1086/316293}, 111, 63

\bibitem[\protect\citeauthoryear{Flewelling}{Flewelling}{2018}]{flewelling_2018_PanSTARRSDataRelease}
Flewelling H.,  2018, in American {{Astronomical Society Meeting Abstracts}} \#231. p. 436.01

\bibitem[\protect\citeauthoryear{Fraser et~al.,}{Fraser et~al.}{2017}]{fraser_2017_EPESSTOSpectroscopicClassification}
Fraser M.,  et~al., 2017, The Astronomer's Telegram, 10747, 1

\bibitem[\protect\citeauthoryear{Fremling}{Fremling}{2023}]{fremling_2023_ZTFTransientDiscovery}
Fremling C.,  2023, Transient Name Server Discovery Report, 2023--1230, 1

\bibitem[\protect\citeauthoryear{Gehrels et~al.,}{Gehrels et~al.}{2004}]{gehrels_2004_SwiftGammaRayBursta}
Gehrels N.,  et~al., 2004, \mn@doi [The Astrophysical Journal] {10.1086/422091}, 611, 1005

\bibitem[\protect\citeauthoryear{Gezari et~al.,}{Gezari et~al.}{2017}]{gezari_2017_IPTFDiscoveryRapid}
Gezari S.,  et~al., 2017, \mn@doi [ApJ] {10.3847/1538-4357/835/2/144}, 835, 144

\bibitem[\protect\citeauthoryear{Graur, French, Zahid, Guillochon, Mandel, Auchettl  \& Zabludoff}{Graur et~al.}{2018}]{graur_2018_DependenceTidalDisruptiona}
Graur O.,  French K.~D.,  Zahid H.~J.,  Guillochon J.,  Mandel K.~S.,  Auchettl K.,   Zabludoff A.~I.,  2018, \mn@doi [ApJ] {10.3847/1538-4357/aaa3fd}, 853, 39

\bibitem[\protect\citeauthoryear{Gromadzki, Rybicki, Fraser, Yaron  \& Knezevic}{Gromadzki et~al.}{2017}]{gromadzki_2017_EPESSTOTransientClassification}
Gromadzki M.,  Rybicki K.,  Fraser M.,  Yaron O.,   Knezevic N.,  2017, Transient Name Server Classification Report, 2017--1001, 1

\bibitem[\protect\citeauthoryear{Guillochon \& {Ramirez-Ruiz}}{Guillochon \& {Ramirez-Ruiz}}{2013}]{guillochon_2013_HydrodynamicalSimulationsDetermine}
Guillochon J.,  {Ramirez-Ruiz} E.,  2013, \mn@doi [The Astrophysical Journal] {10.1088/0004-637X/767/1/25}, 767, 25

\bibitem[\protect\citeauthoryear{Guy et~al.,}{Guy et~al.}{2023}]{guy_2023_SpectroscopicDataProcessing}
Guy J.,  et~al., 2023, \mn@doi [The Astronomical Journal] {10.3847/1538-3881/acb212}, 165, 144

\bibitem[\protect\citeauthoryear{Hahn et~al.,}{Hahn et~al.}{2023}]{hahn_2023_DESIBrightGalaxy}
Hahn C.,  et~al., 2023, \mn@doi [The Astronomical Journal] {10.3847/1538-3881/accff8}, 165, 253

\bibitem[\protect\citeauthoryear{Hammerstein et~al.,}{Hammerstein et~al.}{2023}]{hammerstein_2023_FinalSeasonReimagined}
Hammerstein E.,  et~al., 2023, \mn@doi [The Astrophysical Journal] {10.3847/1538-4357/aca283}, 942, 9

\bibitem[\protect\citeauthoryear{Hawkins}{Hawkins}{2002}]{hawkins_2002_VariabilityActiveGalactic}
Hawkins M.,  2002, \mn@doi [Monthly Notices of the Royal Astronomical Society] {10.1046/j.1365-8711.2002.04939.x}, 329, 76

\bibitem[\protect\citeauthoryear{Hayasaki \& Jonker}{Hayasaki \& Jonker}{2021}]{hayasaki_2021_OriginLatetimeXRay}
Hayasaki K.,  Jonker P.~G.,  2021, \mn@doi [The Astrophysical Journal] {10.3847/1538-4357/ac18c2}, 921, 20

\bibitem[\protect\citeauthoryear{He, Dou, Ai, Shu, Jiang, Wang, Zhang  \& Shen}{He et~al.}{2021}]{he_2021_LongtermXrayEvolution}
He J.~S.,  Dou L.~M.,  Ai Y.~L.,  Shu X.~W.,  Jiang N.,  Wang T.~G.,  Zhang F.~B.,   Shen R.~F.,  2021, arXiv:2106.03692 [astro-ph]

\bibitem[\protect\citeauthoryear{Hills}{Hills}{1975}]{hills_1975_PossiblePowerSource}
Hills J.~G.,  1975, \mn@doi [Nature] {10.1038/254295a0}, 254, 295

\bibitem[\protect\citeauthoryear{Hinkle, Holoien, Shappee  \& Auchettl}{Hinkle et~al.}{2021}]{hinkle_2021_SwiftFixNuclear}
Hinkle J.~T.,  Holoien T. W.-S.,  Shappee {\relax Benjamin}.~J.,   Auchettl K.,  2021, \mn@doi [ApJ] {10.3847/1538-4357/abe4d8}, 910, 83

\bibitem[\protect\citeauthoryear{Hinkle, Shappee  \& Holoien}{Hinkle et~al.}{2023}]{hinkle_2023_CoronalLineEmitters}
Hinkle J.~T.,  Shappee B.~J.,   Holoien T. W.-S.,  2023, Coronal {{Line Emitters}} Are {{Tidal Disruption Events}} in {{Gas-Rich Environments}} (\mn@eprint {arxiv} {2303.05525}), \mn@doi{10.48550/arXiv.2303.05525}

\bibitem[\protect\citeauthoryear{Hung et~al.,}{Hung et~al.}{2021}]{hung_2021_DiscoveryFastIron}
Hung T.,  et~al., 2021, \mn@doi [The Astrophysical Journal] {10.3847/1538-4357/abf4c3}, 917, 9

\bibitem[\protect\citeauthoryear{Hyland \& Allen}{Hyland \& Allen}{1982}]{hyland_1982_InfraredStudyQuasars}
Hyland A.~R.,  Allen D.~A.,  1982, \mn@doi [Monthly Notices of the Royal Astronomical Society] {10.1093/mnras/199.4.943}, 199, 943

\bibitem[\protect\citeauthoryear{Jansen et~al.,}{Jansen et~al.}{2001}]{jansen_2001_XMMNewtonObservatorySpacecraft}
Jansen F.,  et~al., 2001, \mn@doi [A\&A] {10.1051/0004-6361:20000036}, 365, L1

\bibitem[\protect\citeauthoryear{Jiang et~al.,}{Jiang et~al.}{2021}]{jiang_2021_MidinfraredOutburstsNearby}
Jiang N.,  et~al., 2021, \mn@doi [ApJS] {10.3847/1538-4365/abd1dc}, 252, 32

\bibitem[\protect\citeauthoryear{Kauffmann et~al.,}{Kauffmann et~al.}{2003}]{kauffmann_2003_HostGalaxiesActive}
Kauffmann G.,  et~al., 2003, \mn@doi [Monthly Notices of the Royal Astronomical Society] {10.1111/j.1365-2966.2003.07154.x}, 346, 1055

\bibitem[\protect\citeauthoryear{Kent et~al.,}{Kent et~al.}{2016}]{kent_2016_ImpactOpticalDistortions}
Kent S.,  et~al., 2016, in Proceedings of the {{SPIE}}. p. 99088F, \mn@doi{10.1117/12.2232689}

\bibitem[\protect\citeauthoryear{Kesden}{Kesden}{2012}]{kesden_2012_TidaldisruptionRateStars}
Kesden M.,  2012, \mn@doi [Phys. Rev. D] {10.1103/PhysRevD.85.024037}, 85, 024037

\bibitem[\protect\citeauthoryear{Kewley, Dopita, Sutherland, Heisler  \& Trevena}{Kewley et~al.}{2001}]{kewley_2001_TheoreticalModelingStarburst}
Kewley L.~J.,  Dopita M.~A.,  Sutherland R.~S.,  Heisler C.~A.,   Trevena J.,  2001, \mn@doi [The Astrophysical Journal] {10.1086/321545}, 556, 121

\bibitem[\protect\citeauthoryear{Komossa et~al.,}{Komossa et~al.}{2008}]{komossa_2008_DiscoverySuperstrongFading}
Komossa S.,  et~al., 2008, \mn@doi [The Astrophysical Journal] {10.1086/588281}, 678, L13

\bibitem[\protect\citeauthoryear{Komossa et~al.,}{Komossa et~al.}{2009}]{komossa_2009_NTTSpitzerChandra}
Komossa S.,  et~al., 2009, \mn@doi [Astrophysical Journal] {10.1088/0004-637X/701/1/105}, 701, 105

\bibitem[\protect\citeauthoryear{Lacy, Townes  \& Hollenbach}{Lacy et~al.}{1982}]{lacy_1982_NatureCentralParsec}
Lacy J.~H.,  Townes C.~H.,   Hollenbach D.~J.,  1982, \mn@doi [The Astrophysical Journal] {10.1086/160402}, 262, 120

\bibitem[\protect\citeauthoryear{Lodato \& Rossi}{Lodato \& Rossi}{2011}]{lodato_2011_MultibandLightCurves}
Lodato G.,  Rossi E.~M.,  2011, \mn@doi [Monthly Notices of the Royal Astronomical Society] {10.1111/j.1365-2966.2010.17448.x}, 410, 359

\bibitem[\protect\citeauthoryear{Mainzer et~al.,}{Mainzer et~al.}{2011}]{mainzer_2011_NEOWISEOBSERVATIONSNEAREARTH}
Mainzer A.,  et~al., 2011, \mn@doi [ApJ] {10.1088/0004-637X/743/2/156}, 743, 156

\bibitem[\protect\citeauthoryear{Mainzer et~al.,}{Mainzer et~al.}{2014}]{mainzer_2014_INITIALPERFORMANCENEOWISE}
Mainzer A.,  et~al., 2014, \mn@doi [ApJ] {10.1088/0004-637X/792/1/30}, 792, 30

\bibitem[\protect\citeauthoryear{Martin et~al.,}{Martin et~al.}{2005}]{martin_2005_GalaxyEvolutionExplorer}
Martin D.~C.,  et~al., 2005, \mn@doi [ApJ] {10.1086/426387}, 619, L1

\bibitem[\protect\citeauthoryear{Masci et~al.,}{Masci et~al.}{2018}]{masci_2018_ZwickyTransientFacility}
Masci F.~J.,  et~al., 2018, \mn@doi [PASP] {10.1088/1538-3873/aae8ac}, 131, 018003

\bibitem[\protect\citeauthoryear{Miller \& Stone}{Miller \& Stone}{1994}]{miller_1994_KASTDoubleSpectrograph}
Miller J.~S.,  Stone R.~P.,  1994, Lick Observatory Technical Reports, 66

\bibitem[\protect\citeauthoryear{Nagao, Taniguchi  \& Murayama}{Nagao et~al.}{2000}]{nagao_2000_HighIonizationNuclearEmissionLine}
Nagao T.,  Taniguchi Y.,   Murayama T.,  2000, \mn@doi [AJ] {10.1086/301411}, 119, 2605

\bibitem[\protect\citeauthoryear{Newsome, Arcavi, Dgany  \& Pellegrino}{Newsome et~al.}{2022}]{newsome_2022_2022upjZTF22abegjtxDiscovery}
Newsome M.,  Arcavi I.,  Dgany Y.,   Pellegrino C.,  2022, Transient Name Server AstroNote, 236, 1

\bibitem[\protect\citeauthoryear{Nicholl et~al.,}{Nicholl et~al.}{2020}]{nicholl_2020_OutflowPowersOptical}
Nicholl M.,  et~al., 2020, \mn@doi [Monthly Notices of the Royal Astronomical Society] {10.1093/mnras/staa2824}, 499, 482

\bibitem[\protect\citeauthoryear{Onori et~al.,}{Onori et~al.}{2022}]{onori_2022_NuclearTransient2017ggea}
Onori F.,  et~al., 2022, \mn@doi [Monthly Notices of the Royal Astronomical Society] {10.1093/mnras/stac2673}, 517, 76

\bibitem[\protect\citeauthoryear{Palaversa, Gezari, Sesar, Stuart, Wozniak, Holl  \& Ivezi{\'c}}{Palaversa et~al.}{2016}]{palaversa_2016_REVEALINGNATUREEXTREME}
Palaversa L.,  Gezari S.,  Sesar B.,  Stuart J.~S.,  Wozniak P.,  Holl B.,   Ivezi{\'c} {\v Z}.,  2016, \mn@doi [ApJ] {10.3847/0004-637X/819/2/151}, 819, 151

\bibitem[\protect\citeauthoryear{Phinney}{Phinney}{1989}]{phinney_1989_CosmicMergerMania}
Phinney E.~S.,  1989, \mn@doi [Nature] {10.1038/340595a0}, 340, 595

\bibitem[\protect\citeauthoryear{Rees}{Rees}{1988}]{rees_1988_TidalDisruptionStars}
Rees M.~J.,  1988, \mn@doi [Nature] {10.1038/333523a0}, 333, 523

\bibitem[\protect\citeauthoryear{Schlafly \& Finkbeiner}{Schlafly \& Finkbeiner}{2011}]{schlafly_2011_MeasuringReddeningSloan}
Schlafly E.~F.,  Finkbeiner D.~P.,  2011, \mn@doi [Astrophysical Journal] {10.1088/0004-637X/737/2/103}, 737

\bibitem[\protect\citeauthoryear{Shingles et~al.,}{Shingles et~al.}{2021}]{shingles_2021_ReleaseATLASForced}
Shingles L.,  et~al., 2021, Transient Name Server AstroNote, 7

\bibitem[\protect\citeauthoryear{Short et~al.,}{Short et~al.}{2023}]{short_2023_DelayedAppearanceEvolution}
Short P.,  et~al., 2023, \mn@doi [Monthly Notices of the Royal Astronomical Society] {10.1093/mnras/stad2270}, 525, 1568

\bibitem[\protect\citeauthoryear{Silverman et~al.,}{Silverman et~al.}{2012}]{silverman_2012_BerkeleySupernovaIaa}
Silverman J.~M.,  et~al., 2012, \mn@doi [Monthly Notices of the Royal Astronomical Society] {10.1111/j.1365-2966.2012.21270.x}, 425, 1789

\bibitem[\protect\citeauthoryear{Skrutskie et~al.,}{Skrutskie et~al.}{2006}]{skrutskie_2006_TwoMicronAll}
Skrutskie M.~F.,  et~al., 2006, \mn@doi [AJ] {10.1086/498708}, 131, 1163

\bibitem[\protect\citeauthoryear{Smartt et~al.,}{Smartt et~al.}{2015}]{smartt_2015_PESSTOSurveyDescription}
Smartt S.~J.,  et~al., 2015, \mn@doi [A\&A] {10.1051/0004-6361/201425237}, 579, A40

\bibitem[\protect\citeauthoryear{Smith et~al.,}{Smith et~al.}{2020}]{smith_2020_DesignOperationATLAS}
Smith K.~W.,  et~al., 2020, \mn@doi [Publications of the Astronomical Society of the Pacific] {10.1088/1538-3873/ab936e}, 132, 85002

\bibitem[\protect\citeauthoryear{Stern et~al.,}{Stern et~al.}{2012}]{stern_2012_MIDINFRAREDSELECTIONACTIVE}
Stern D.,  et~al., 2012, \mn@doi [ApJ] {10.1088/0004-637X/753/1/30}, 753, 30

\bibitem[\protect\citeauthoryear{Stokes, Evans, Viggh, Shelly  \& Pearce}{Stokes et~al.}{2000}]{stokes_2000_LincolnNearEarthAsteroida}
Stokes G.~H.,  Evans J.~B.,  Viggh H.~E.,  Shelly F.~C.,   Pearce E.~C.,  2000, \mn@doi [Icarus] {10.1006/icar.2000.6493}, 148, 21

\bibitem[\protect\citeauthoryear{Tody}{Tody}{1986}]{tody_1986_IRAFDataReduction}
Tody D.,  1986, \mn@doi [Instrumentation in Astronomy VI] {10.1117/12.968154}, 627, 733

\bibitem[\protect\citeauthoryear{Tonry et~al.,}{Tonry et~al.}{2018}]{tonry_2018_ATLASHighcadenceAllsky}
Tonry J.~L.,  et~al., 2018, \mn@doi [Publications of the Astronomical Society of the Pacific] {10.1088/1538-3873/aabadf}, 130, 64505

\bibitem[\protect\citeauthoryear{Ulmer}{Ulmer}{1999}]{ulmer_1999_FlaresTidalDisruption}
Ulmer A.,  1999, \mn@doi [ApJ] {10.1086/306909}, 514, 180

\bibitem[\protect\citeauthoryear{Veilleux \& Osterbrock}{Veilleux \& Osterbrock}{1987}]{veilleux_1987_SpectralClassificationEmissionLine}
Veilleux S.,  Osterbrock D.~E.,  1987, \mn@doi [The Astrophysical Journal Supplement Series] {10.1086/191166}, 63, 295

\bibitem[\protect\citeauthoryear{Wang, Zhou, Wang, Lu  \& Xu}{Wang et~al.}{2011}]{wang_2011_TransientSuperstrongCoronala}
Wang T.-G.,  Zhou H.-Y.,  Wang L.-F.,  Lu H.-L.,   Xu D.,  2011, \mn@doi [ApJ] {10.1088/0004-637X/740/2/85}, 740, 85

\bibitem[\protect\citeauthoryear{Wang, Zhou, Komossa, Wang, Yuan  \& Yang}{Wang et~al.}{2012}]{wang_2012_EXTREMECORONALLINE}
Wang T.-G.,  Zhou H.-Y.,  Komossa S.,  Wang H.-Y.,  Yuan W.,   Yang C.,  2012, \mn@doi [The Astrophysical Journal] {10.1088/0004-637X/749/2/115}, 749, 115

\bibitem[\protect\citeauthoryear{Wang et~al.,}{Wang et~al.}{2022}]{wang_2022_MidinfraredOutburstsNearby}
Wang Y.,  et~al., 2022, \mn@doi [The Astrophysical Journal Supplement Series] {10.3847/1538-4365/ac33a6}, 258, 21

\bibitem[\protect\citeauthoryear{Wright et~al.,}{Wright et~al.}{2010}]{wright_2010_WIDEFIELDINFRAREDSURVEY}
Wright E.~L.,  et~al., 2010, \mn@doi [The Astronomical Journal] {10.1088/0004-6256/140/6/1868}, 140, 1868

\bibitem[\protect\citeauthoryear{Yang, Wang, Ferland, Yuan, Zhou  \& Jiang}{Yang et~al.}{2013}]{yang_2013_LONGTERMSPECTRALEVOLUTION}
Yang C.-W.,  Wang T.-G.,  Ferland G.,  Yuan W.,  Zhou H.-Y.,   Jiang P.,  2013, \mn@doi [The Astrophysical Journal] {10.1088/0004-637X/774/1/46}, 774, 46

\bibitem[\protect\citeauthoryear{York et~al.,}{York et~al.}{2000}]{york_2000_SloanDigitalSky}
York D.~G.,  et~al., 2000, \mn@doi [AJ] {10.1086/301513}, 120, 1579

\bibitem[\protect\citeauthoryear{{van Velzen} et~al.,}{{van Velzen} et~al.}{2021}]{vanvelzen_2021_SeventeenTidalDisruption}
{van Velzen} S.,  et~al., 2021, \mn@doi [ApJ] {10.3847/1538-4357/abc258}, 908, 4

\makeatother
\end{thebibliography}

\clearpage
\appendix

\onecolumn
\section{Object Summary Information}
\label{Appendix:Object_Summary_Info}

In this appendix, we provide summary information on the properties of the ECLE sample used within this work and detail the parameters of the spectroscopic observations utilised in the analysis.

\FloatBarrier

\begin{table*}
\caption{Summary information for the ECLE sample used in this paper}
\label{Tab:Object_Summary_Info}
\begin{tabular}{lcccccc}
\hline
Object & Short Name & RA (J2000) & Dec (J2000) & Redshift $z$ & Other Host Name & Classification \\ \hline
SDSS~J0748+4712 & SDSS~J0748 & 07:48:20.6668 & +47:12:14.2648 & 0.062 & 2MASS J07482067+4712138 & TDE light echo \\
SDSS~J0938+1353 & SDSS~J0938 & 09:38:01.6376 & +13:53:17.0423 & 0.101 & 2MASX J09380164+1353168 & AGN \\
SDSS~J0952+2143 & SDSS~J0952 & 09:52:09.5629 & +21:43:13.2979 & 0.079 & 2MASS J09520955+2143132 & TDE light echo \\
SDSS~J1055+5637 & SDSS~J1055 & 10:55:26.4177 & +56:37:13.1010 & 0.074 & 2MASS J10552641+5637129 & AGN \\
SDSS~J1241+4426 & SDSS~J1241 & 12:41:34.2561 & +44:26:39.2636 & 0.042 & LEDA 2244532 & TDE light echo \\
SDSS~J1342+0530 & SDSS~J1342 & 13:42:44.4150 & +05:30:56.1451 & 0.037 & 2MASX J13424441+0530560 & TDE light echo \\
SDSS~J1350+2916 & SDSS~J1350 & 13:50:01.4946 & +29:16:09.6460 & 0.078 & 2MASS J13500150+2916097 & TDE light echo \\ \hline
\end{tabular}
\end{table*}

\begin{landscape}
\begin{table}
\centering
\caption{Details on the combined optical spectroscopic dataset available for each ECLE.}
\label{Tab:Spectra_Info}
\begin{adjustbox}{width=1\linewidth}
\begin{tabular}{lcccccccccccccccl}
\hline
Object & Date & Telescope & Instrument & Resolution (R) & Source & \makecell{[Fe~VII]\\$\lambda$6088} & \makecell{[Fe~X]\\$\lambda$6376} &\makecell{[Fe~XI]\\$\lambda$7894} & \makecell{[Fe~XIV]\\$\lambda$5304}& H$\alpha$ & H$\beta$ & \makecell{[O~III]\\$\lambda$4959} & \makecell{[O~III]\\$\lambda$5007} & \makecell{He~II\\$\lambda$4686} & Broad Features & Notes \\ \hline
SDSS~J0748 & 2004 Feb 20 & SDSS & - & 1500 -- 2500 & 1 &  N &  Y &  Y &  Y &  Y &  Y &  Y &  Y &  Y & H + He & $\square$ \\
SDSS~J0748 & 2011 Dec 26 & MMT & BCS$\dag$ & 1730 Blue, 1430 Red  & 2 &  N &  N & - &  N &  Y &  Y &  Y &  Y &  N & None &\\
SDSS~J0748 & 2019 Oct 31 & MMT & BCS & 1730 Blue, 1430 Red & 3 &  N &  N &  N &  N &  Y &  Y &  Y &  Y &  N & None &\\
 &  &  &  &  &  &  &  &  &  &  &  &  &  &  &  &\\
SDSS~J0938 & 2006 Dec 23 & SDSS & - & 1500 -- 2500 & 1 &  Y &  Y &  Y &  Y &  Y &  Y &  Y &  Y &  Y & None & $\triangle$\\
SDSS~J0938 & 2011 Dec 26 & MMT & BCS & 1730 Blue, 1430 Red & 2 &  Y &  Y & - &  Y &  Y &  Y &  Y &  Y &  Y & None & Sey 2 / SF Composite, * \\
SDSS~J0938 & 2021 Apr 09 & NTT & EFOSC2 & & 3 &  Y &  Y &  Y &  Y &  Y &  Y &  Y &  Y &  Y & Unresolved &\\
SDSS~J0938 & 2022 Mar 22 & Mayall & DESI & 1500 -- 4000 & 3 &  Y &  Y &  Y &  Y &  Y &  Y &  Y &  Y &  Y & None \\
 &  &  &  &  &  &  &  &  &  &  &  &  &  &  &  &\\
SDSS~J0952 & 2005 Dec 30 & SDSS & - & 1500 -- 2500 & 1 &  Y &  Y &  Y &  Y &  Y &  Y &  Y &  Y &  Y & H & $\square$ \\
SDSS~J0952 & 2011 Dec 26 & MMT & BCS & 1730 Blue, 1430 Red & 2 &  Y &  Y & - &  Y &  Y &  Y &  Y &  Y &  Y & None & Fe and He II lines have faded significantly\\
SDSS~J0952 & 2021 May 11 & NTT & EFOSC2 & 310 & 3 &  N &  N &  N &  N &  Y &  N &  Y &  Y &  N & None & Low S/N, Observed with GR13\\
SDSS~J0952 & 2021 Nov 18 & Mayall & DESI & 1500 -- 4000 & 3 &  N &  N &  N &  N &  Y &  Y &  Y &  Y &  N & None \\
 &  &  &  &  &  &  &  &  &  &  &  &  &  &  &  &\\
SDSS~J1055 & 2002 Apr 09 & SDSS & - & 1500 -- 2500 & 1 &  Y &  Y &  Y &  Y &  Y &  Y &  Y &  Y &  Y & H & $\triangle$\\
SDSS~J1055 & 2011 Dec 26 & MMT & BCS & 1730 Blue, 1430 Red & 2 &  Y &  Y & - &  Y &  Y &  Y &  Y &  Y &  Y & H & Sey 1, * \\
SDSS~J1055 & 2021 May 18 & Shane 3m & Kast & $\approx 800$ & 3 &  Y &  Y &  Y &  Y &  Y &  Y &  Y &  Y &  U & H &\\
 &  &  &  &  &  & &  &  &  &  &  &  &  &  &  &  \\
SDSS~J1241 & 2004 Feb 27 & SDSS & - & 1500 -- 2500 & 1 &  Y &  Y &  Y &  Y &  Y &  Y &  Y &  Y &  Y & None & There is a unique feature blueward of $\Fex$, $\square$ \\
SDSS~J1241 & 2011 Dec 26 & MMT & BCS & 1730 & 2 & Y & - & - & - & - &  Y & - & - &  Y & None & Blue spectral region only, * \\
SDSS~J1241 & 2021 Jul 16 & Shane 3m & Kast & $\approx 800$ & 3 &  Y &  U &  N &  N &  Y &  Y &  Y &  Y &  Y & None \\
 &  &  &  &  &  & &  &  &  &  &  &  &  &  &  &\\
SDSS~J1342 & 2002 Apr 09 & SDSS & - & 1500 -- 2500 & 1 &  N &  Y &  Y &  Y &  Y &  Y &  Y &  Y &  Y & None & $\square$ \\
SDSS~J1342 & 2011 Dec 26 & MMT & BCS & 1730 Blue, 1430 Red & 2 &  Y &  N & - &  N &  Y &  Y &  Y &  Y &  Y & None &\\
SDSS~J1342 & 2021 Mar 06 & Mayall & DESI & 1500 -- 4000 & 3 &  Y &  N &  N &  N &  Y &  Y &  Y &  Y &  Y & None & \\
SDSS~J1342 & 2021 Mar 21 & NTT & EFOSC2 & 390, 440 & 3 &  U &  U &  U &  U &  Y &  Y &  Y &  Y &  Y & Unresolved & Low S/N, Observed with GR11 and GR 16 OG530\\
 &  &  &  &  &  &  & &  &  &  &  &  &  &  &  &\\
SDSS~J1350 & 2006 Apr 23 & SDSS & - & 1500 -- 2500 & 1 &  N &  Y &  Y &  Y &  Y &  Y &  Y &  Y &  Y & H + He & $\square$ \\
SDSS~J1350 & 2011 Dec 26 & MMT & BCS & 1730 Blue, 1430 Red & 2 &  Y &  N &  N &  N &  Y &  Y &  Y &  Y &  Y & None & \\
SDSS~J1350 & 2021 Apr 03 & NTT & EFOSC2 & 390, 440 & 3 &  N &  N &  N &  N &  Y &  Y &  Y &  Y &  Y & Unresolved & Observed with GR11 and GR 16 OG530\\
\hline
\end{tabular}
\end{adjustbox}
\begin{flushleft}
\textbf{Sources:}\\
1: \cite{wang_2011_TransientSuperstrongCoronala}\\
2: \cite{yang_2013_LONGTERMSPECTRALEVOLUTION} \\
3: This work\\
\smallskip
\textbf{Feature Classification:}\\
Y = Line is present, N = Line is not present, U = Presence of line is unclear.\\
- Indicates a line not in the wavelength coverage of a particular spectrum.\\
\smallskip
\textbf{Notes:}\\
* Identified by \cite{yang_2013_LONGTERMSPECTRALEVOLUTION} as non-variable.\\
$\dag$ Blue Channel Spectrograph.\\
$\square$ Spectrum included in the variable ECLE template \\
$\triangle$ Spectrum included in the non-variable ECLE template \\
Quoted resolutions were obtained from instrument specifications, or from the published source of the data\\
\end{flushleft}
\end{table}

\end{landscape}

\twocolumn
\section{ECLE Spectral Templates}
\label{Appendix:Spectral_Templates}

In order to look for observational signatures in the spectra of ECLEs that could be used to better distinguish between TDE and AGN related ECLEs based on a single spectroscopic observation in the future, and to provide reference spectra to be used in comparisons to identify new ECLEs in the future, we have used the original SDSS spectral sample to produce two median-combined ECLE template spectra. The following section outlines the construction process of these templates and their similarities and differences.

\subsection{Spectral Template Construction and Comparison}
\label{subsec:Spectral_Template_Analysis}

\begin{figure*}
    \centering
    \includegraphics[width=\textwidth]{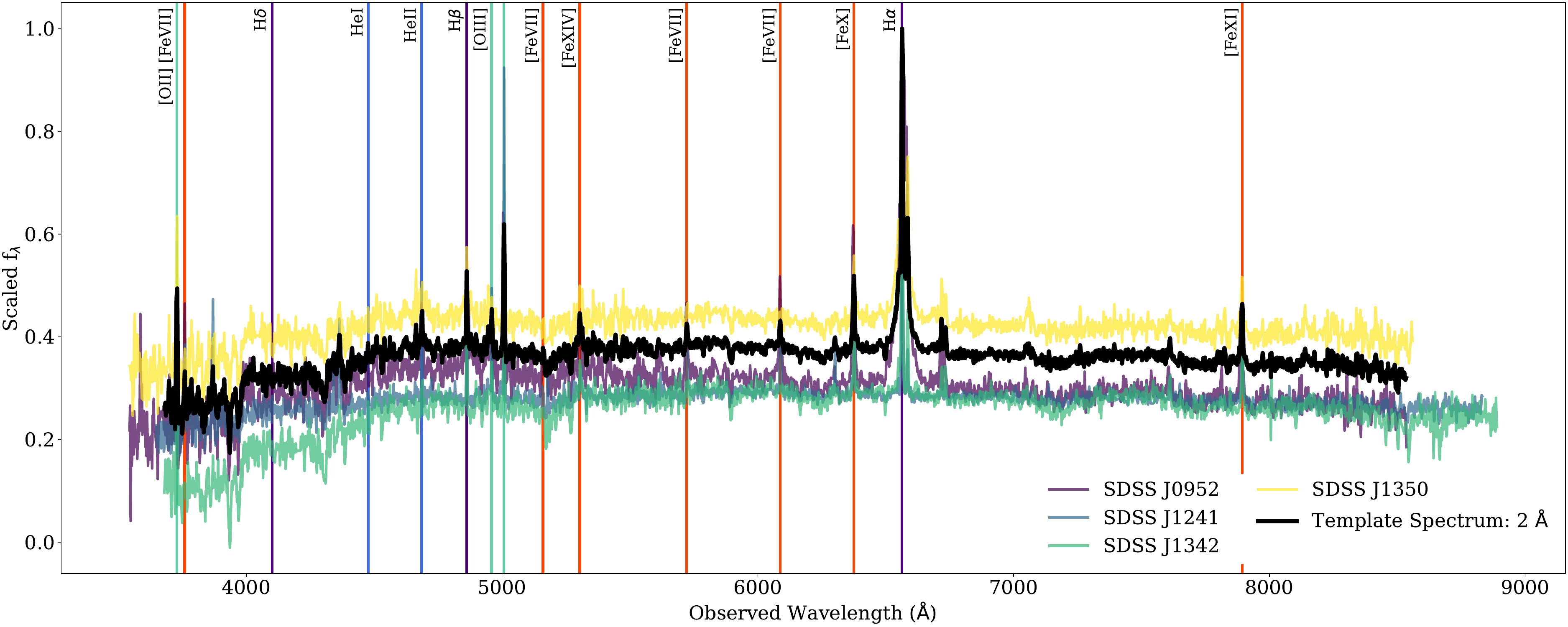}
    \caption{Variable ECLE template spectrum and constituent spectra. The Fe coronal lines from the stronger lines of \Fevii--\Fexiv\ are clearly present (in particular, \Fex\ and \Fexi) along with narrow \Heii\ emission and strong H$\alpha$ emission relative to \Oiii. }
    \label{fig:ECLE_Template_Construction_Variable}
\end{figure*}

\begin{figure*}
    \centering
    \includegraphics[width=\textwidth]{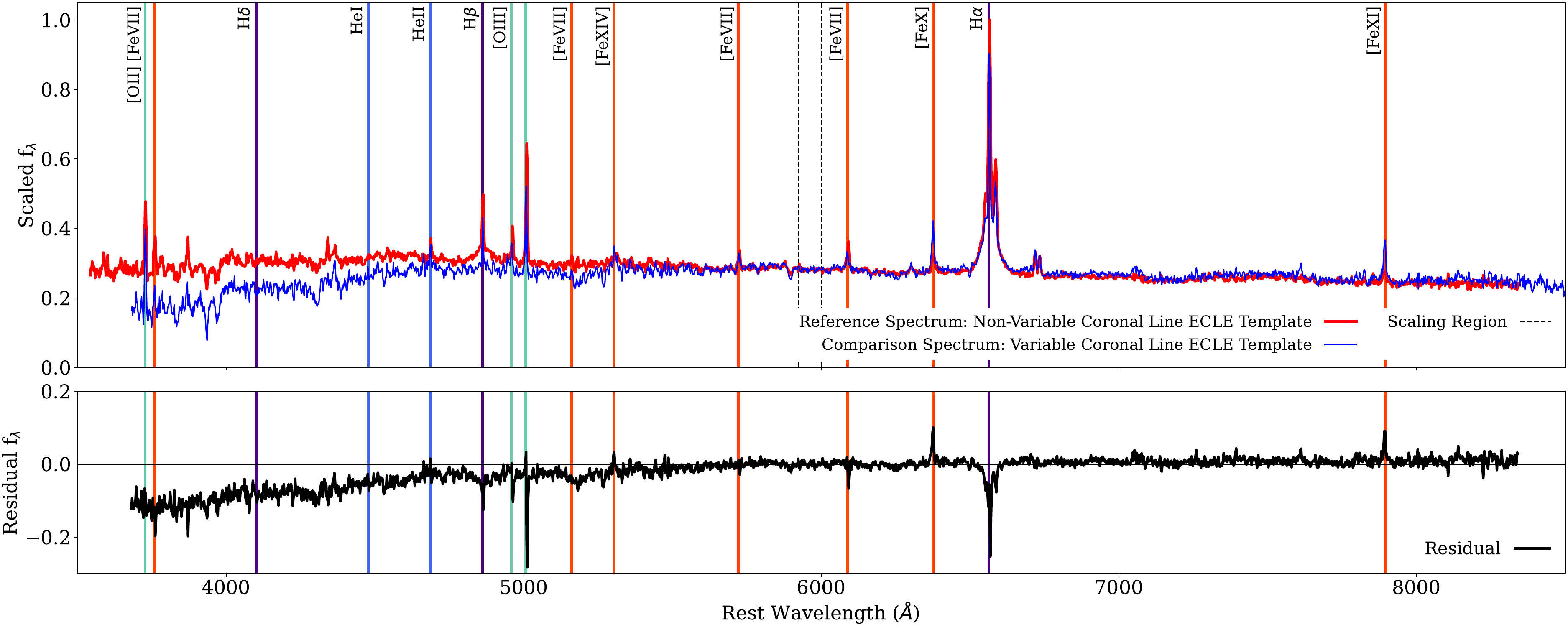}
    \caption{Comparison between the variable ECLE template spectrum and the template constructed from the non-variable ECLE SDSS spectra. Displayed residual produced through subtraction of the non-variable ECLE template from the variable template spectrum. Both spectra are displayed at a resolution of 2~\AA.}
    \label{fig:ECLE_Variable_Template_Comapred_to_Non_Variable_Template}
\end{figure*}

The first of these templates is composed of those objects showing variable coronal lines (those related to transient events rather than AGN activity), though excluding SDSS~J0748 as it is the only object with significant contamination from broad features produced by the still active TDE, with the resulting spectrum (and the spectra utilised in its construction) shown in Figure~\ref{fig:ECLE_Template_Construction_Variable}. Similarly, we have constructed a second template spectrum from the SDSS sample of ECLEs with non-variable coronal lines (i.e., SDSS~J0938 and J1055) with the comparison between the two ECLE templates shown in Figure~\ref{fig:ECLE_Variable_Template_Comapred_to_Non_Variable_Template}. The template spectra have been corrected for redshift and Milky Way extinction, but have not had additional underlying spectral components (e.g., non-thermal AGN activity) removed as we are most interested in comparisons between directly observed spectra for future immediate candidate classification purposes. Each of the included spectra is weighted equally in the comparison following normalisation and then median combination after the mean offset in the clean spectra region between the rest-frame wavelengths 5925--6000~\AA\ is taken into account. 

As the known sample of ECLEs is limited, the template spectra are composed of spectra from objects at different stages of evolution. The process of template construction will become more robust as more ECLEs are identified and observed with a faster cadence. We also note that our combined template has a broadened H$\alpha$ feature due to the inclusion of objects with residual broad TDE features. We consider the inclusion of these objects to be an acceptable compromise given the small number of objects overall, and the lack of more general continuum contamination. Whilst the two template spectra have similar overall spectral shapes, the variable ECLE template spectrum is redder than the non-variable template; this difference in shape is clear at wavelengths blueward of $\sim 5500$~\AA.

Narrow oxygen emission lines are of comparable relative strength in both template spectra. In contrast, Balmer emission features are seen to be both stronger and broader in objects with non-variable coronal lines. Whilst there is no clear distinction in the behaviour of the \Fevii\ emission lines between the spectral categories (evidenced by the minimal residual profiles at these line locations), \Fex\ and \Fexi\ emission are much more pronounced in objects with variable coronal lines. The same could also be said for \Fexiv\ emission when the continuum difference between the two template spectra in this region is considered, though the difference is not as clear.

We also use these template spectra to compare both ECLE categories to the SDSS cross-correlation template spectra of a range of galaxy classes, including quiescent galaxies, quasistellar objects (QSOs), and star-forming galaxies. We note here that Galaxy templates 1--3 represent increments on the continuum between fully quiescent galaxies (the `Early-Type Galaxy' template) and those with high star-formation rates (the `Late-Type Galaxy' template).\footnote{The templates used in this analysis were obtained from \url{https://classic.sdss.org/dr7/algorithms/spectemplates/index.html}} The best-matching comparison was determined using the Akaike information criterion \citep[AIC;][]{akaike_1974_NewLookStatistical}. 

\begin{figure*}
    \centering
    \includegraphics[width=\textwidth]{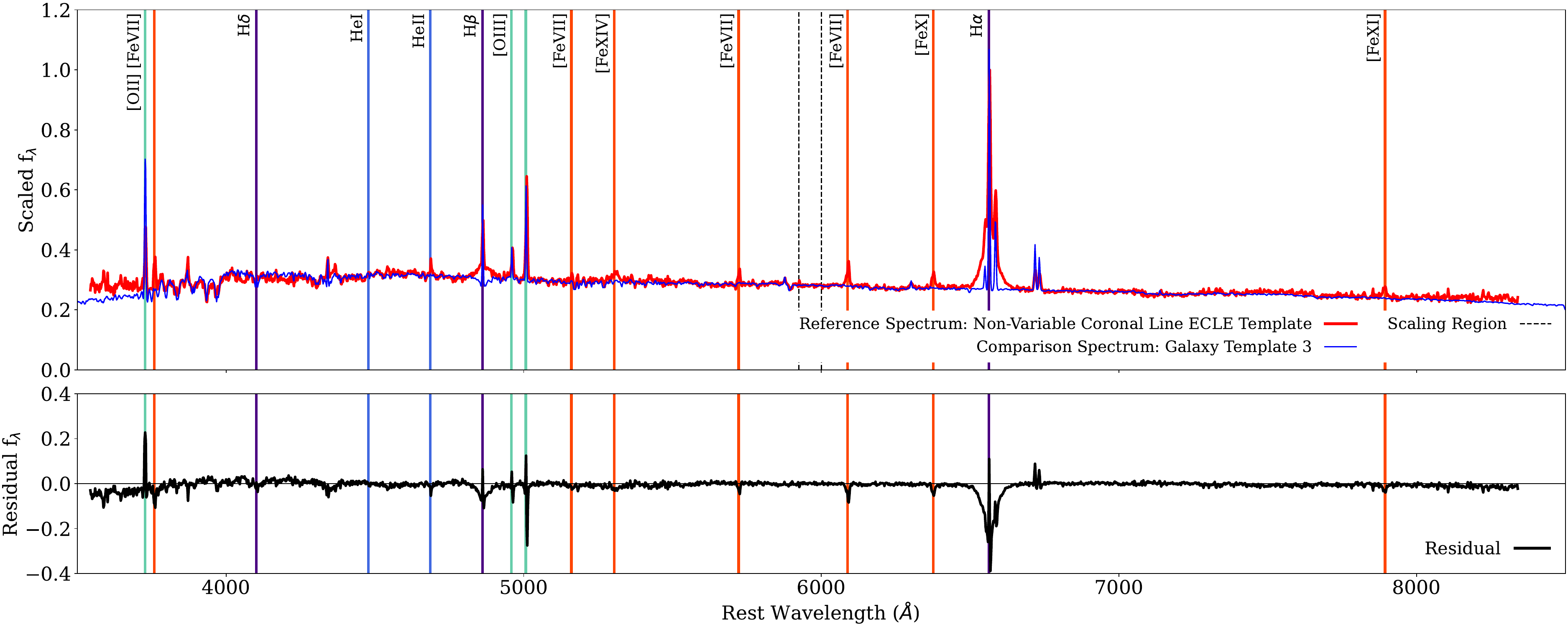}
    \caption{Comparison between the non-variable coronal line ECLE template and the best overall matched SDSS cross-correlation template spectrum: `Galaxy Template 3 / ID 26'.}
    \label{fig:ECLE_and_SDSS_Template_Comparison_Non_Variable}
\end{figure*}

\begin{figure*}
    \centering
    \includegraphics[width=\textwidth]{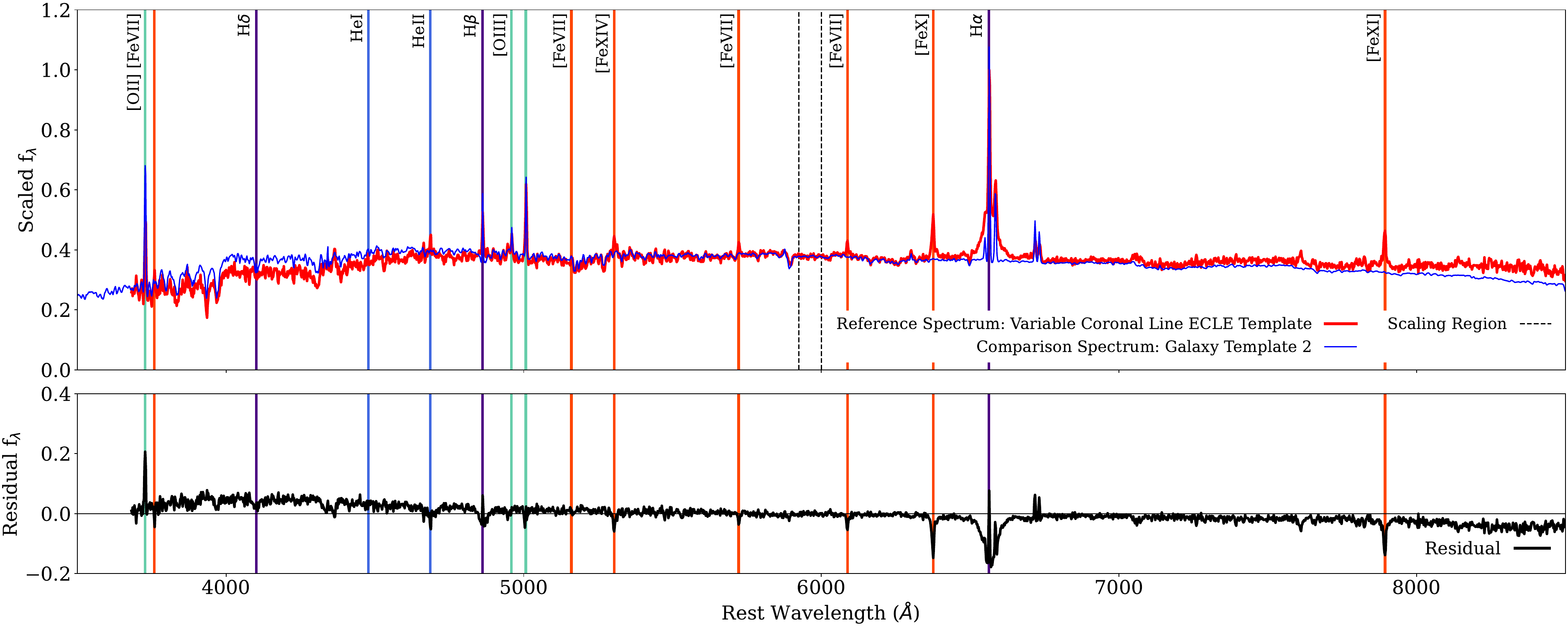}
    \caption{Comparison between the variable coronal line ECLE template and the best overall matched SDSS cross-correlation template spectrum: `Galaxy Template 2 / ID 25'.}
    \label{fig:ECLE_and_SDSS_Template_Comparison_Variable}
\end{figure*}

We explore the fit in the `blue' and `red' spectral regions, separated at 6000~\AA, to provide a more nuanced comparison , and in particular compare the spectra with and without the significantly differing blue continua and H$\alpha$ complexes. We present the comparison for the best overall match to the non-variable coronal line ECLE template spectrum in Figure~\ref{fig:ECLE_and_SDSS_Template_Comparison_Non_Variable} and the corresponding comparison for the variable ECLE template spectrum in Figure~\ref{fig:ECLE_and_SDSS_Template_Comparison_Variable}. This comparison was made using the templates rebinned to a range of resolutions (2, 5, 10, and 20~\AA) to explore how the use of lower resolution spectra would affect the comparisons and to determine if coronal lines would be observable in such spectra. We find that even at 20~\AA\ resolution the coronal lines  are still clearly distinguishable in our template spectra and in the residual patterns resulting from the comparisons. Note that the construction of the template boosts the S/N ratio of recurrent spectral features. Hence, the unambiguous presence of coronal-line signatures in similar low-resolution spectra of single objects is much less certain. 

\begin{figure*}
    \centering
    \includegraphics[width=\textwidth]{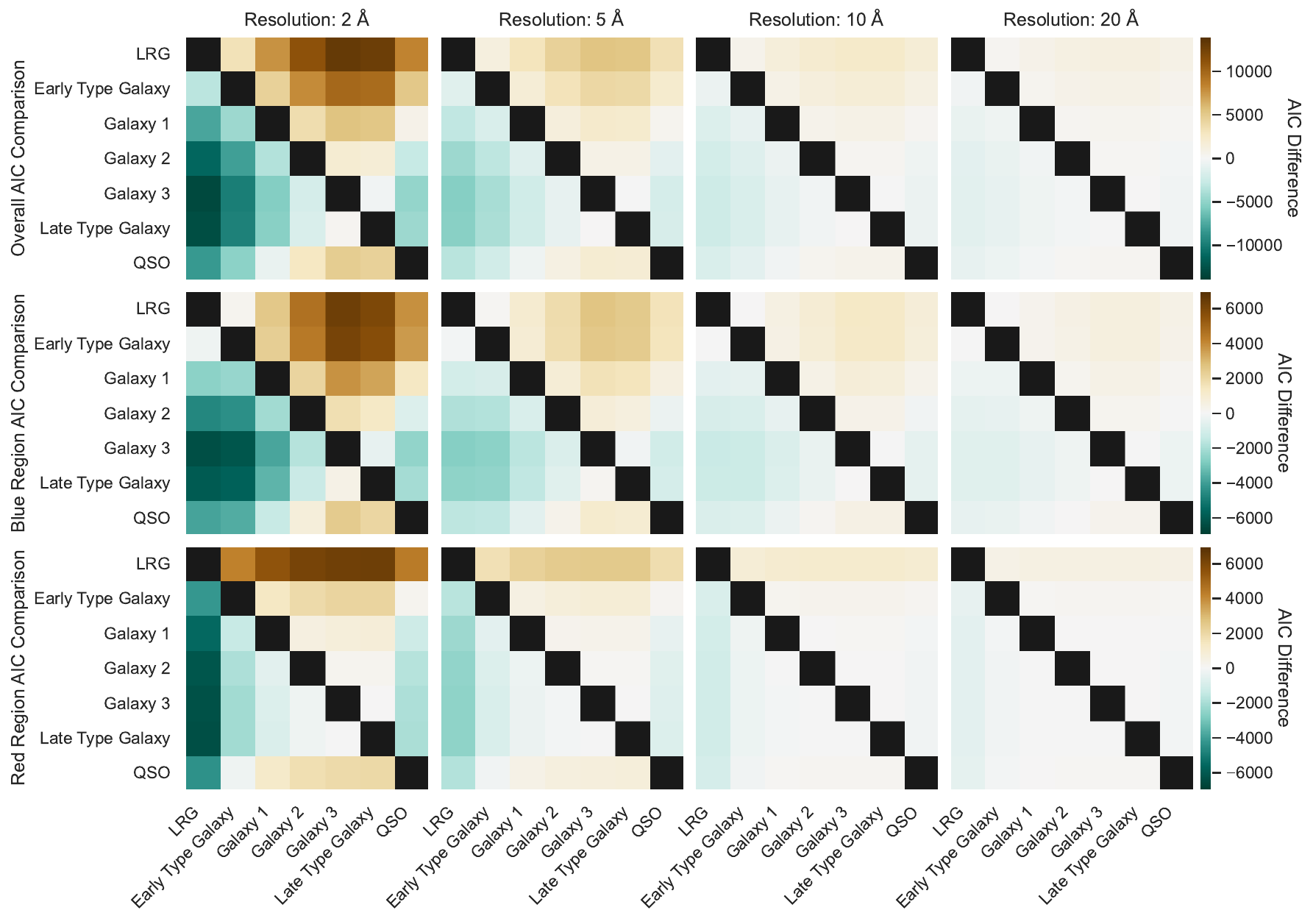}
    \caption{Difference matrix of the calculated AIC values for the non-variable ECLE template spectrum compared to the SDSS galaxy template set. When looking across a row, a green hue indicates that this template is a closer match than the other templates in the same column. Brown indicates the reverse. More intense colouration displays a higher preference for one model. The top panels are for the full spectrum comparison, middle panels for the blue ($<6000$~\AA) spectral region, and bottom panels for the red ($\geq 6000$~\AA) spectral region.}
    \label{fig:AIC_NonVariable}
\end{figure*}

\begin{figure*}
    \centering
    \includegraphics[width=\textwidth]{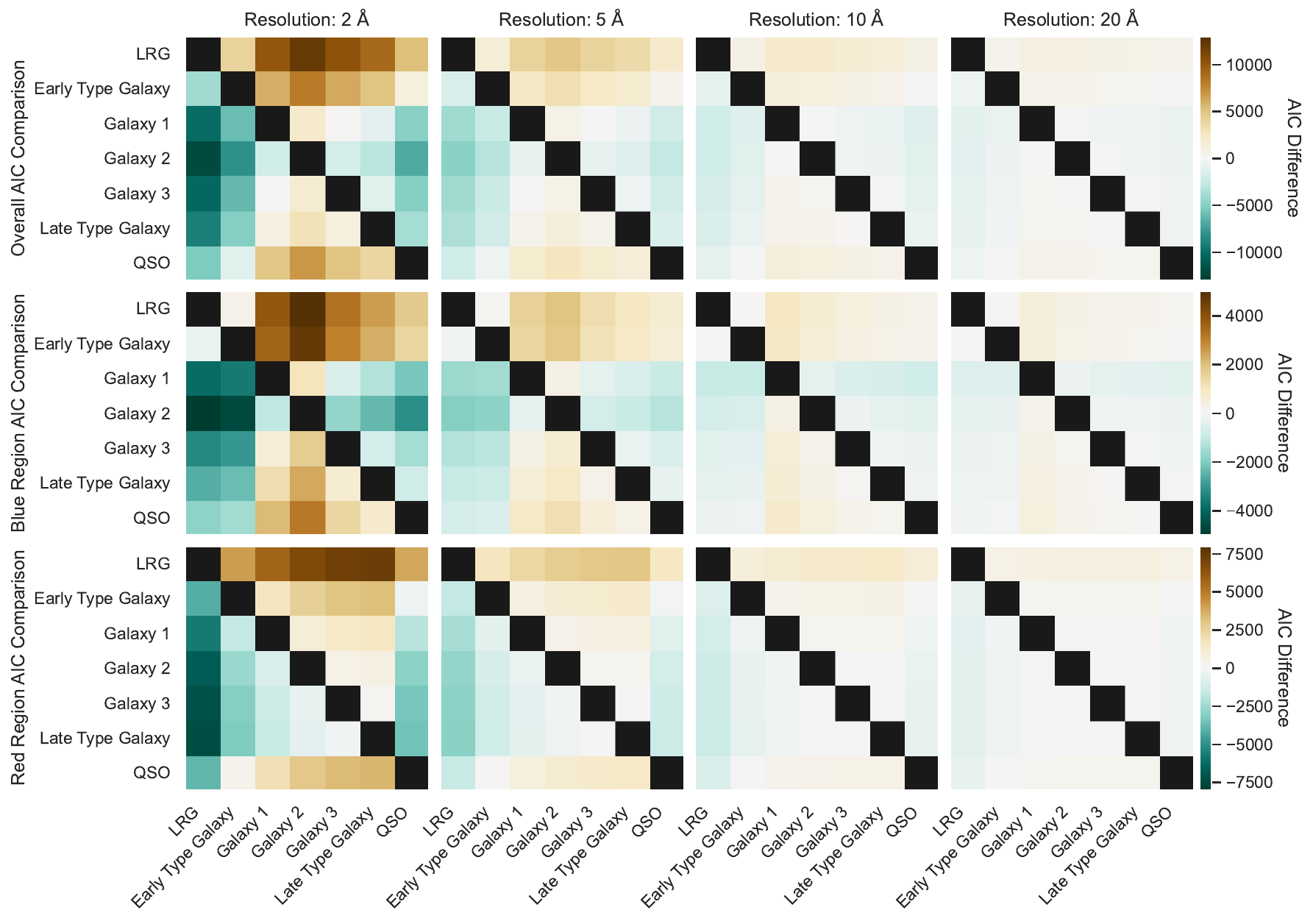}
    \caption{Difference matrix of the calculated AIC values for the variable ECLE template spectrum compared to the SDSS galaxy template set. When looking across a row, a green hue indicates that this template is a closer match than the other templates in the same column. Brown indicates the reverse. More intense colouration displays a higher preference for one model. The top panels are for the full spectrum comparison, middle panels for the blue ($<6000$~\AA) spectral region, and bottom panels for the red ($\ge 6000$~\AA) spectral region.}
    \label{fig:AIC_Variable}
\end{figure*}

The full comparison difference matrix for the ECLE template analysis are provided in Figure~\ref{fig:AIC_NonVariable} for the non-variable template, and likewise for the variable template comparison in  Figure~\ref{fig:AIC_Variable}. Both ECLE templates are found to have the best overall matches and blue-region matches to `intermediate' galaxies between the `Early' and `Late-type' galaxy spectra. The best overall and blue match to the non-variable ECLE template is found to be `Galaxy 2,' whilst the variable ECLE template is found to be most similar to `Galaxy 3.' The best comparison match in the red spectral region for both ECLE templates is the `Late-type' comparison template. This difference is driven by the improved match to the broad H$\alpha$ complex and better match to the red region's continuum shape. The poorest fits to both ECLE templates (both overall and when subdivided) are found with the `Luminous Red Galaxy' (LRG) and `Early-type' galaxy spectra. These galaxy types have significantly lower relative fluxes in the bluest and reddest regions whilst also lacking the strong Balmer and oxygen features observed in the ECLE spectra.

Whilst the differences observed in the best-matching template spectra could be taken to suggest a difference in the underlying stellar populations present in both groups of ECLE, the differences in overall spectral shape will also be influenced by the presence or absence of an AGN-generated spectral component and any residual broad features from the TDE flare. However, the difference between the generated templates does provide an additional tool for the classification of these objects as TDE- or AGN-driven based on a single spectroscopic observation, rather than through long-term follow-up observations. Additional analysis will be required to expand on this, and to refine the ECLE templates themselves with new observations of objects at different, and well-defined, stages of evolution.

As expected when rebinned to lower resolution, it becomes more difficult to identify the best-matching template, as distinguishing features are blurred by the lowered resolution. Additionally, whilst the best-matching templates remain `intermediate' galaxies, there is some variation as to what specific spectrum is preferred at the varying resolutions, highlighting the need for spectra to be obtained at as high a resolution as possible.

\FloatBarrier

\onecolumn
\section{MIR Power-Law Fitting Parameters}
\label{Appndix:MIR_Power_Law}

\FloatBarrier

\begin{figure*}
    \includegraphics[width=\textwidth]{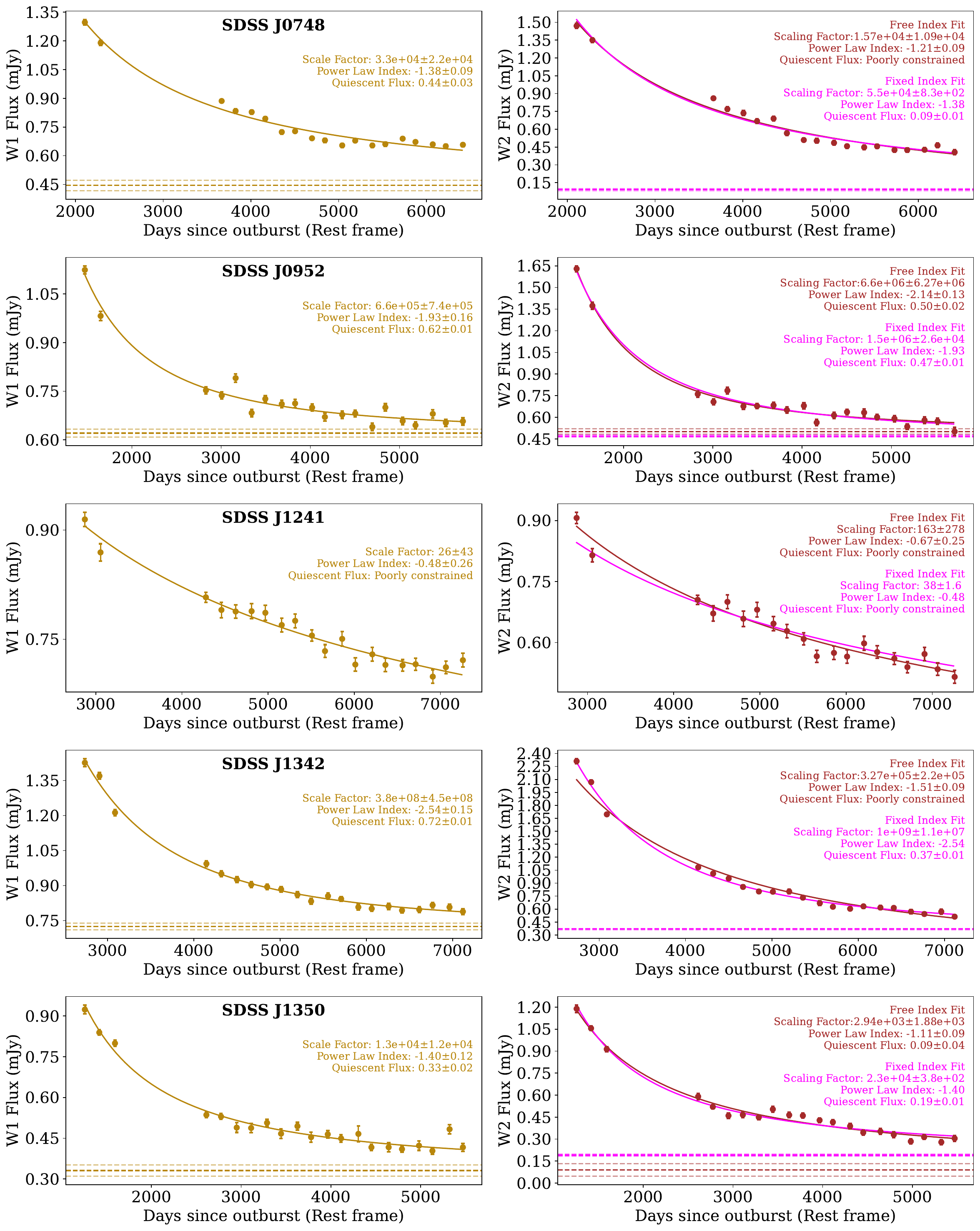}
    \caption{Power-law fits to the \textit{W1} {\it (left)} and \textit{W2} {\it (right)} photometry using Equation~\ref{Equation1}. Quiescent-flux values ($C$) are included when constrained by the fitting ($C > 0$ and $\Delta C < 0.15$) and shown by the dashed lines accompanied by the 1$\sigma$ uncertainties.}
    \label{fig:ECLE_MIR_Power_Law_Fit}
\end{figure*}

Here we present the results of the power-law fits to the MIR data of each of the objects with variable coronal lines. The results are detailed in Table~ \ref{tab:MIR_Power_Law_Fits} and presented visually with comparison to the raw data points in Figure~\ref{fig:ECLE_MIR_Power_Law_Fit}.

\begin{table*}
\caption{MIR power law fitting parameters}
\label{tab:MIR_Power_Law_Fits}
\begin{adjustbox}{width=0.9\textwidth}
\begin{tabular}{lllllllllll}
\hline
\multicolumn{11}{c}{Model : $f(t) = At\textsuperscript{$B$} + C$} \\ \hline
Object & $A$\textsubscript{W1} & $B$\textsubscript{$W1$} & $C$\textsubscript{$W1$} &  & $A$\textsubscript{$W2$ Free} & $B$\textsubscript{$W2$ Free} & $C$\textsubscript{$W2$ Free} &  & $A$\textsubscript{$W2$ Fixed} & $C$\textsubscript{$W2$ Fixed} \\ \hline
SDSS J0748 & 3.29e+04$\pm$2.25e+04 & -1.38$\pm$0.09 & 0.44$\pm$0.03 &  &1.57e+04$\pm$1.09e+04 & -1.21$\pm$0.09 & 0.00* & &5.51e+04$\pm$826 & 0.09$\pm$0.01 \\
SDSS J0952 & 6.58e+05$\pm$7.41e+05 & -1.93$\pm$0.16 & 0.62$\pm$0.01 &  &6.6e+06$\pm$6.27e+06 & -2.14$\pm$0.13 & 0.50$\pm$0.02 & &1.54e+06$\pm$2.63e+04 & 0.47$\pm$0.01 \\
SDSS J1241 & 25.9$\pm$42.9 & -0.48$\pm$0.26 & 0.34$\pm$0.25* &  &163$\pm$278 & -0.67$\pm$0.25 & 0.11$\pm$0.21* & &38.5$\pm$1.6 & 0.00* \\
SDSS J1342 & 3.84e+08$\pm$4.46e+08 & -2.54$\pm$0.15 & 0.72$\pm$0.01 &  &3.27e+05$\pm$2.2e+05 & -1.51$\pm$0.09 & 0.00* & &1.04e+09$\pm$1.14e+07 & 0.37$\pm$0.01 \\
SDSS J1350 & 1.35e+04$\pm$1.15e+04 & -1.40$\pm$0.12 & 0.33$\pm$0.02 &  &2.94e+03$\pm$1.88e+03 & -1.11$\pm$0.09 & 0.09$\pm$0.04 & &2.26e+04$\pm$382 & 0.19$\pm$0.01 \\\hline

\hline
\end{tabular}
\end{adjustbox}
\begin{flushleft}
\textbf{Notes:}\\
* Indicates a poorly constrained quiescent flux value.\\
For the `$W2$ Fixed' parameters the value of $B$ was set to match that determined by the $W1$ fitting.
\end{flushleft}
\end{table*}

\bsp	
\label{lastpage}
\end{document}